\documentclass[fleqn,usenatbib]{mnras}

\pdfoutput=1

\usepackage{newtxtext,newtxmath}

\usepackage[T1]{fontenc}
\usepackage{ae,aecompl}

\usepackage{graphicx}	
\usepackage{amsmath}	
\usepackage{amssymb}

\usepackage{appendix}
\usepackage{subfig}

\title[Double peaked meteor light curves]{Properties of meteors with double peaked light curves}

\author[D. Subasinghe and M. Campbell-Brown]{
Dilini Subasinghe,$^{1}$\thanks{E-mail: dsubasi@uwo.ca}
Margaret Campbell-Brown,$^{1}$
\\
$^{1}$Department of Physics and Astronomy, The University of Western Ontario, London, Ontario, Canada, N6A 3K7
}

\date{Accepted XXX. Received YYY; in original form ZZZ}

\pubyear{2019}

\begin{document}
\label{firstpage}
\pagerange{\pageref{firstpage}--\pageref{lastpage}}
\maketitle

\begin{abstract}
Twenty-one meteors showing double peaked light curves were analysed with observations collected with the Canadian Automated Meteor Observatory tracking system. Each event has orbital information, photometry, and at least one high-resolution observation. Two distinct light curve shapes were found: sudden double peaked curves, and smooth double peaked curves. The sudden peaked curves were produced by objects on asteroidal orbits and mostly showed noticeable fragmentation, while the smooth peaked curves were produced by cometary meteoroids and predominantly showed little to no visible fragmentation. An attempt to model these meteors as single bodies with two chemical components was unsuccessful, implying that fragmentation must be included in meteoroid ablation models.

\end{abstract}

\begin{keywords}
meteors, meteoroids
\end{keywords}



\section{Introduction}
It is well known that meteoroids do not behave as the classical ablation model predicts, mainly because of fragmentation \citep{Jacchia1955,Verniani1965,Weryk2012,Subasinghe2016}. Observations are used as constraints in ablation models to improve our understanding of the processes occurring. The shape of a meteor light curve (the light produced as a function of time or height) is sometimes used to infer the grain size and distribution of fragments within the meteoroid \citep[e.g.][]{Beech2003,CampbellBrown2004,Borovicka2007}. The meteoroid origin, determined from its orbit, can suggest physical properties such as strength and porosity, and high-resolution optical observations independently confirm fragmentation processes. An analysis of these three types of observations (photometry, orbit, and high-resolution optical observations) was completed by \cite{Subasinghe2016} to investigate meteoroid behaviour. The results of that study found that meteoroids do not behave as expected: more than 90\% of meteors observed in the high-resolution camera showed obvious fragmentation in some form; meteoroids that showed little to no visible fragmentation had mostly symmetric light curves rather than the expected classical late peaked light curve; and about a third (but likely fewer) of the meteor light curves showed unusual shapes. Asteroidal and cometary meteoroids were found to behave very similarly, in terms of light curve shape and fragmentation behaviour. \newline

Despite the many constraints provided by these observations, it is still difficult to model meteoroid ablation. The classical meteoroid ablation model predicts the behaviour of a solid, single-bodied object that does not fragment: this is the simplest model and it does not work well \citep[e.g.][]{Jacchia1955, Koten2004} because most meteoroids fragment \citep{Weryk2013a,Subasinghe2016}. Even when fragmentation is included in models and a good match is obtained for the light curve and deceleration, high-resolution observations from the Canadian Automated Meteor Observatory (CAMO) tracking system show the model limitations: \citet{CampbellBrown2013} modelled ten meteors with two different fragmentation models and both predicted wakes which were both too long and too bright compared to the observed wakes. \cite{CampbellBrown2017} attempted to model an apparently single-body meteoroid which was one of the ten from \citet{CampbellBrown2013}, and found that a single-body ablation model was not able to reproduce the observations: the modelled light curve was too bright over too wide a range of heights compared to the observed curve. An ablation model in which the meteoroid fragments with many small bursts was more successful at reproducing both the light curve and the high resolution behaviour, indicating that it is possible to have fragmentation which does not produce a long visible wake. \newline

The dustball model was proposed in 1975 by \citeauthor{Hawkes1975} who suggested that meteoroids are composed of two fundamental parts: grains with a high boiling temperature, and an organic glue with a lower boiling temperature, that holds the grains together. It has been suggested that the glue may be organic \citep{Murray2000,Roberts2014}, or may contain volatile sodium \citep{Borovicka2007}, and in some cases is assumed not to contribute to the observed optical emission. The dustball model has been modified to include different fragmentation mechanisms and has been used successfully in recent years to reproduce meteor observations \citep{CampbellBrown2004,Borovicka2007}. \newline

Many studies have found that meteors show mainly symmetric light curves \citep[e.g.][]{Flemming1993,Koten2004,Subasinghe2016}. Symmetric light curves can be explained by fragmentation with the dustball model. However, single-peaked light curves are not exclusively observed: some meteors have multi-peaked curves. The mechanisms behind double peaked curves are not well studied. \cite{Subasinghe2016} found that 18\% of their analysed meteor light curves showed a double peaked shape, but suggest that the real contribution is lower due to false inflection points at the edges of the light curves due to the fitting method. Twenty-one double peaked meteor light curves were studied by \cite{Roberts2014}, whose modelling work suggested that the light curves could be reproduced using a dustball model where some grains are released before intensive ablation begins, followed by a later release of larger grains. The range of grain masses used was 10$^{-12}$ to 10$^{-5}$ kg. Many of the events studied by \cite{Roberts2014} showed a rounded first peak, followed by a sharp increase in luminosity for the second peak. The model used in that work produced an almost instantaneous increase in brightness at the onset of the second peak. In this work, we look at meteors with double peaked light curves to study their physical properties.

\section{Observations}
\subsection{Equipment}
The observations for this work were collected with the Canadian Automated Meteor Observatory (CAMO) tracking system \citep{Weryk2013a}. This is a two-station system, with a narrow-field and wide-field camera at each station. The two stations are located at Elginfield, Ontario, Canada (43.193$^\circ$N, 81.316$^\circ$W) and Tavistock, Ontario, Canada (43.265$^\circ$N, 80.772$^\circ$W) and are about 45 km apart. The wide-field cameras have a field of view of 26$^\circ$ $\times$ 19$^\circ$, and run at 80 frames per second. The narrow-field cameras have a field of view of 1.5$^\circ$ $\times$ 1$^\circ$, and run at 110 frames per second. In late October 2017 the Elginfield cameras were upgraded, and the field of view for the wide-field camera is now 34$^\circ$ $\times$ 34$^\circ$, and the narrow-field is 2$^\circ$ $\times$ 2$^\circ$. \newline

From the wide-field observations the meteor light curve can be obtained, as well as the meteoroid trajectory (and orbit). The narrow-field cameras have resolutions as high as 3 metres per pixel at a range of 100 km, and are capable of capturing the fragmentation behaviour of the meteoroid.

\subsection{Data}
We used both wide and narrow-field observations from the CAMO system. The All Sky and Guided Automatic Realtime Detection (ASGARD) software detects meteors in the wide-field camera in real time, and directs a pair of mirrors to guide the meteor light into the narrow-field camera. ASGARD also provides a solution (trajectory and photometry) for each recorded wide-field meteor event. This is done by comparing pixels frame by frame, searching for clusters where the intensity exceeds five standard deviations above the mean background. If in at least three consecutive frames, a minimum of six pixels in an 8 $\times$ 8 pixel region have intensities greater than the predefined threshold, a meteor has been detected. ASGARD uses centroiding to select the head of the meteor in each frame and these position picks are used in MILIG, a least squares method trajectory solver \citep{Borovicka1990}. ASGARD also masks out meteor pixels with intensities greater than the background to determine the meteor photometry. The meteor photometry and astrometry are calibrated using manually made photometric and astrometric plates, in which a user selects typically more than 20 stars in the field of view, which are calibrated against those from the Sky2000v4 catalogue \citep{Myers2001}. More than 7000 ASGARD reductions were searched for meteors with double peaked light curves. Our initial search fit a curve to the observed wide-field photometry data, and selected those with three turning points.  \newline

\begin{figure}
    \centering
    \includegraphics[width=\columnwidth]{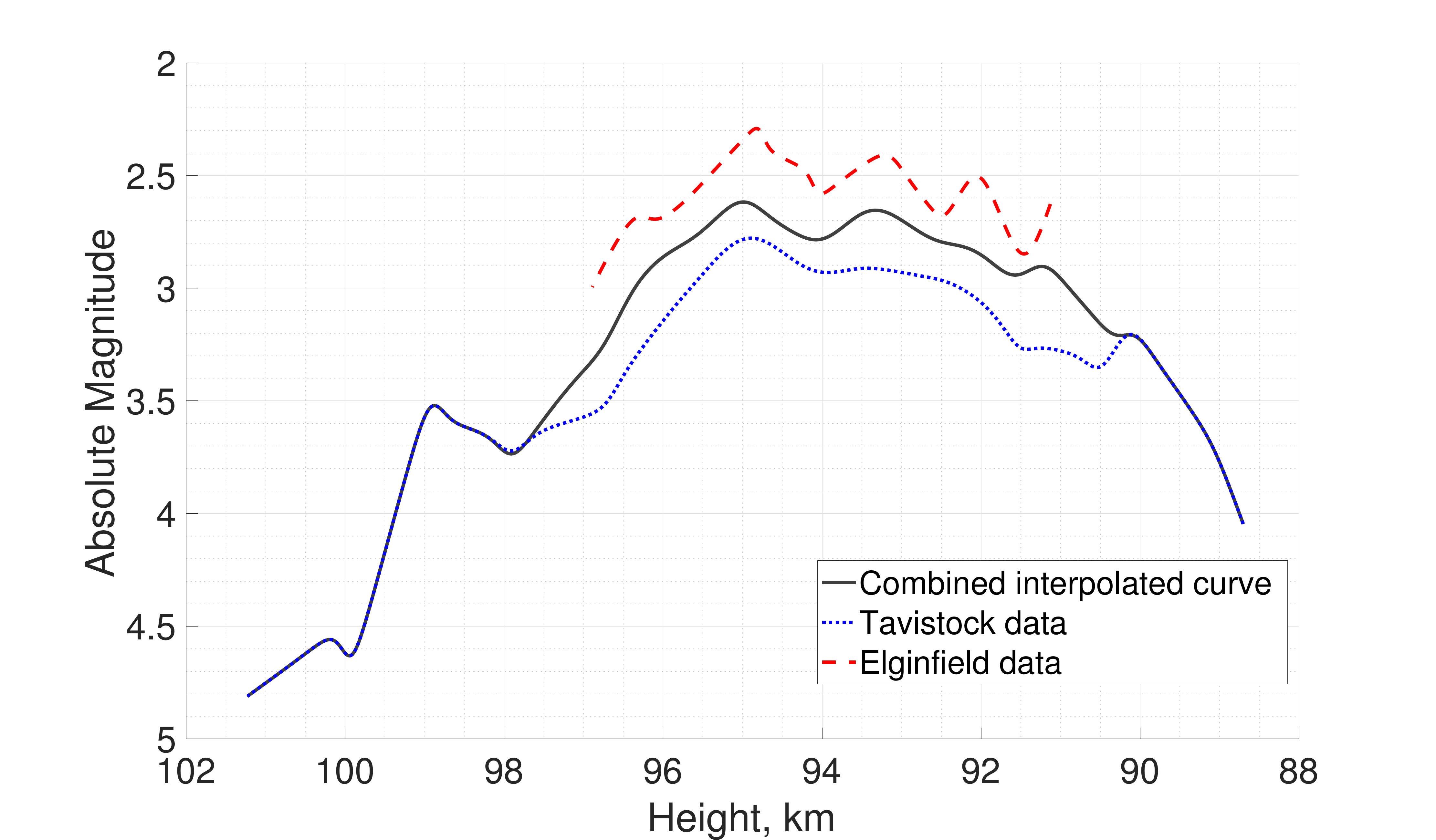}
    \includegraphics[width=\columnwidth]{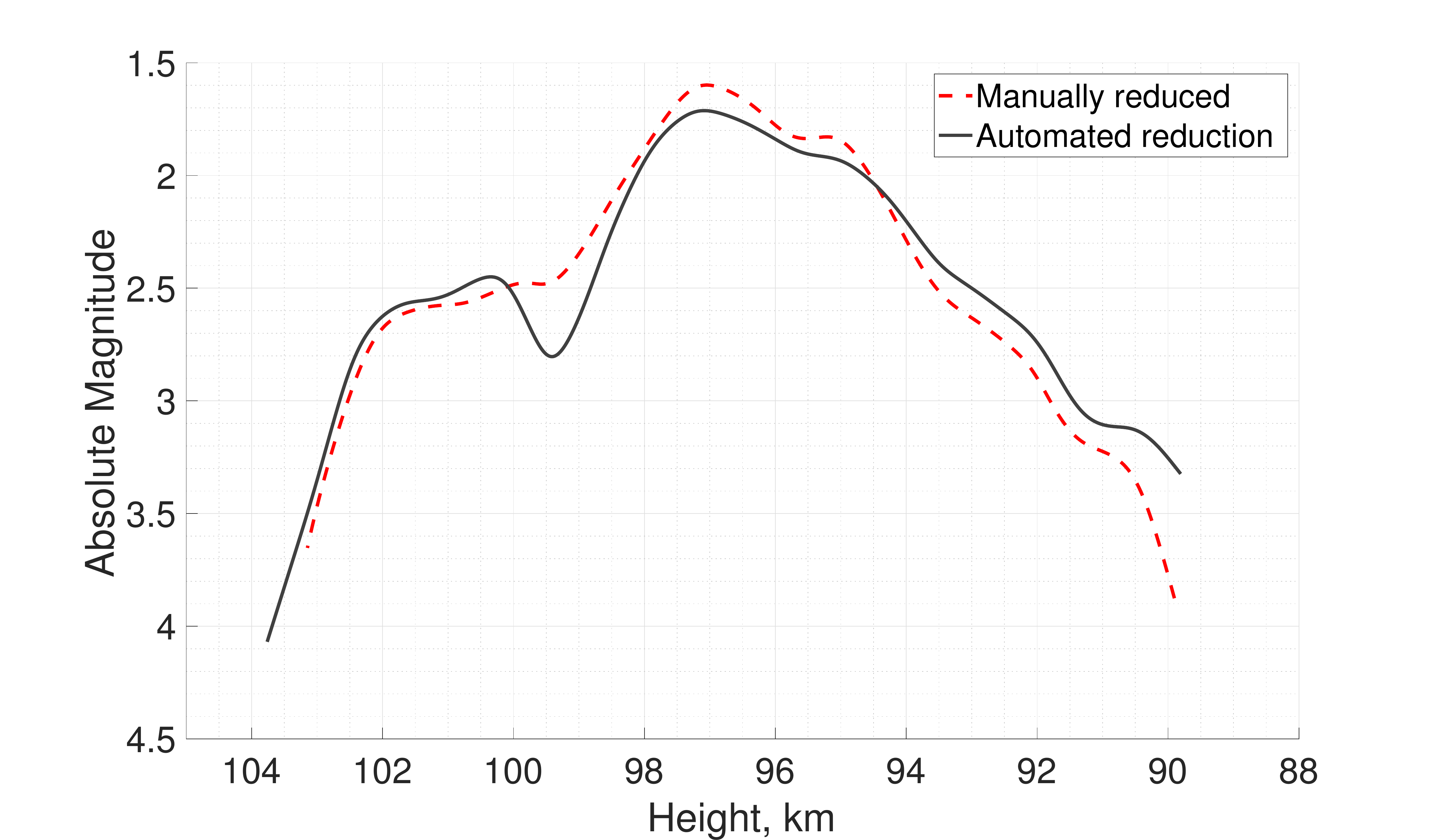}
    \caption{Examples of mis-categorized meteor light curves based on automatic ASGARD reductions. The top figure shows the Tavistock and Elginfield interpolated curves based on the observed data points, in red and blue. The two station observations disagree by up to half a magnitude. The gray curve is the interpolated curve based on both observations, and shows 9 turning points. Many of those turning points are within 0.5 magnitude and are likely due to noise in the system. The bottom figure shows two curves: the solid gray line shows the interpolated curve based on both station observations determined by ASGARD, and the red curve shows the manually reduced interpolated curve, showing that the trough was due to poor ASGARD photometry. }
    \label{fig:bad_picks}
\end{figure}

After this search, the light curves of the potential candidates (1257 out of 7160 events) were manually checked to remove events whose automatic reductions were either erroneously flagged as double peaked due to noise or false inflection points at the start or end of the curve due to the fitting method, or events whose double peak was somewhat ambiguous. The remaining potential events (53 total) were then reduced using the software program METAL (METeor AnaLysis) \citep{Weryk2012}. METAL works in a similar way to ASGARD, but needs a human operator, which avoids the errors sometimes made by ASGARD in accidentally selecting a hot pixel, or nearby star. Any event that showed a double peak in its manually reduced wide-field photometry data (rather than oscillations in the light curve due to noise in the system, clouds, or poor photometric calibrations) had its narrow-field observation further reduced in the software program mirfit (33 events total). Two events are shown in Fig.~\ref{fig:bad_picks}: one shows an ASGARD reduction initially classified as multi-peaked, but was then rejected on the grounds that the two stations did not agree (the oscillations were due to noise);  and a second event that was classified as double peaked based on the automated light curve, but proved to be single peaked when reduced manually.

The software program mirfit allows observations collected with the high-resolution narrow-field cameras to be reduced. Meteors are manually reduced in mirfit in a similar way to METAL: the position of the head of the meteor is selected and the light from the meteor is masked out. Both the position of the meteor in the field and the position of the mirrors at the time of exposure are used in determining the angular position of the meteor in the sky. The meteor trajectory is determined using MILIG, and the log of the sum of the pixel brightness is determined. The absolute brightness of the meteor cannot be determined in mirfit due to the small field of view: there are typically not enough visible stars to compute a photometric calibration. The mirfit brightness values can be calibrated against the manually obtained absolute magnitude METAL light curve to determine the absolute magnitude mirfit light curve. In this work, we were not concerned with the mirfit photometry, so much as we were with the meteoroid fragmentation behaviour: did the meteoroid break into many small pieces, or a few large pieces? This can provide insight into the strength of the object and the processes it undergoes. The combined METAL and mirfit photometry for each event was manually assessed to determine whether to include the event in the final analysis (visually, how strong is the double peak? Is there good overlap between the wide and narrow-field observations? How well tracked was the event in the narrow-field camera?). This reduced our set of 33 events to 21. \newline

These twenty-one double peaked events with their relevant parameters are listed in Table~\ref{tab:events}. Not all meteor events had two station narrow-field observations, meaning some fragmentation behaviour analysis was completed on single-station data only. 

\begin{table*}
\caption{Parameters of the double peaked meteor events. The given Tisserand parameter is determined with respect to Jupiter, the peak magnitudes and related heights are based on curves interpolated to the observations, and $z$ represents the zenith angle}.
\label{tab:events}
\begin{tabular}{ccccccccccc}
\hline
Event & v$_{i}$ & $\alpha_{g}$ & $\delta_{g}$ &  T$_{\rm{J}}$ & Peak magnitudes & Heights at peak magnitudes & $a$ & $e$ & $i$ & $z$\\
& km s$^{-1}$ & ($^\circ$) & ($^\circ$) & && km & AU & & ($^\circ$) &($^\circ$) \\
\hline
20100920\_084608 & 60.30 & 50.10  & 19.01 & 0.8  & 1.1, 1.0  & 97.9, 92.7  & 3.7   & 0.94 & 178.00  & 24.9\\
20101009\_060641 & 17.82 & 17.30  & 7.86  & 5.1  & 3.4, 4.7  & 81.8, 77.8  & 1.2   & 0.48 & 0.23  & 32.7\\
20110830\_071519 & 59.88 & 27.44  & 43.59 & 0.4  & 1.7, 1.7  & 103.7, 99.0 & 5.8   & 0.88 & 119.86 & 15.9\\
20111005\_084734 & 66.36 & 125.55 & 35.06 & 0.4  & 1.0, 0.8  & 101.5, 96.5 & 4.0   & 0.81 & 150.40 & 47.0\\
20111104\_082320 & 69.77 & 131.65 & 36.16 & -0.6 & 1.0, 0.6  & 190.8, 96.4 & 13.9  & 0.93 & 149.37 & 34.5\\
20120912\_063437 & 38.01 & 14.54  & 20.28 & 4.1  & 3.2, 2.0  & 94.6, 90.9  & 1.4   & 0.92 & 36.34 & 23.5 \\
20130816\_081853 & 35.79 & 347.53 & -8.28 & 3.1  & 3.6, 2.1  & 93.0, 89.9  & 2.0   & 0.91 & 5.50  & 53.8\\
20131114\_095829 & 57.55 & 151.20 & 53.30 & 1.5  & 1.0, 2.4  & 105.6, 96.9 & 2.8   & 0.66 & 110.95 & 21.4\\
20140421\_051749 & 40.74 & 244.33 & 20.64 & 1.5  & 1.8, 0.6  & 92.4, 88.9  & 5.3   & 0.89 & 57.27 & 38.4\\
20140904\_070414 & 31.94 & 8.17   & -10.65 & 4.1 & 2.1, 0.8  & 93.1, 90.1  & 1.5   & 0.83 & 19.80 & 52.9\\
20141102\_102214 & 55.69 & 146.98 & 51.69 & 2.9  & 2.4, 2.1  & 100.1, 95.0 & 1.6   & 0.36 & 113.91 & 22.3 \\
20150121\_053147 & 23.20 & 115.56 & 15.41 & 3.0  & 0.8, 1.1  & 83.4, 79.6  & 2.5   & 0.74 & 3.68 & 27.0\\
20150121\_084050 & 38.37 & 147.88 & 39.77 & 1.0  & 2.8, 4.4  & 88.0, 81.1  & 12.7  & 0.97 & 35.96 & 15.4\\
20150522\_071644 & 62.73 & 314.90 & 16.24 & 0.3  & 2.3, 1.5  & 101.4, 96.2 & 5.7   & 0.83 & 124.58 & 48.5\\
20150925\_083237 & 69.10 & 99.38  & 37.03 & -0.1 & 1.5, 1.8  & 101.8, 97.0 & 5.2   & 0.81 & 155.67 & 37.3\\
20151210\_082756 & 66.59 & 151.30 & 28.54 & -0.5 & 0.9, 2.4  & 106.2, 98.7 & 16.6  & 0.96 & 146.83 & 26.6\\
20160808\_083545 & 60.41 & 42.69  & 57.70 & -0.4 & 2.8, -0.6 & 97.6, 91.6  & 167 & 0.99 & 112.23 & 27.4\\
20170824\_023951 & 24.68 & 329.09 & -3.97 & 3.1  & 2.0, 2.2  & 84.9, 80.0  & 2.4   & 0.75 & 5.85 & 55.8\\
20170924\_070328 & 35.63 & 1.20   & 24.30 & 0.9  & 2.6, 4.5  & 92.7, 83.1  & 31.9  & 0.99 & 27.48 & 28.0\\
20171016\_093056 & 40.87 & 47.22  & 20.57 & 2.6  & 3.0, 4.7  & 91.8, 80.9  & 2.3   & 0.97 & 9.44 & 38.8\\
20171201\_103027 & 66.98 & 149.74 & 18.00 & 1.2  & 0.6, 0.8  & 101.4, 96.6 & 2.4   & 0.67 & 169.69 & 26.4 \\
\hline
\end{tabular}
\end{table*}

\section{Results and Analysis}
\subsection{Light Curves}
\begin{figure}
\includegraphics[width=\columnwidth]{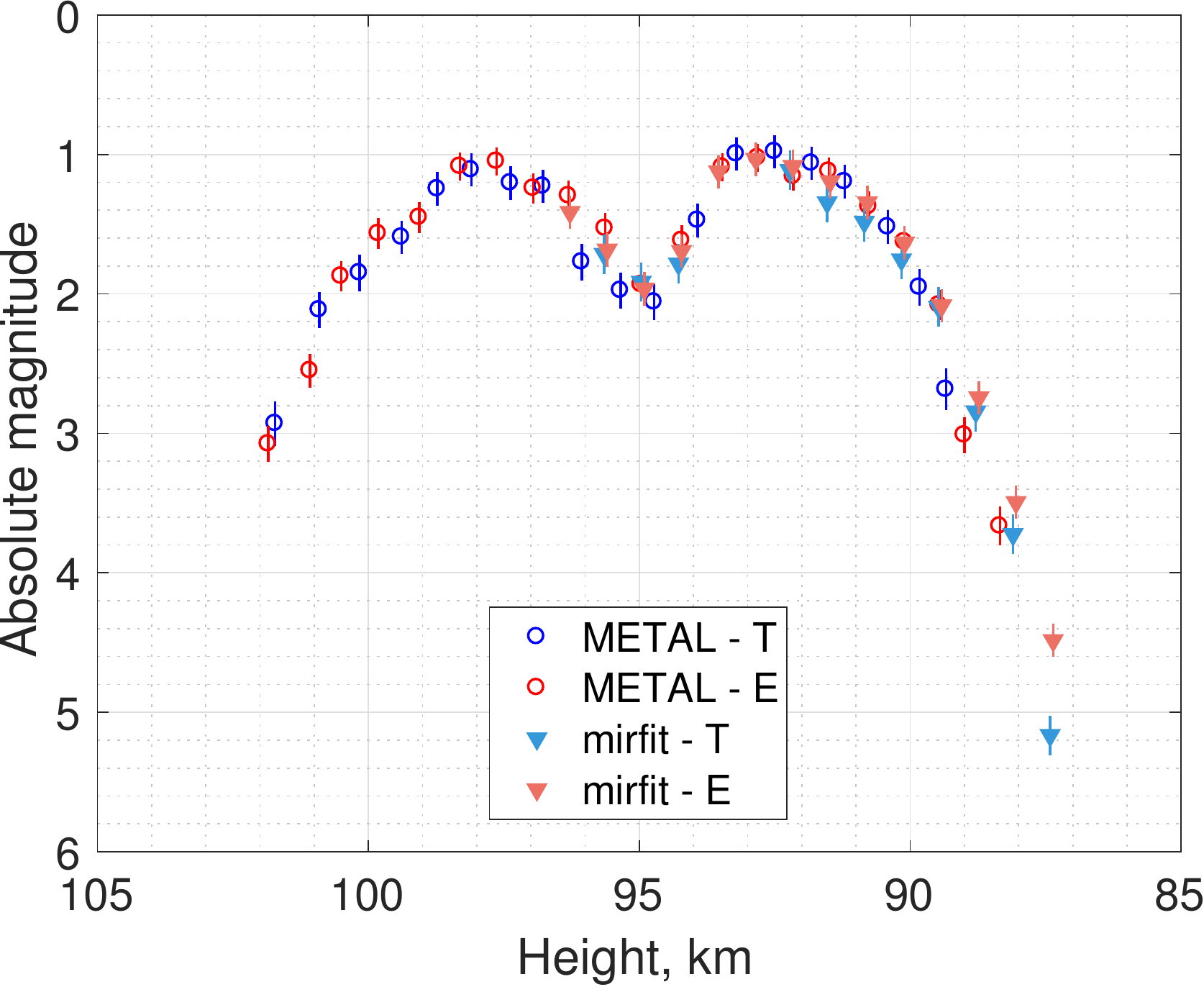}
\caption{A typical `double peaked' curve. This event was recorded on September 20, 2010 at 08:46:08 UTC.}
\label{fig:double}
\end{figure}

\begin{figure}
\includegraphics[width=\columnwidth]{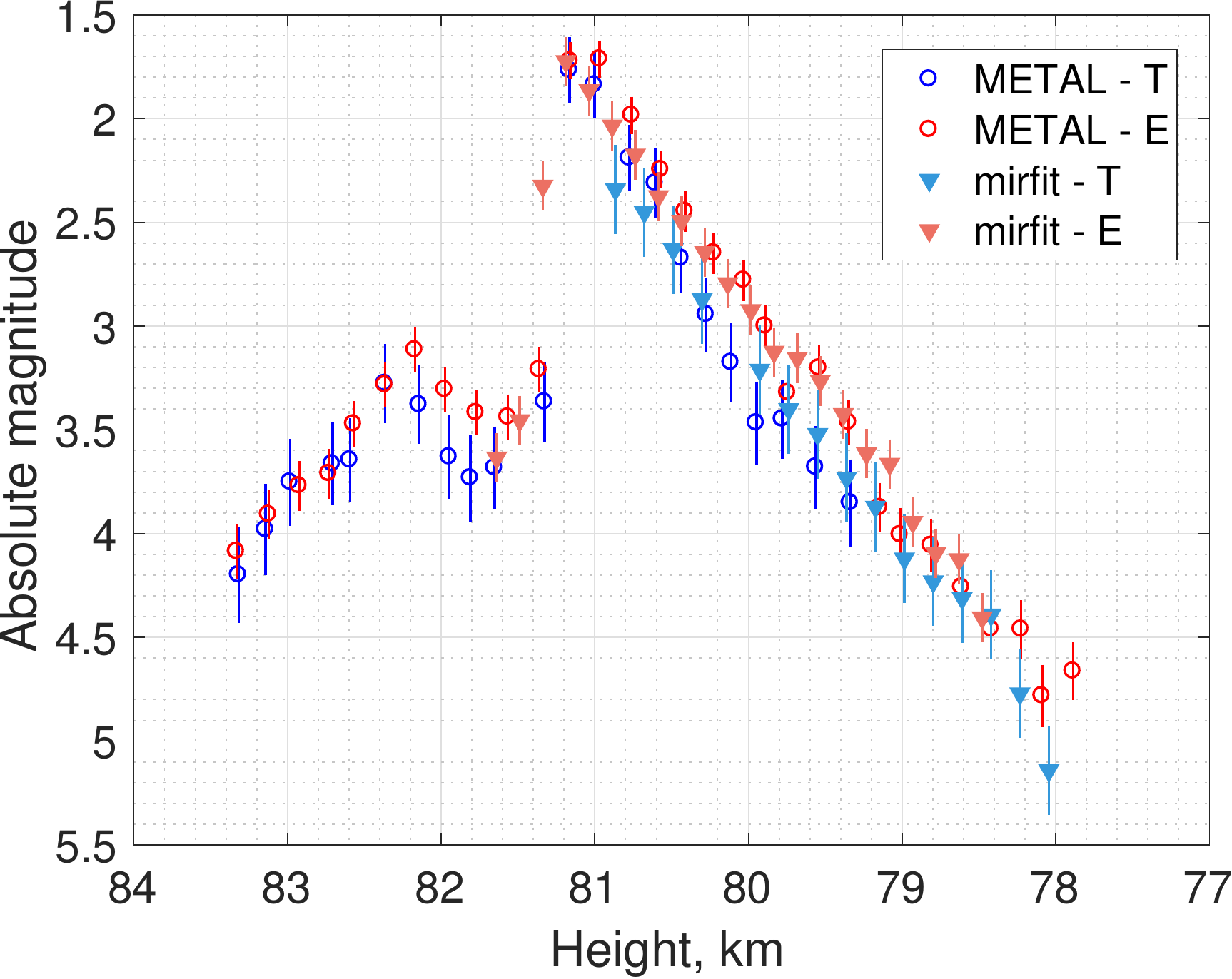}
\caption{A typical `sudden peaked' curve, recorded on October 9, 2010 at 06:06:41 UTC.}
\label{fig:sudden}
\end{figure}

Two different categories of multi-peaked light curves became obvious during analysis: smooth double peaked, and sudden double peaked curves. Smooth peaked curves typically showed a shape similar to that shown in Fig.~\ref{fig:double}, increasing gradually in brightness in both peaks, while sudden peaked curves showed a much steeper, more instantaneous increase in brightness for the second peak, as shown in Fig.~\ref{fig:sudden}.

The steepness of the second peak was used to classify the light curves as either smooth double peaked, or sudden double peaked. Any event with a second peak rising slope greater than 1 mag/km was sorted into the sudden peaked group. The sudden peaked meteor events are listed in Table~\ref{tab:sudden}, along with a brief description of the fragmentation behaviour observed in the high-resolution narrow-field cameras. The smooth double peaked events are listed in Table~\ref{tab:double}.

\subsection{Narrow-field observations}
The fragmentation behaviour of each meteor was studied with the high-resolution narrow-field observations. Two station narrow-field observations were not available for all meteors, either due to poor weather or tracking issues at one of the stations. At a single station, the meteor events had at least 8 and sometimes more than 50 frames of narrow-field observations. A visual analysis of the observations was performed, and a description of the fragmentation behaviour is listed in Tables ~\ref{tab:sudden} and~\ref{tab:double}. \newline

\begin{figure}
    \centering
    \includegraphics{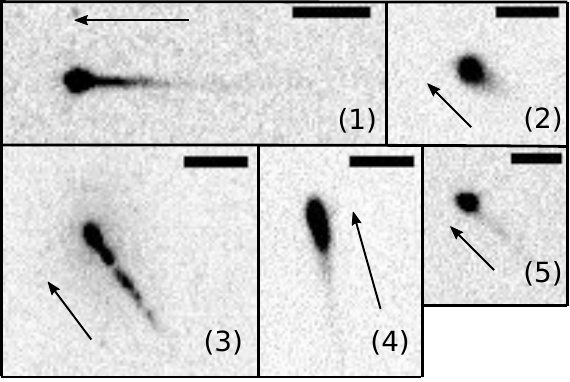}
    \caption{The five morphologies that were observed in the meteor events. Each example is from a different event, showing a single frame of the narrow-field observation, with the colour inverted to show detail. (1) shows a meteor with a long visible wake; (2) shows a meteor with a short faint wake; (3) shows a meteor with fragmentation along the line of motion; (4) shows a meteor with an elongated head and faint wake; and (5) shows a meteor that has faint wake and looks almost single bodied. The scale bar in each frame is 100 m, and the arrow indicates the direction of motion.}
    \label{fig:nf_images}
\end{figure}

The meteor events were visually classified into one of five fragmentation types: long visible wake; short faint wake; line of motion fragmentation; elongated head with faint wake; or faint wake (single body). The categories used here are similar but not identical to those used in \cite{Subasinghe2016}: our groups here are more restrictive. Our categories of short faint wake and faint wake (single body) are similar to the \cite{Subasinghe2016} category of `round head/little to no wake/no distinct fragments'. Our categories of elongated head with faint wake and long visible wake are similar to the \cite{Subasinghe2016} category of `smeared head/distinct wake/no fragmentation', but without the distinct wake for the events with elongated heads. Events in our line of motion fragmentation group would fall under the `one of more noticeable fragments' group. Examples of each category are shown in Fig.~\ref{fig:nf_images}.

\begin{table*}
 \caption{Sudden peaked meteor events. Column H$_{x}$ describes the amount of light produced by the head at station 1 or 2, and T$_{x}$ describes the amount of light produced by the wake, at one of the two stations. Column ht$_{avg,x}$ lists the average height at which the light produced values were calculated. The narrow-field description and the amount of light produced values do not necessarily correspond, since the light produced values are averaged over a number of frames, and the meteor appearance changes greatly during ablation. These sudden peaked events had a second peak rising slope greater than 1 mag/km, with their second peak slope value given in column $m_2$.}
 \label{tab:sudden}
 \begin{tabular}{ccccccccccccc}
  \hline
  Event & v$_{i}$ &  T$_{\rm{J}}$ & Narrow-field description & H$_{1}$& T$_{1}$ & ht$_{avg,1}$& H$_{2}$ & T$_{2}$ &  ht$_{avg,2}$ & $m_2$\\
   & km s$^{-1}$ & & & \%& \% & km & \%& \% km & & mag km$^{-1}$\\
  \hline
20101009\_060641 & 17.82 & 5.1 & long visible wake                 & 52  & 48 & 79 & 57 & 43 & 80   & 3.2\\
20120912\_063437 & 38.01 & 4.1 & faint wake (single body)         &     &    &    & 67 & 33  & 88  & 1.2\\
20130816\_081853 & 35.79 & 3.1 & line of motion fragmentation     & 34  & 66 & 86 &    &     &     & 1.6\\
20140904\_070414 & 31.94 & 4.1 & line of motion fragmentation     & 52  & 48 & 91 &    &     &     & 1.2\\
20170824\_023951 & 24.68 & 3.1 & line of motion fragmentation     & 49  & 51 & 80 & 50 & 50  & 80  & 3.7\\
  \hline
 \end{tabular}
\end{table*}

\begin{table*}
 \caption{Smooth double peaked meteor events. Column H$_{x}$ describes the amount of light produced by the head at station 1 or 2, and T$_{x}$ describes the amount of light produced by the wake, at one of the two stations. Column ht$_{avg,x}$ lists the average height at which the light produced values were calculated. The amount of light produced by the head and wake of the meteor may not agree with the narrow-field description, as these are values that are averaged over many frames, during which the meteor appearance changed. These smooth double peaked events had a second peak rising slope less than 1 mag/km, with their second peak slope value given in column $m_2$.}
 \label{tab:double}
 \begin{tabular}{ccccccccccc}
  \hline
  Event & v$_{i}$ &  T$_{\rm{J}}$  & Narrow-field description & H$_{1}$& T$_{1}$ & ht$_{avg,1}$ &H$_{2}$ & T$_{2}$& ht$_{avg,2}$ & $m_2$\\
   & km s$^{-1}$  & & & \%& \% & km & \%& \% & km & mag km$^{-1}$\\
  \hline
20100920\_084608 & 60.30 & 0.8  & long visible  wake                   & 79 & 21 & 91 & 72 & 28 & 92   & 0.5\\
20110830\_071519 & 59.88 & 0.4  & short faint wake                    & 77 & 23 & 98 & 75 & 25 & 99   & 0.2\\
20111005\_084734 & 66.36 & 0.4  & elongated head with faint wake      &    &    &    & 81 & 19 & 97   & 0.8\\
20111104\_082320 & 69.77 & -0.6 & faint wake (single body)            & 80 & 20 & 100 & 67 & 33 & 105 & 0.5\\
20131114\_095829 & 57.55 & 1.5  & faint wake (single body)            & 86 & 14 & 98 &    &    &      & 0.5\\
20140421\_051749 & 40.74 & 1.5  & long visible  wake                   & 56 & 44 & 98 &    &    &      & 0.9\\
20141102\_102214 & 55.69 & 2.9 & faint wake (single body)             & 77 & 23 & 93 & 73 & 27 & 91   & 0.6 \\
20150121\_053147 & 23.20 & 3.0  & long visible  wake                   &    &    &    & 64 & 36 & 82   & 0.5\\
20150121\_084050 & 38.37 & 1.0  & faint wake (single body)            &    &    &    & 70 & 30 &  87  & 0.6\\
20150522\_071644 & 62.73 & 0.3  & short faint wake                    & 100 & 0 & 91 & 48 & 52 & 99   & 0.6\\
20150925\_083237 & 69.10 & -0.1 & faint wake (single body)            & 70 & 30 & 93 & 66 & 34 & 97   & 0.3\\
20151210\_082756 & 66.59 & -0.5 & long visible  wake                   &    &    &    & 87 & 13 & 102  & 0.3\\
20160808\_083545 & 60.41 & -0.4 & elongated head with faint wake      &    &    &    & 82 & 18 & 100  & 0.8\\
20170924\_070328 & 35.63 & 0.9  & line of motion fragmentation        & 56 & 44 & 90 & 66 & 34 & 90   & 0.2\\
20171016\_093056 & 40.87 & 2.6  & long visible  wake                   &    &    &    & 79 & 21 & 87   & 0.4\\
20171201\_103027 & 66.98 & 1.2  & short faint wake                    & 90 & 10 & 93 & 82 & 18 & 93   & 0.3\\
\hline
\end{tabular}
\end{table*}

\subsection{Trail brightness profiles}
\begin{figure}
    \centering
    \includegraphics[width =\columnwidth]{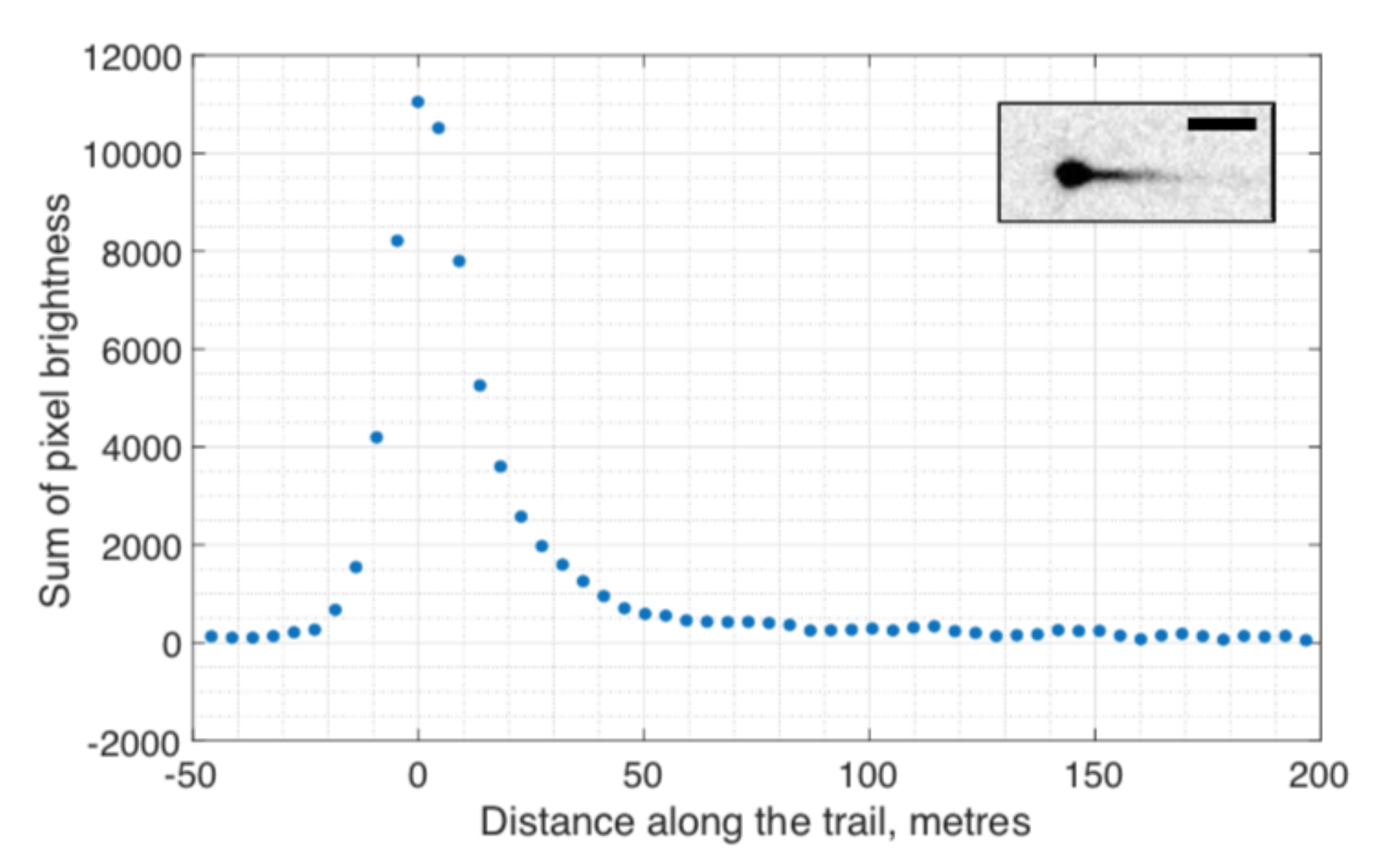}
    \caption{The brightness profile for meteor event 20100920\_084608, Tavistock, frame 96. Each sum of pixel brightness value corresponds to a transverse slice through the meteor. The brightness profile is inset with the corresponding frame taken from the narrow-field observation. The scale bar is 100 m, and the colours have been inverted.}
    \label{fig:brightness}
\end{figure}

\begin{figure}
    \centering
    \includegraphics[scale=0.5]{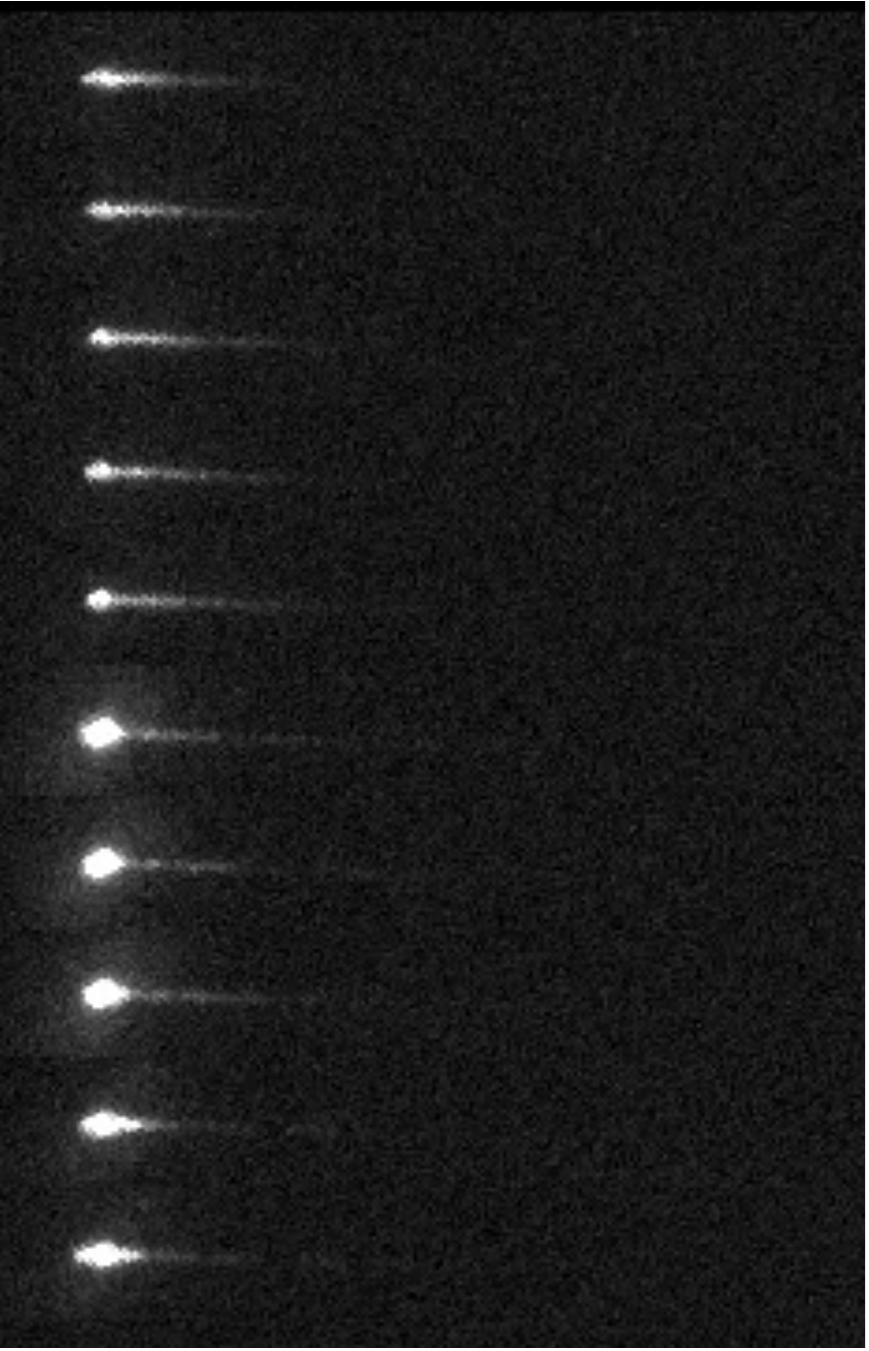}
    \caption{Frames from meteor 20171016\_093056, which have been cropped, rotated and stacked to show the evolution. Time increases downwards. A circle was fit to the head of the meteor, with reference to the brightness profile. It is clear that the radius of that circle would increase, in this example, as ablation progresses. For simplicity, the mean radius from all frames with a meteor position pick was used to determine the size of the meteor head. }
    \label{fig:profile}
\end{figure}

\begin{figure*}
    \centering
    \includegraphics[width=0.8\textwidth]{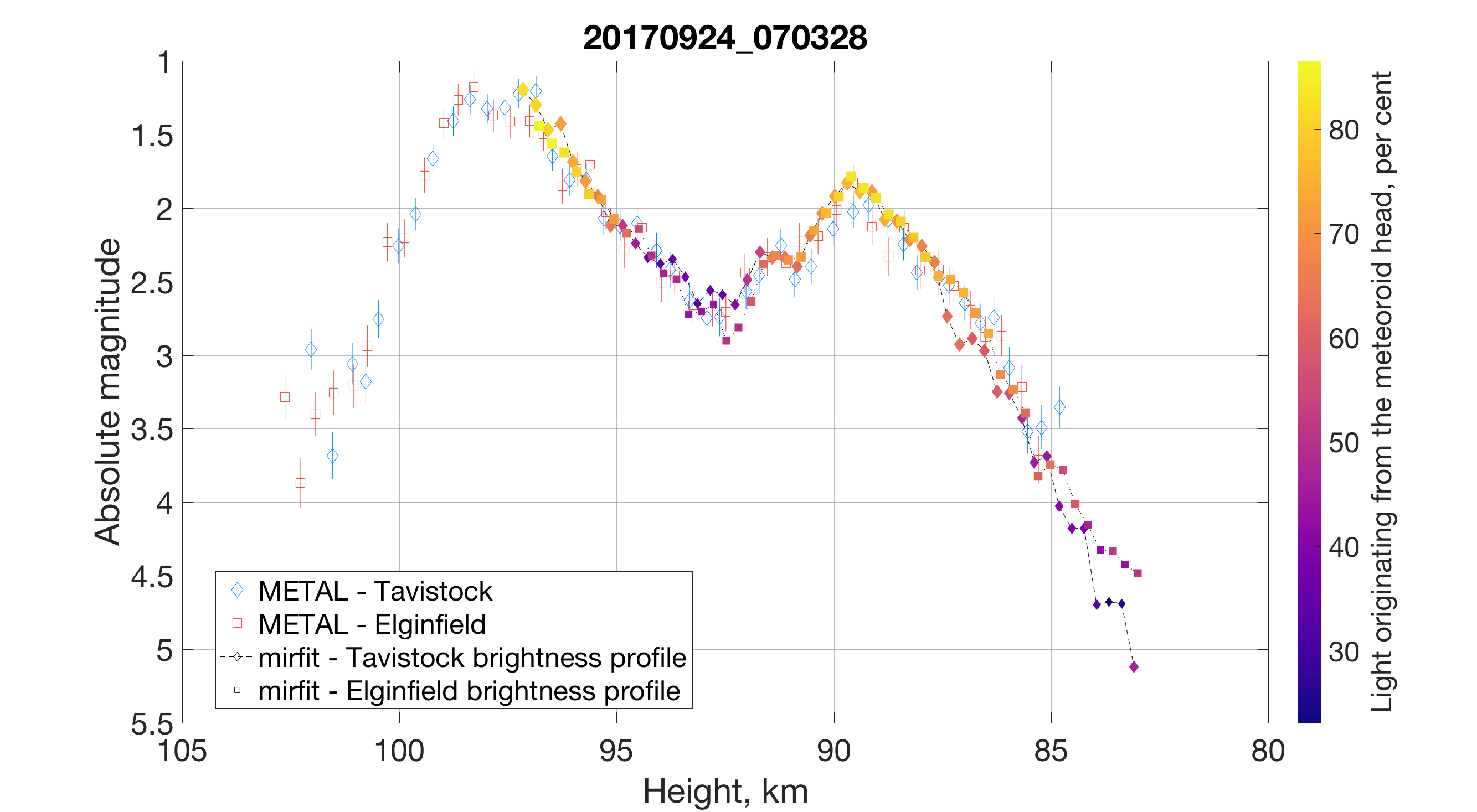}
    \includegraphics[width=0.8\textwidth]{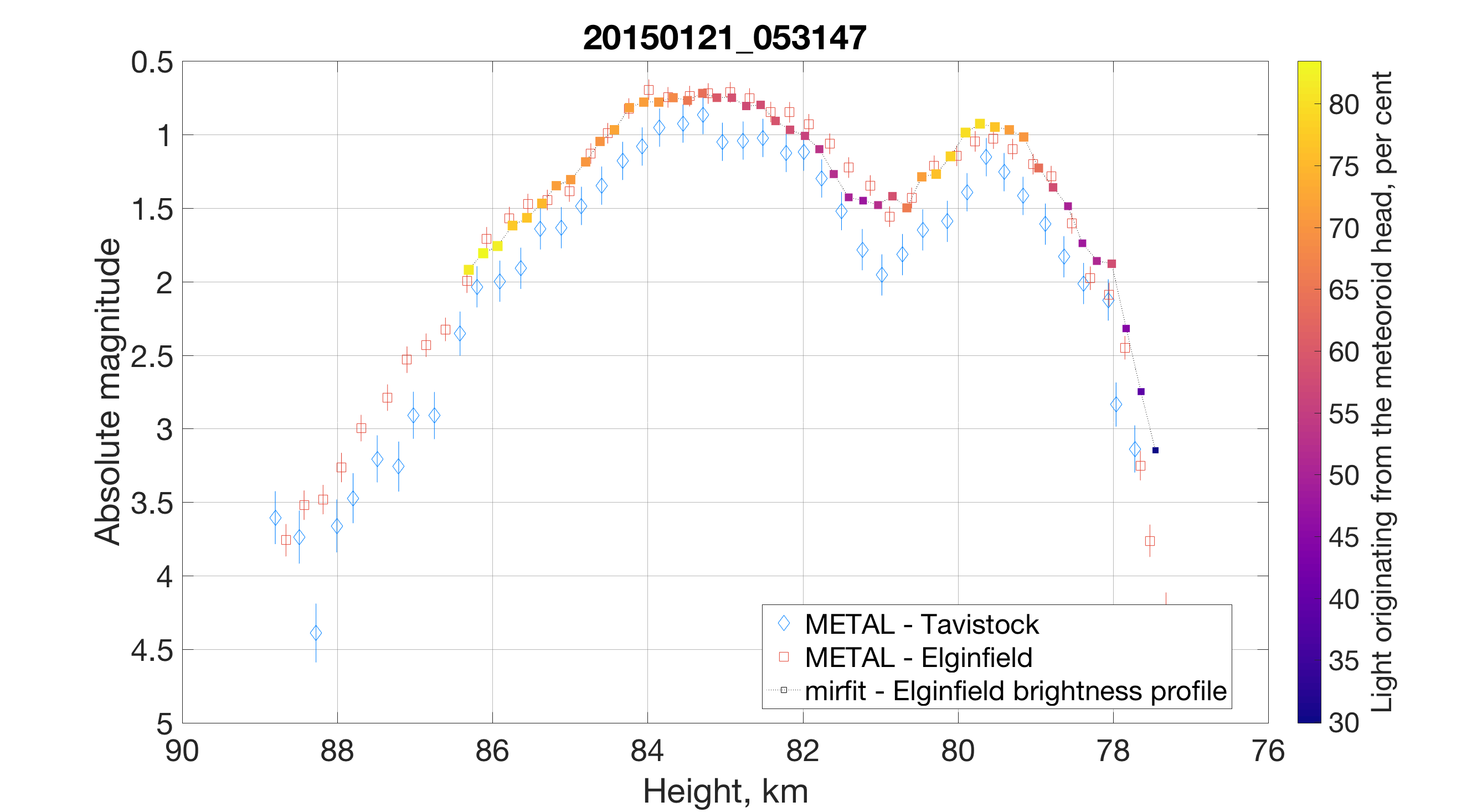}
    \caption{Two meteor events with their narrow-field observations plotted such that the size and colour of the symbol indicates the percentage of light originating from the meteoroid head. The top figure shows what the majority of the brightness profile observations look like: the greatest amount of light coming from the meteoroid head occurs near the peaks, and the amount decreases with fainter magnitudes. The bottom figure shows a brightness profile only seen on one event: the greatest amount of light coming from the head occurred before the first peak, and the first peak shows more equal amounts of light coming from the head and wake. The second peak shows the greatest amount of light from the head occurring at the peak. }
    \label{fig:lc_brightness_profile}
\end{figure*}

In order to quantify the different fragmentation behaviour of the meteors, the fraction of light produced at the head of the meteor compared to the wake was analysed. The software program mirfit extracts for each selected meteor position in a frame, the sum of pixel brightness values for each transverse slice of the meteor as a function of distance along the trail. This gives a brightness profile for each frame, and can be used to study the contribution of light originating at the head of the meteor versus the wake. The brightness profile from a single frame is shown in Fig.~\ref{fig:brightness}. To determine the amount of light coming from the head of the meteor, the size of each meteor head was determined in the image analysis program ImageJ \citep{Schneider2012}. Each head was assumed to be circular, and an average radius in pixels was manually determined. To determine the size of the head of the meteor in a given frame, a circle was fit to the meteor with the diameter equal to that of a transverse slice through the widest part of the meteoroid. Variations in brightness as well as smearing in the tracking system caused the size of meteor head to be different from meteor to meteor, and even from frame to frame of the same meteor event. An example of the variation in meteor appearance is given in Fig.~\ref{fig:profile}. For each frame, the fraction of light originating from the head of the meteor was determined, and the mean value from all the frames at a station was calculated. Values for the light coming from the meteoroid head and wake are given as percentages in Tables~\ref{tab:sudden} and~\ref{tab:double}. These are mean values determined based on all frames at a given station which had meteor position picks in the narrow-field cameras. Tables~\ref{tab:sudden} and~\ref{tab:double} also give the mean height for these mean calculated brightness values. Fig.~\ref{fig:lc_brightness_profile} shows examples of the observed wide-field light curve data, over-plotted with the narrow-field observations, where the colour of the narrow-field observation describes the fraction of light originating from the head of the meteoroid. These brightness profile-light curves can be found for all meteor events in the Appendix.

\subsection{Modelling}
The cause of the two peaks is not clear, as the above observations show a variety of behaviour from these meteoroids in origin, speed, and fragmentation. About half of these observed events show very faint wakes, which implies that bursts of fragmentation, as suggested by \cite{Roberts2014}, are not a main contributor. Here, we attempted to model double peaked meteors without fragmentation, using a two-component single body ablation code. This model is described in detail in \cite{CampbellBrown2017}, and a summary is given here. This inhomogeneous implementation of the single-body meteoroid model allows a meteoroid to have components with two or more chemical compositions and properties. Each component has its own mass, density, heat of ablation, boiling and melting temperatures, specific heat, condensation coefficient, molar mass, thermal conductivity, and luminous efficiency. The components are assumed to be well mixed, so that each can ablate independently. The meteoroid heats uniformly, and is able to have a non-zero porosity, which can change if one component ablates faster than the other, and can drop to zero at a specified temperature, to simulate the meteoroid fully melting and becoming a non-porous droplet. This model is able to produce non-classical light curves, without fragmentation. This is important for this work as many of the meteors studied here show little to no visible wake (eleven of our twenty-one events were classified as having faint wake or appearing single-body like). \newline

We considered a two component meteoroid, in which each component could have one of three possible masses (which affects the total mass of the meteoroid, and the ratio of the masses of the two components); one of five possible densities; and one of five heats of ablation. All other parameters were kept constant and identical between the two components due to their weaker influence on the resulting meteor light curve. The values that were varied (mass, density, heat of ablation) are given in Table~\ref{tab:heats}, and the other parameter values are given in Table~\ref{tab:parameters}. The initial speed of each meteoroid was set to 30~km s$^{-1}$. Each combination of parameters was used to simulate a meteor, resulting in 5625 simulations. The results are discussed in \ref{modellingresults}

\begin{table}
 \caption{The parameters that were varied in the simulations, and their possible values.}
 \label{tab:heats}
 \begin{tabular}{ccc}
  \hline
  Mass & Density & Heat of Ablation \\
  kg & kg m$^{-3}$ & 10$^{6}$ J kg$^{-1}$ \\
  \hline
10$^{-4}$   & 500   & 0.5   \\
10$^{-5}$   & 1000  & 1     \\
10$^{-6}$   & 2000  & 2     \\
            & 3000  & 4     \\
            & 8000  & 6     \\
  \hline
 \end{tabular}
\end{table}

\begin{table*}
 \caption{The simulation parameters that were not varied. Both meteoroid components had the following identical parameters for every simulation.}
 \label{tab:parameters}
 \begin{tabular}{cccccccccccc}
  \hline
  Porosity & Boiling & Melting  & Specific & Condensation & Molar & Thermal & Luminous & Shape  & Emissivity & Zenith\\
  & Point & Point & heat & Coefficient & Mass & Conductivity & Efficiency & Factor&&Angle\\
  \hline
    & K & K & J kg$^{-1}$K & psi & atomic units & J m$^{-1}$s$^{-1}$K$^{-1}$ & per cent & & &degrees\\
  \hline
  \hline
0& 2100 & 2000 & 1000 & 0.5 & 40 & 3 & 0.7 & 1.21 & 0.9 & 45  \\

  \hline
 \end{tabular}
\end{table*}

\section{Discussion}
The observed meteor light curves show two distinct shapes: a smooth double peaked curve, which reaches a maximum, decreases in brightness, and then smoothly repeats the process, and a sudden double peaked light curve, which typically shows a smooth increase in magnitude, followed by a slight decrease and a sharp linear increase in brightness. This sudden change is on average an increase of 2.2 mag per 1.0 km height decrease. The first peak for two of the sudden peaked events is not clear due to noise and the faintness of the event (see \ref{fig:A3} and \ref{fig:A4} in the Appendix). \newline

The origin of a small Solar System body can be inferred using its Tisserand parameter. A Tisserand parameter greater than three is suggestive of an asteroidal origin; values between two and three are associated with Jupiter family comets; and values less than two are often related to long period comets. The boundaries between these groups are not concrete: objects may show asteroidal properties but have cometary Tisserand values. Despite this, the Tisserand parameter provides a method for associating a meteoroid with a broad class of parent body. 
\newline

We find that the sudden peaked meteor events have Tisserand values greater than 3, associating them with asteroidal orbits. The smooth peaked meteor events have Tisserand parameter values less than 3 (except for one event, with a Tisserand of 3.0), associating them with cometary bodies. \newline

\begin{figure*}
    \centering
    \includegraphics[width=0.8\textwidth]{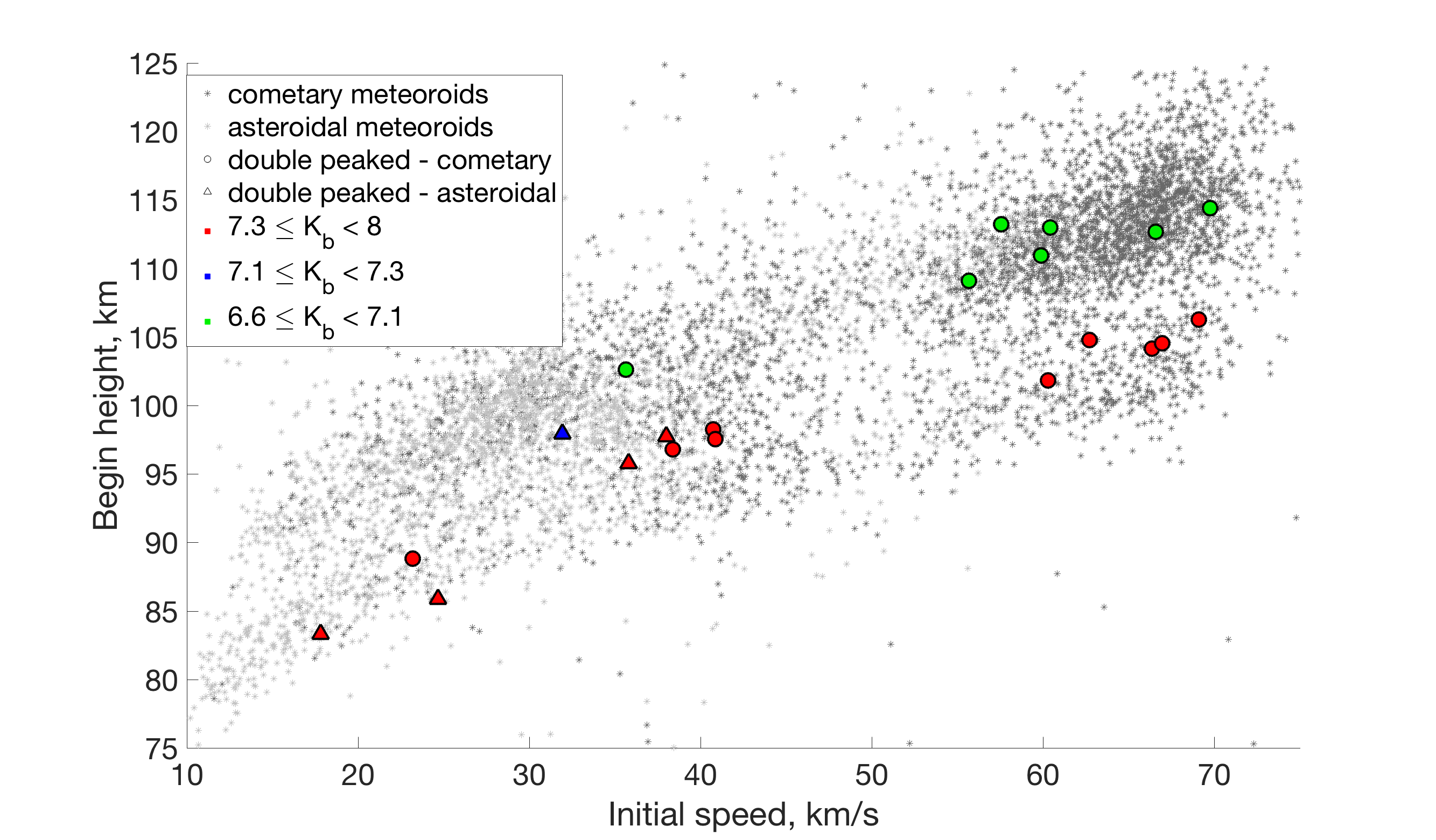}
    \caption{ASGARD reduced meteors are shown as asterisks: dark gray for cometary meteoroids, and light gray for asteroidal meteoroids based on their Tisserand values. The double peaked events in this study are plotted as circles if they are cometary in origin, or triangles for those that are asteroidal. The double peaked meteors are then coloured by their K$_b$ values. For interpretation of the references to colour in this figure, the reader is referred to the web version of this paper.}
    \label{fig:kb}
\end{figure*}

We also considered the K$_b$ parameter of each double peaked meteor. The K$_b$ parameter, defined by \cite{Ceplecha1958} is used to classify meteors based on their strength, and can be used to assign a density to a meteoroid. It assumes that all meteoroids begin their luminous phase at the same surface temperature. This implies that the meteor begin height is strongly related to the meteoroid composition. A K$_b$ value less than 6.6 is associated with weak cometary material; values between 6.6 and 7.1 are associated with regular cometary objects; dense cometary objects have values between 7.1 and 7.3; carboneous chondrites are associated with values between 7.3 and 8; and values greater than or equal to 8 describe chondritic asteroidal material. To investigate our double peaked meteors and their K$_b$ parameters, we plotted the begin height and initial speed of a large set (7160 events) of meteors which were automatically reduced by ASGARD. These meteors were coloured according to their Tisserand parameter (cometary or asteroidal). We then overplotted our double peaked meteor events, with shapes that describe their origin based on their Tisserand parameter (cometary or asteroidal) and with colours that describe their K$_b$ parameter. This is shown in Figure \ref{fig:kb}. All the double peaked meteors from this study fall within the larger set of meteors, and roughly follow the split of cometary meteoroids having higher begin heights and speeds, and asteroidal meteoroids having lower begin heights and speeds, with no unusual values. In terms of K$_b$ parameter, we have no double peaked events with values associated with the two extreme categories (weak cometary material, and chondritic asteroidal material). \newline

The morphology of each meteor was analysed and manually categorized based on the extent of the visible wake; the shape of the meteor head; and whether gross fragmentation was occurring. The meteors were assigned to one of five categories, and examples are shown in Fig.~\ref{fig:nf_images}. Many of the meteors in the smooth double peaked category (63\%) show faint wakes or single body like behaviour, whereas only one fifth of the sudden peaked events show similar behaviour. The majority of the sudden peaked events show fragmentation, mainly in the form of gross fragmentation along the line of motion. \newline

Asteroidal meteoroids are often assumed to be stronger than cometary meteoroids due to their origin. However,  \cite{Subasinghe2016} found that asteroidal meteoroids fragment as often as cometary meteoroids (i.e. they found that asteroidal meteoroids showed a distinct trail or wake 87.2\% of the time, compared to 86.8\% for cometary meteoroids). We find similar results here: our sudden peaked meteors are all asteroidal in origin and 80\% of the events show significant fragmentation. The smooth double peaked curves are all cometary in origin and about 70\% of those events show fragmentation (five of the sixteen events were classified as looking single body like, and those events showed less wake than the other eleven events). We do note that this study has considerably fewer events to analyse. \newline

The contribution of light for the sudden peaked events comes almost equally from the head and wake of the meteor for 3 of the 5 events. This is expected for a fragmenting object. For an object that is more single bodied and not fragmenting, the majority of the light is expected to be produced in the head of the meteor. However, the size of the meteor head changes over time, and because we used a mean size for the head, light from the meteor head may be incorrectly attributed to the wake, or vice versa. The smooth double peaked meteor events show mostly higher light contribution from the meteor head compared to the wake. The values from the two stations agree quite well with each other, and when they don't (see 20150522\_071644 for example), the disagreement is due to the height at which the observations were recorded (that is, the two stations observed different parts of the ablation profile). Appendix A shows each wide-field meteor light curve, with the narrow-field observations plotted as symbols in which the colour and size correspond to the percentage of light originating from the meteoroid head. Two of these events are reproduced in Fig.~\ref{fig:lc_brightness_profile}. For many events (18 of the 21), the brightness profiles clearly show that the amount of light coming from the head of the meteoroid is greatest at the peaks and minimal in the trough and decreases away from the peaks. One of the remaining three events has too few data points to make a conclusion (20140421\_051749) and one event (20100920\_084608) shows disagreement between the two stations at the end of the observation, where the absolute magnitude is fainter than +4. The remaining event (20150121\_053137) shows that the greatest amount of light coming from the head occurred before the first peak, and decreased as the first maximum meteor brightness was observed. The meteoroid may be breaking up prior to the first peak, and the amount of light coming from the head, which was initially high, is now (at the first peak) coming more equally from the head and wake.

\subsection{Possible mechanisms}
\cite{Roberts2014} studied a set of double peaked meteor light curves observed with the CAMO tracking system. Some of the light curves of those meteors were successfully modelled using a dustball model. More details were obtained from the unpublished thesis \citep{Roberts2014a}. In \cite{Roberts2014}, both peaks were successfully modelled for nine events, and the second peak was successfully modelled for one event. Their implementation of the dustball model has a release of grains before the light curve is observed. As these grains ablate, the meteor increases in brightness, creating the first peak. When the meteor reaches a certain height (determined manually, and corresponding to the end of the first peak), another group of grains is released all at once. The luminosity of these grains produces the second peak. The first group of released grains were referred to as outer grains, and the second group as core grains; and they found that the core grains were more massive than the outer grains. There is no analysis of corresponding narrow-field observations for the group of 21 meteors studied in \cite{Roberts2014}. \newline

We find that a simultaneous release of many grains is not adequate to explain the second peak, as many of our events appear to be single body like (or show very faint wake). The \cite{Roberts2014} dustball model predicts an instantaneous linear increase in magnitude for the second peak. This model may be able to reproduce our sudden peaked light curves, which show mainly (80\% of events) noticeable fragmentation. This model is unlikely to work on our other events which show minimal fragmentation at CAMO's resolution, though they may be undergoing some form of continuous fragmentation, similar to that modelled in \cite{CampbellBrown2017}. If a meteoroid were to fragment into identical grains, no wake would be expected, but this is an unlikely scenario. \newline

Another potential mechanism for producing a multi-peaked light curve is differential ablation. In this process, more volatile species (sodium, magnesium, etc) ablate first, producing the first peak, followed by the more refractory species (e.g. calcium) which may produce the second peak. Differential ablation has been detected in meteor observations \citep{Borovicka1999,Bloxam2017}. The work done by \cite{Bloxam2017} using a four camera spectral system, with CAMO, found that 94\% of meteors showing differential ablation showed fragmentation in the corresponding narrow-field observation. Two of our double peaked meteor events fall into the collection period of the \cite{Bloxam2017} work, however no spectral data was observed for those events: the September 9 2015 event recorded very few frames (< 7) with the magnesium and sodium filters, which were affected by clouds; the December 10 2015 event was outside the field of view of all of the spectral cameras. \newline

The physical structure of a meteoroid may also contribute to the object showing a multi-peaked light curve. This would be a type of differential ablation where the meteoroid was a set of concentric spheres, with the outer layers being composed of volatile material, and the inner layers being more refractory. The layers would ablate at different rates, creating a double peaked light curve. Armoured chondrules are a possible meteoroid structure that could produce a multi-peaked light curve. In these structures, a chondrule is encased in a layer of iron sulfide or iron nickel. Armoured chondrules have been observed in meteorites \citep[e.g.][]{Wood1963,Wasson2010}, but always in a matrix, and never as a stand-alone meteorite; this would require some mechanical separation during a collision or other processing event.

\subsection{Modelling work}
\label{modellingresults}
An investigation into the mechanism responsible for the second peak was completed using a two component single-body ablation model. Here we study how likely the differential ablation process is compared to continuous fragmentation. Our attempt at modelling these double peaked meteors with a two-component single-body ablation model (a form of differential ablation) was not successful. This model is able to produce double peaked light curves, however the simulated curves are too wide in terms of height. Some form of continuous fragmentation is likely needed in the model to reduce the width of the simulated light curves.\newline

Our search of the mass/density/heat of ablation parameter space (5625 meteors total) yielded 142 simulations (71 unique solutions) that produced double peaked curves in which the magnitude differences from the first peak to trough and trough to second peak were at least 0.5 magnitude (called "full double peaked curves"). When we loosened our definition of a double peaked light curve such that only one of the first or second peak to the trough magnitude difference is at least 0.5 mag (called "half double peaked curves"), we found an additional 312 simulations (156 unique solutions) that satisfied this condition,  in addition to the 142 events that had both magnitude differences greater than 0.5. From our set of 21 observed events, 19 are full double peaked curves, and 2 are half double peaked curves, based on the Tavistock station absolute magnitude wide-field observations. We removed symmetric simulated events, i.e. a simulated meteor with $\rho_1$ = 500 kg/m$^3$ and $\rho_2$ = 1000 kg/m$^3$ produces an identical curve to a simulated meteor with $\rho_1$ = 1000 kg/m$^3$ and $\rho_2$ = 500 kg/m$^3$, with all other parameters identical, so one of these events would be removed, as there is only one unique event. \newline

\begin{figure}
    \centering
    \includegraphics[width=\columnwidth]{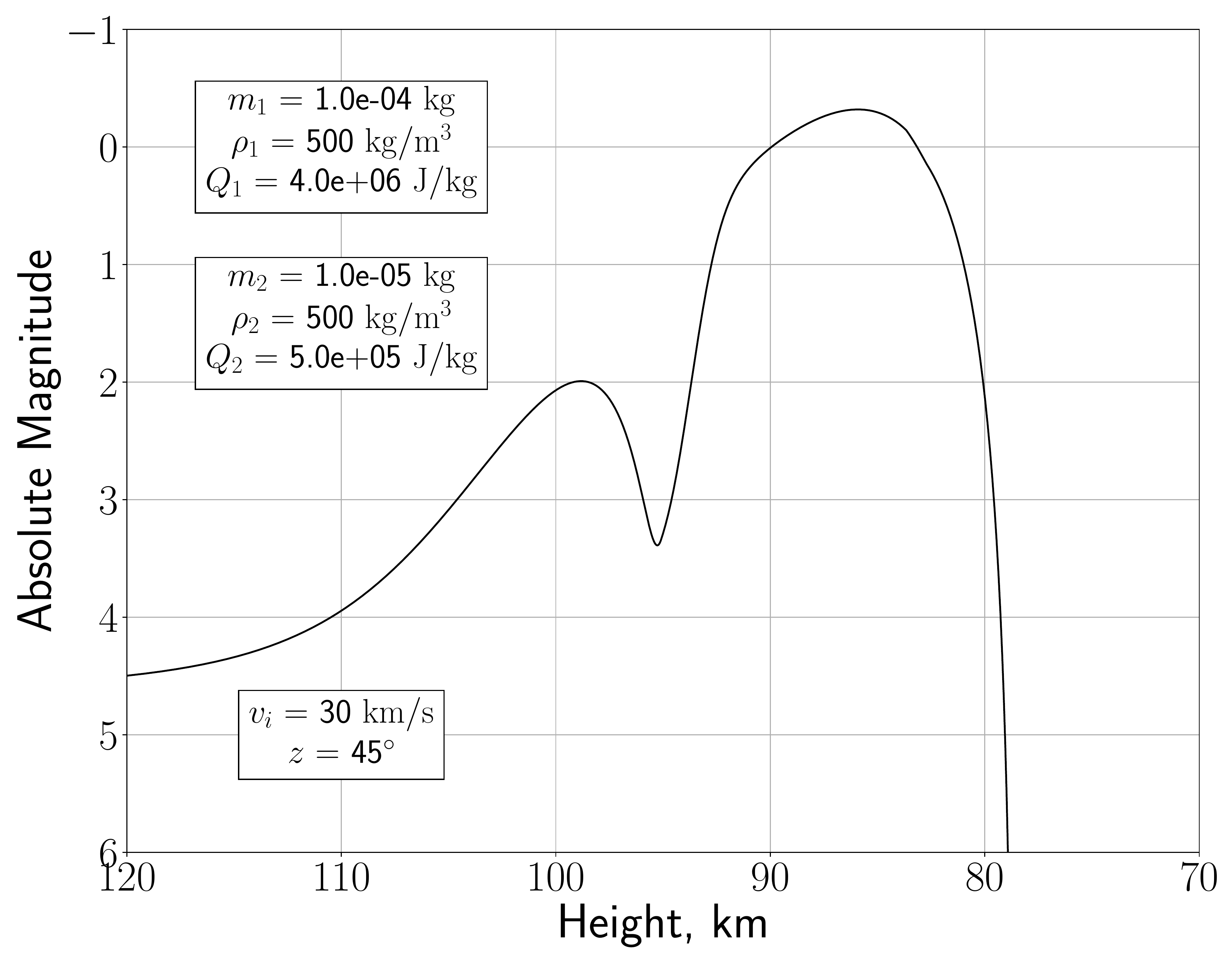}
    \includegraphics[width=\columnwidth]{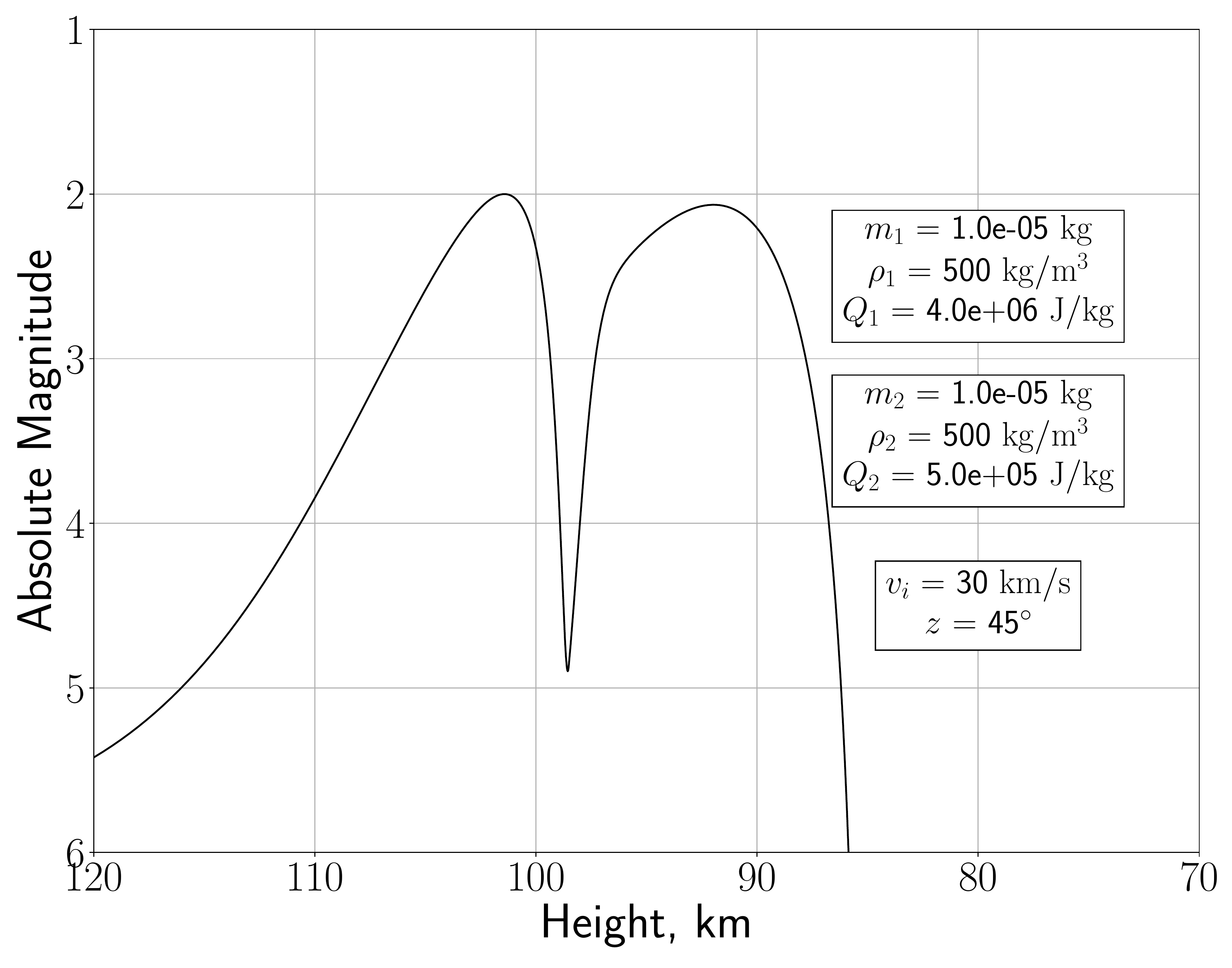}
    \includegraphics[width=\columnwidth]{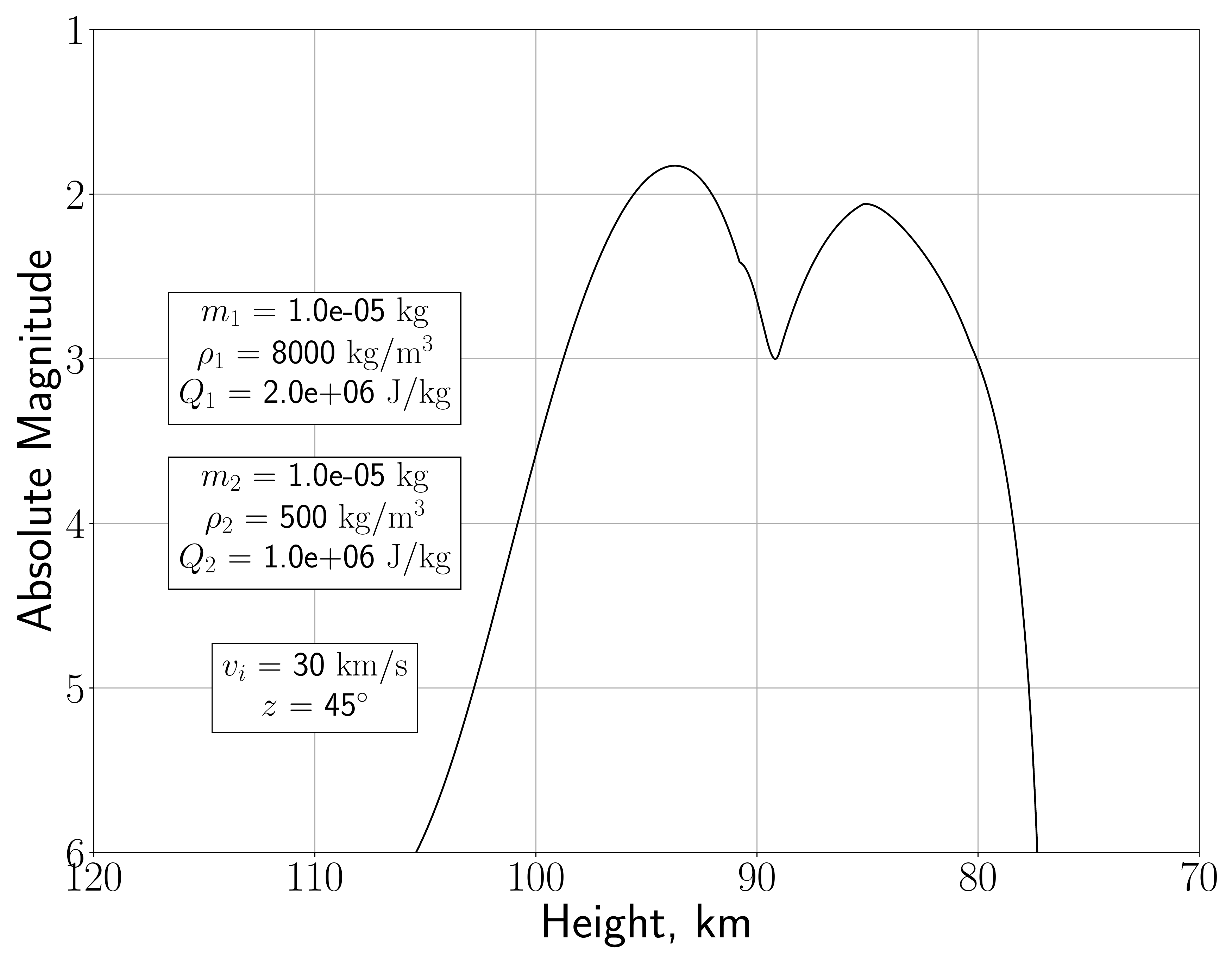}
    \caption{Three simulated meteor events, showing the variety of shapes produced with this ablation model. }
    \label{fig:simulated}
\end{figure}

Some of the simulated double peaked meteors are shown in Fig.~\ref{fig:simulated}, with the masses, densities, and heats of ablation listed. None of the simulated meteors showed shapes similar to the sudden peaked light curves. \newline

We considered the width of the simulated meteors in terms of height (h$_{\rm{begin}}$ - h$_{\rm{end}}$), with the beginning and ending heights determined where the absolute magnitudes crossed +4 (this value was selected as many of our observed meteor beginning and/or ending heights are close to that brightness). In our set of 21 studied meteors, the greatest width was 24.6 km, and the average width was 15 km. Of the 71 unique simulated full double peaked meteors, 53 had meteor widths between 5 - 25 km, at magnitudes brighter than +4. Forty-four of the fifty-three simulated meteors had one peak entirely fainter than +4, and the calculated width corresponded to the width of a single peak only (these 44 simulated events thus do not correspond our observations). Ignoring the fact that most of the simulated events have a peak that is fainter than we could confidently observe with our system, we found that none of the fifty-three simulated meteors with widths that agreed with our observations showed shapes that match our observations: the simulated trough is often too sharp, and the simulated rising and falling slopes of the peaks are often too steep. \newline

\begin{figure}
    \centering
    \includegraphics[width=\columnwidth]{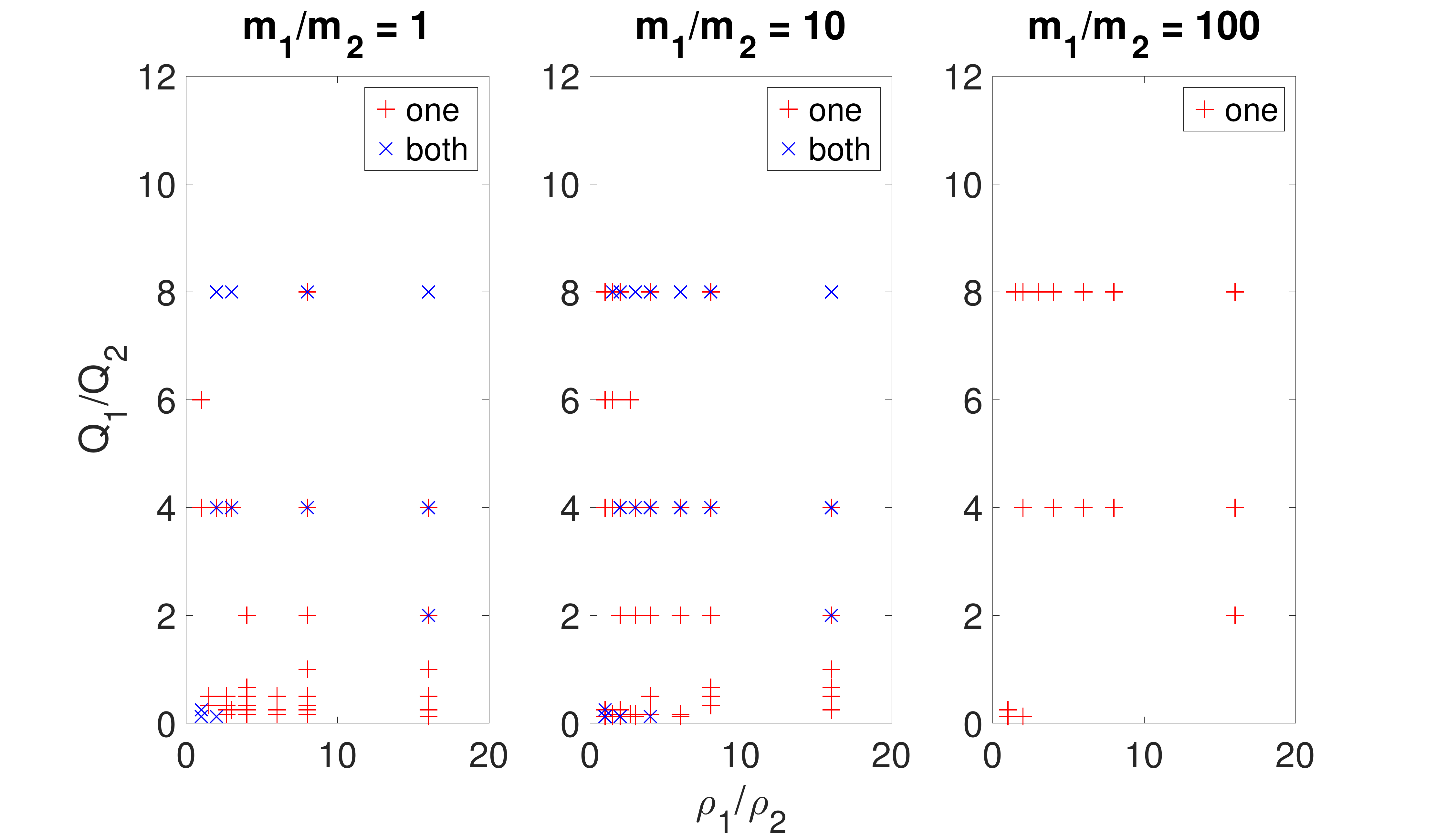}
    \caption{Each plot shows the ratio of heats of ablation versus the ratio of meteoroid densities, for the three possible mass ratios. The red plus symbols (+) indicate simulations which produced a double peaked meteor in which only a single magnitude difference (either the first peak to trough, or the trough to second peak) was greater than 0.5 magnitude (a "half double peak"). The blue crosses (x) indicate simulations in which both magnitude differences were at least 0.5 mag (a "full double peak"). The meteors in these simulations had entry speeds of 30 km/s and zenith angles of 45$^\circ$.}
    \label{fig:ratios}
\end{figure}

We looked closely at the simulated double peaked events, in particular, the heats of ablation and density values needed to produce these shapes. These parameters, along with mass, cause the greatest changes in the simulated light curve shape, and are responsible for the observed double peaks. In Fig.~\ref{fig:ratios}, we plot the unique meteor events that produced double peaked curves, in terms of the ratio of densities vs the ratio of heats of ablation, separated by the ratio of masses. \newline

\begin{figure}
    \centering
    \includegraphics[width=\columnwidth]{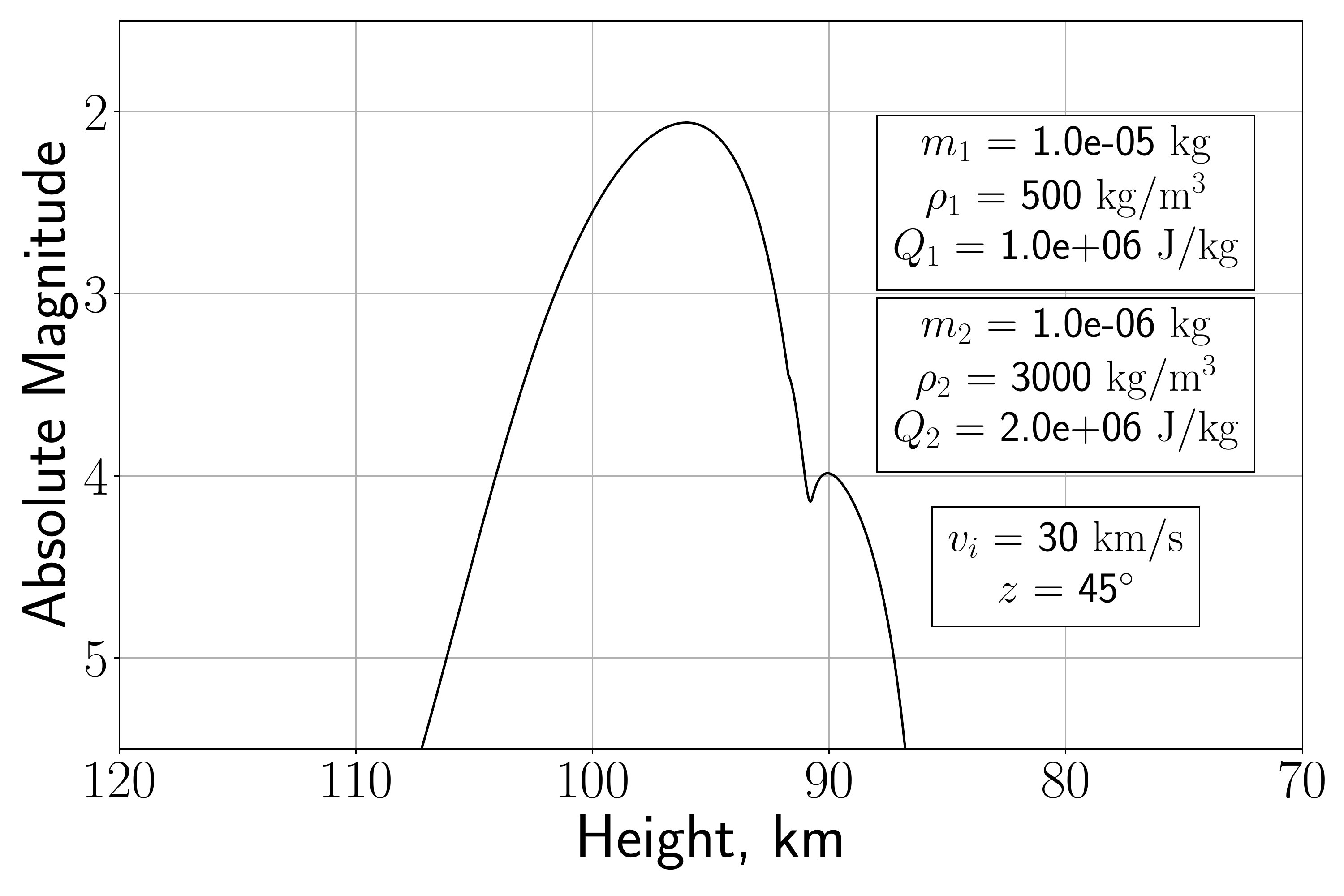}
    \caption{A simulated meteor that was classified as half double peaked, due to the magnitude difference between the first peak and the trough being greater than 0.5 mag. The second `peak' looks more like a blip, and due to noise in the camera system wouldn't be noticed, and this would be classified as a single peaked light curve.}
    \label{fig:METSIM_3208}
\end{figure}

The lack of simulated events showing double peaked light curves with a mass ratio of 100 is explained by the fact that with a large mass ratio, one peak (created by the less massive component) will be overshadowed by the other peak (created by the more massive component). Simulated events with a mass ratio of 10 show more double peaked curves than the other two mass ratio options, likely because one component is able to ablate completely before the other, which produces the dip in luminosity (and the mass ratio is low enough that the less massive component is visible). For simulated events with a mass ratio of 1, the two components are possibly ablating at the same time, but the differing compositions may allow double peaks to be produced. Many of the simulated events in this group (mass ratio of 1) are not truly double peaked and look similar to the simulated event shown in Fig.~\ref{fig:METSIM_3208}. The varying frequency of simulated double peaked events at various density and heat of ablation ratios, for the different mass ratios, indicates that the meteoroid densities and heats of ablation could be important for producing double peaked light curves.\newline

\begin{figure*}
    \centering
    \includegraphics[scale =0.5]{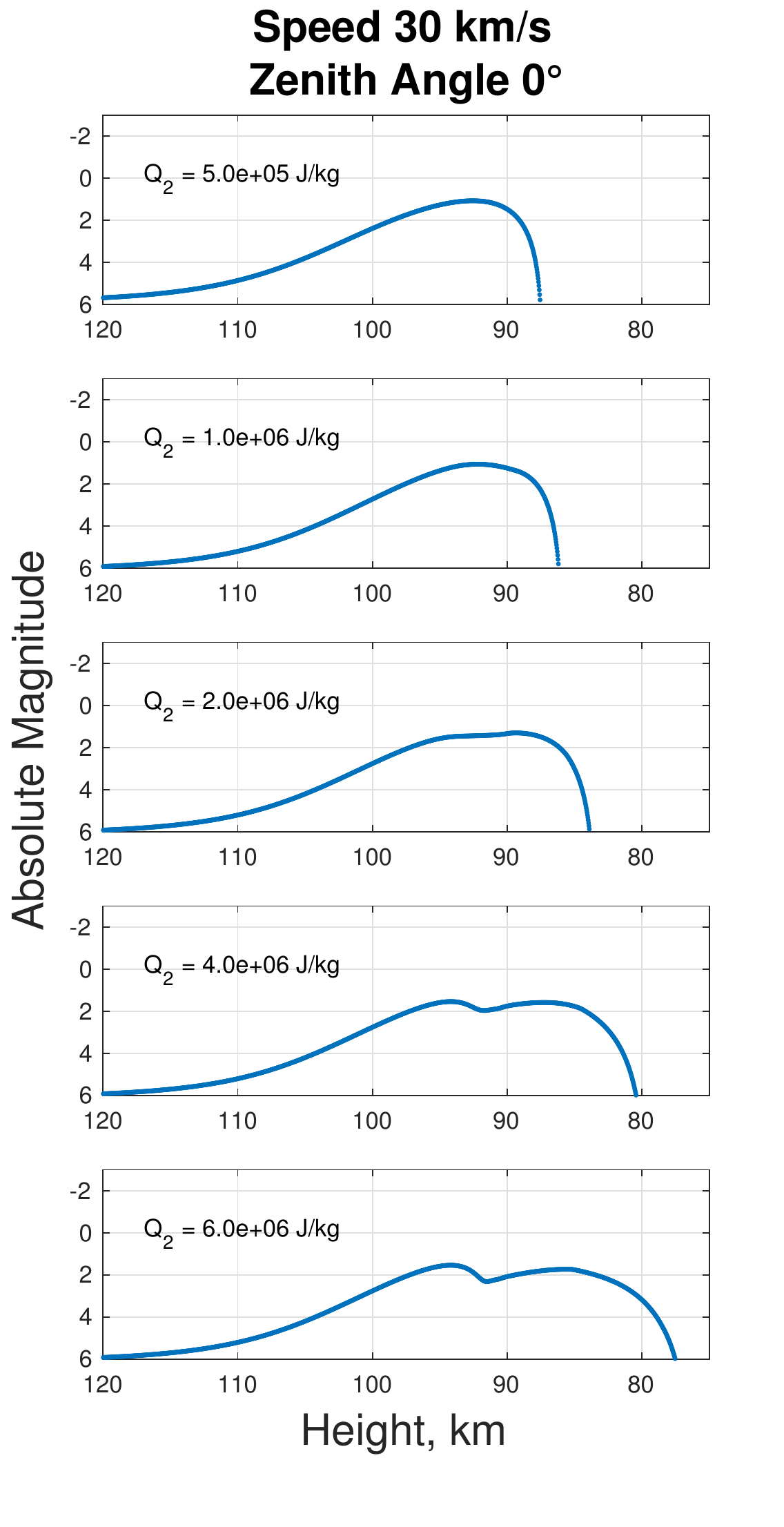}
    \includegraphics[scale=0.5]{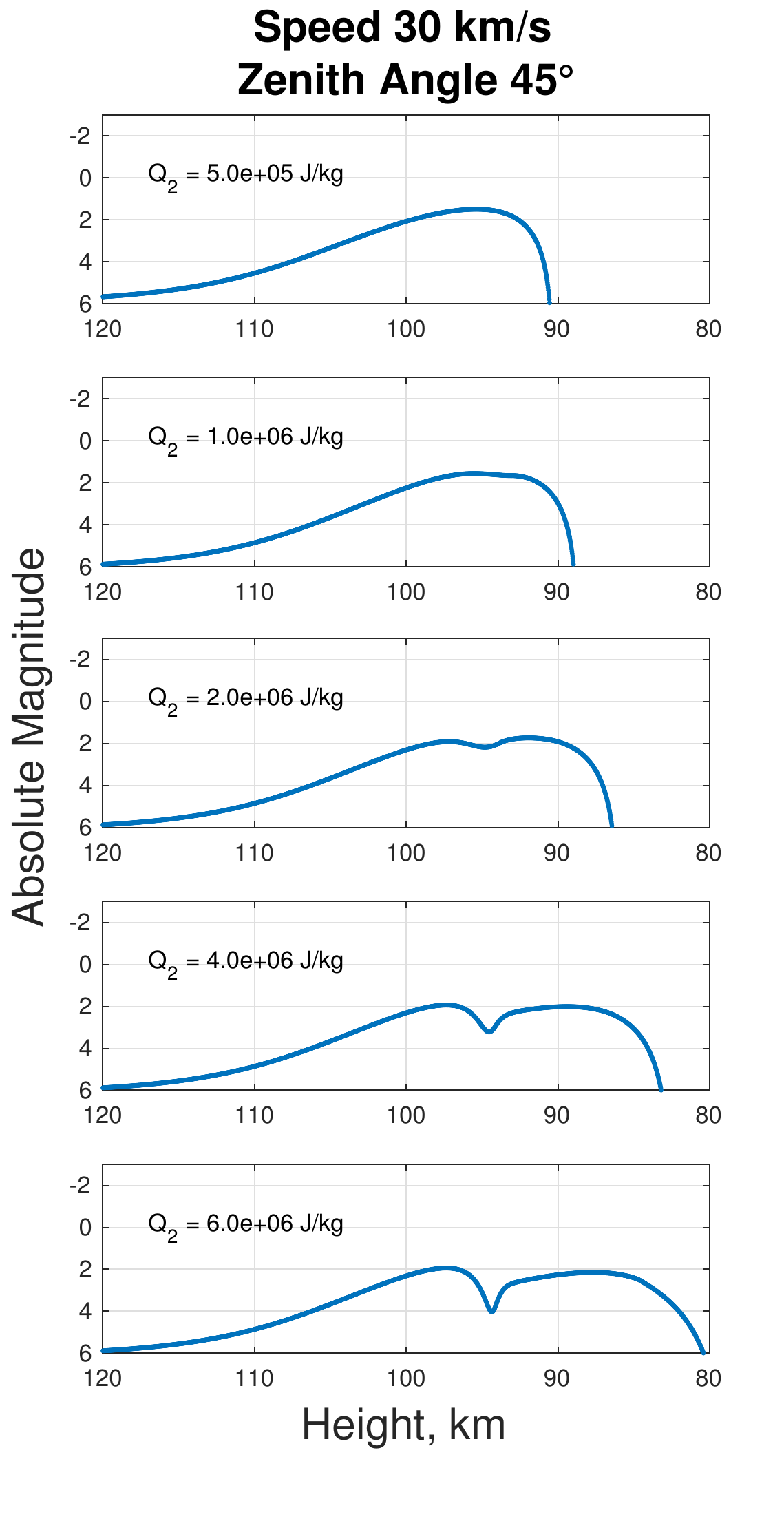}
    \includegraphics[scale=0.5]{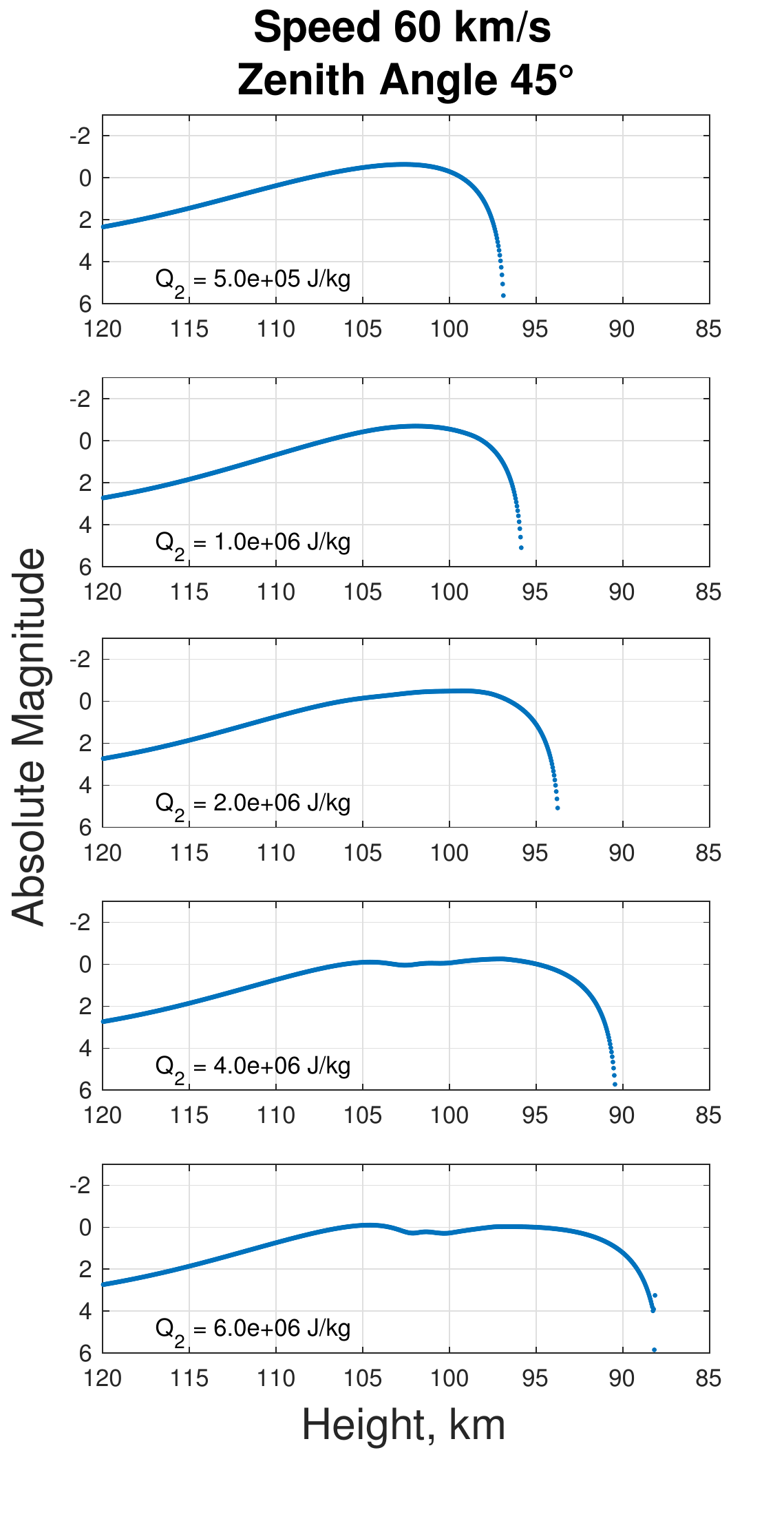}
    
    \caption{Three figures showing the effect of entry speed, zenith angle, and heat of ablation. The leftmost figure shows (from the top down) the effect of increasing the heat of ablation value for one component in our two-component single body ablation model, for a meteor with an entry speed of 30 km/s and zenith angle of 0$^\circ$. The middle figure shows the same effect, but for a zenith angle of 45$^\circ$. The rightmost figure shows the results for an entry speed of 60 km/s and zenith angle of 45$^\circ$. The parameters for each set of figures were: both component masses were 10$^{-5}$kg; both component densities were 1000 kg/m$^3$; and one heat of ablation was 5 X 10$^5$ J/kg.}
    \label{fig:Q2_evolution}
\end{figure*}

\begin{figure*}
    \centering
    \includegraphics[scale=0.5]{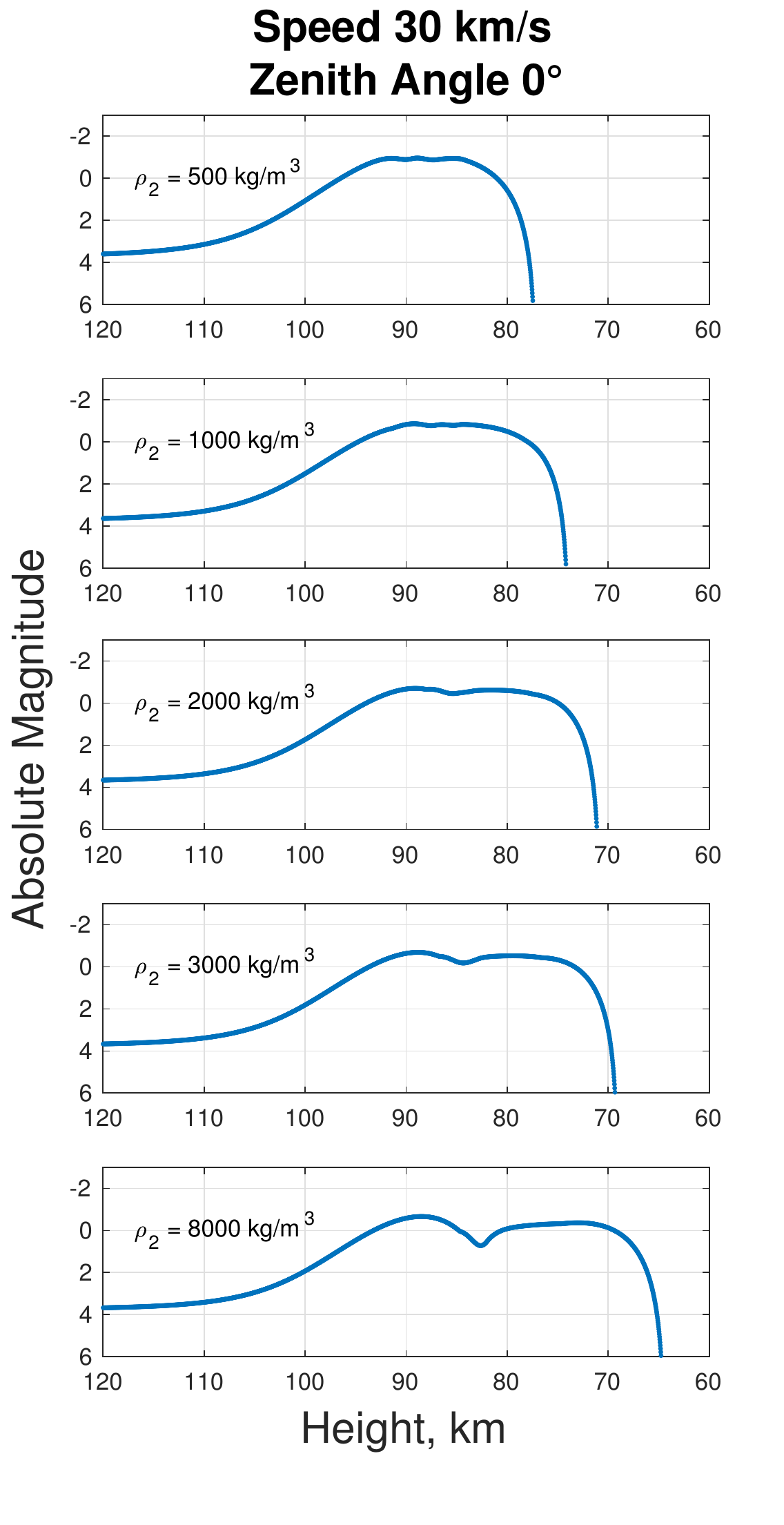}
    \includegraphics[scale=0.5]{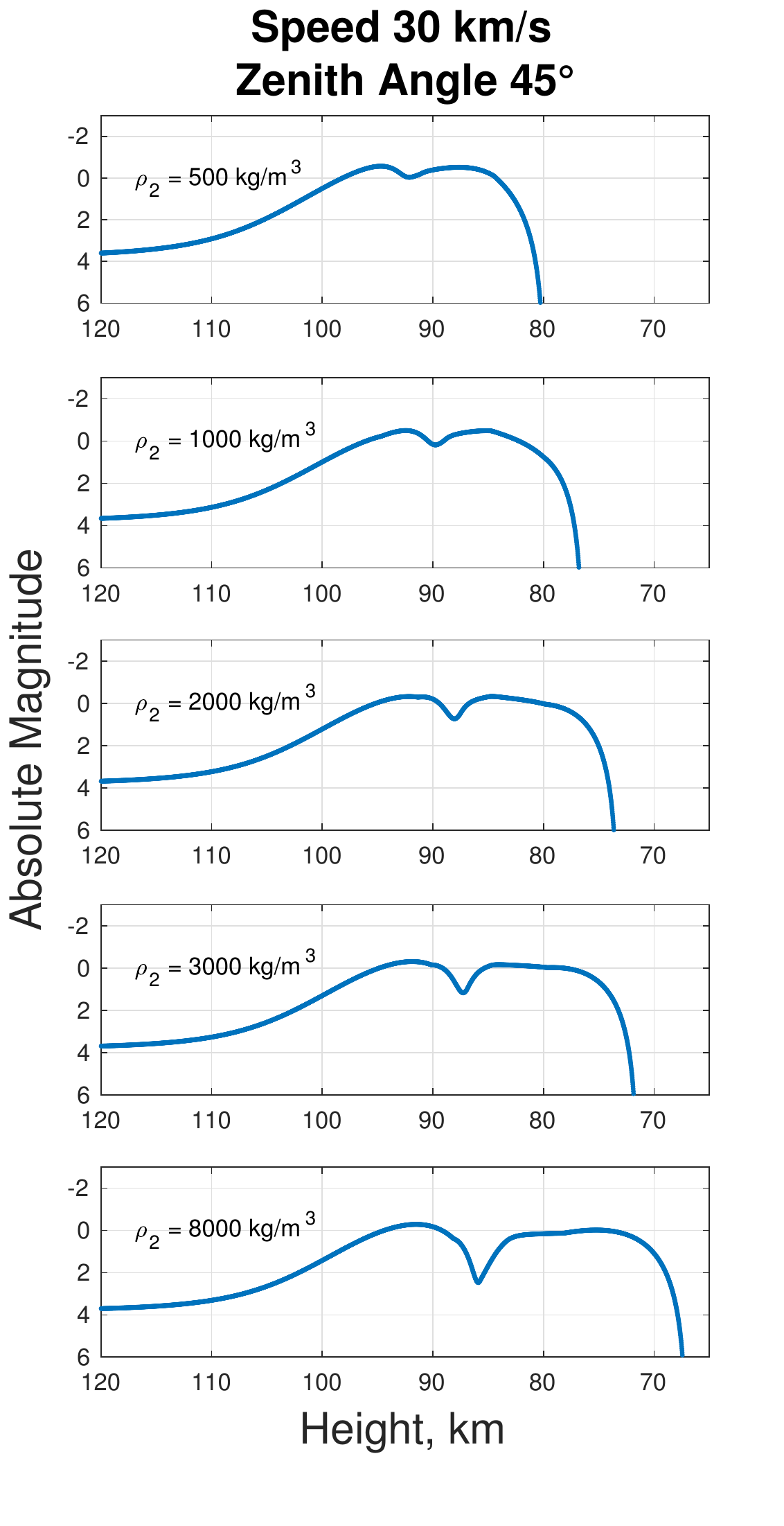}
    \includegraphics[scale=0.5]{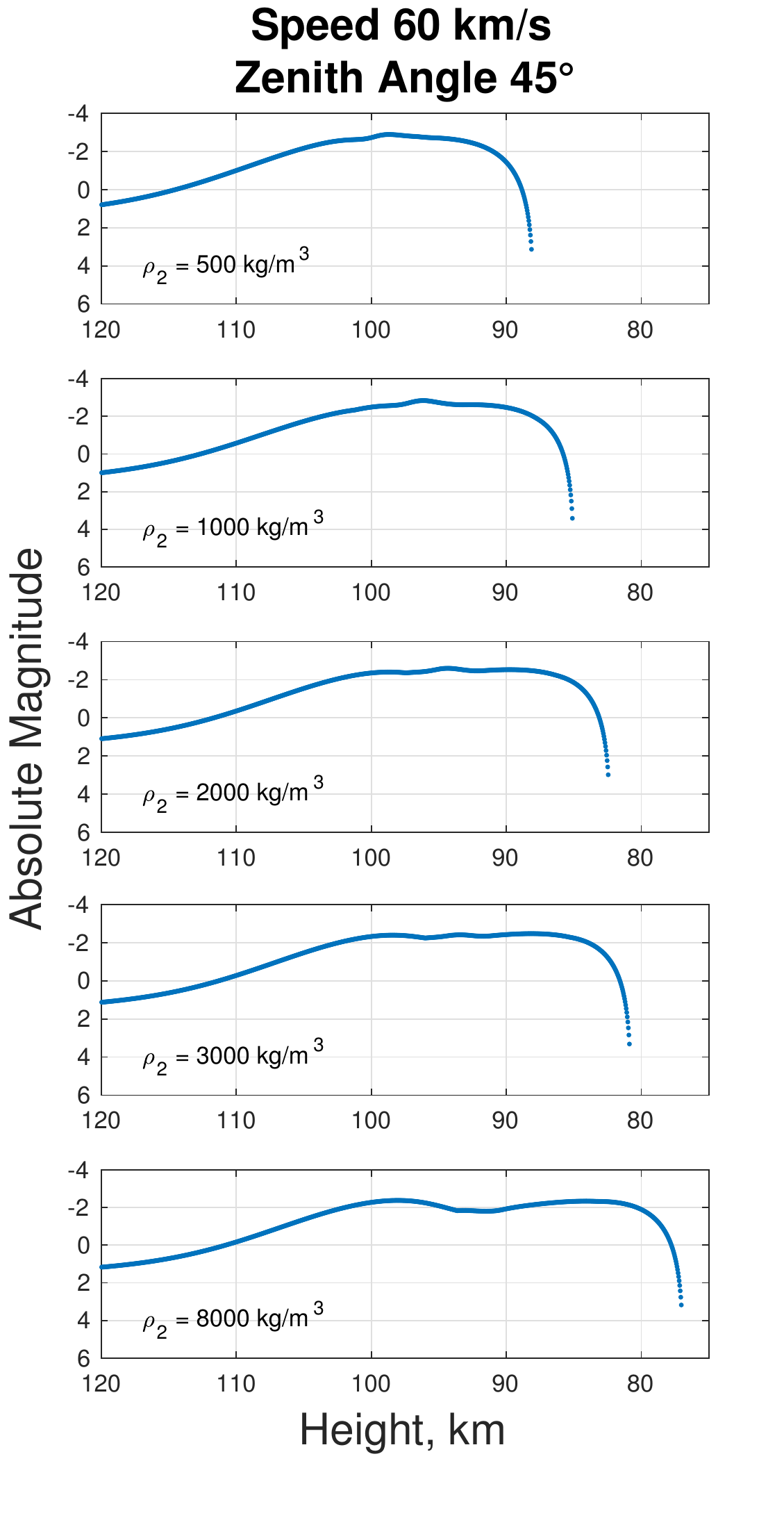}
    \caption{Three figures showing the effect of entry speed, zenith angle, and meteoroid density. The leftmost figure shows (from the top down) the effect of increasing the density value for one component in our two-component single body ablation model, for a meteor with an entry speed of 30 km/s and zenith angle of 0$^\circ$. The middle figure shows the same effect, but for a zenith angle of 45$^\circ$. The rightmost figure shows the results for an entry speed of 60 km/s and zenith angle of 45$^\circ$. The parameters for each set of figures were: both component masses were 10$^{-4}$kg; one component density was 500 kg/m$^3$; the heats of ablation were 5 X 10$^5$ J/kg and 4 X 10$^6$ J/kg.}
    \label{fig:rho2_evolution}
\end{figure*}

To investigate the role that heat of ablation and density play in producing double peaked meteor light curves, we kept all other values constant and varied heat of ablation or density for the second component through the full range of values. We also investigated the effect of the meteor zenith angle. These results are shown in Fig.~\ref{fig:Q2_evolution} and~\ref{fig:rho2_evolution}. \newline

\begin{figure*}
    \centering
    \includegraphics[scale=0.2]{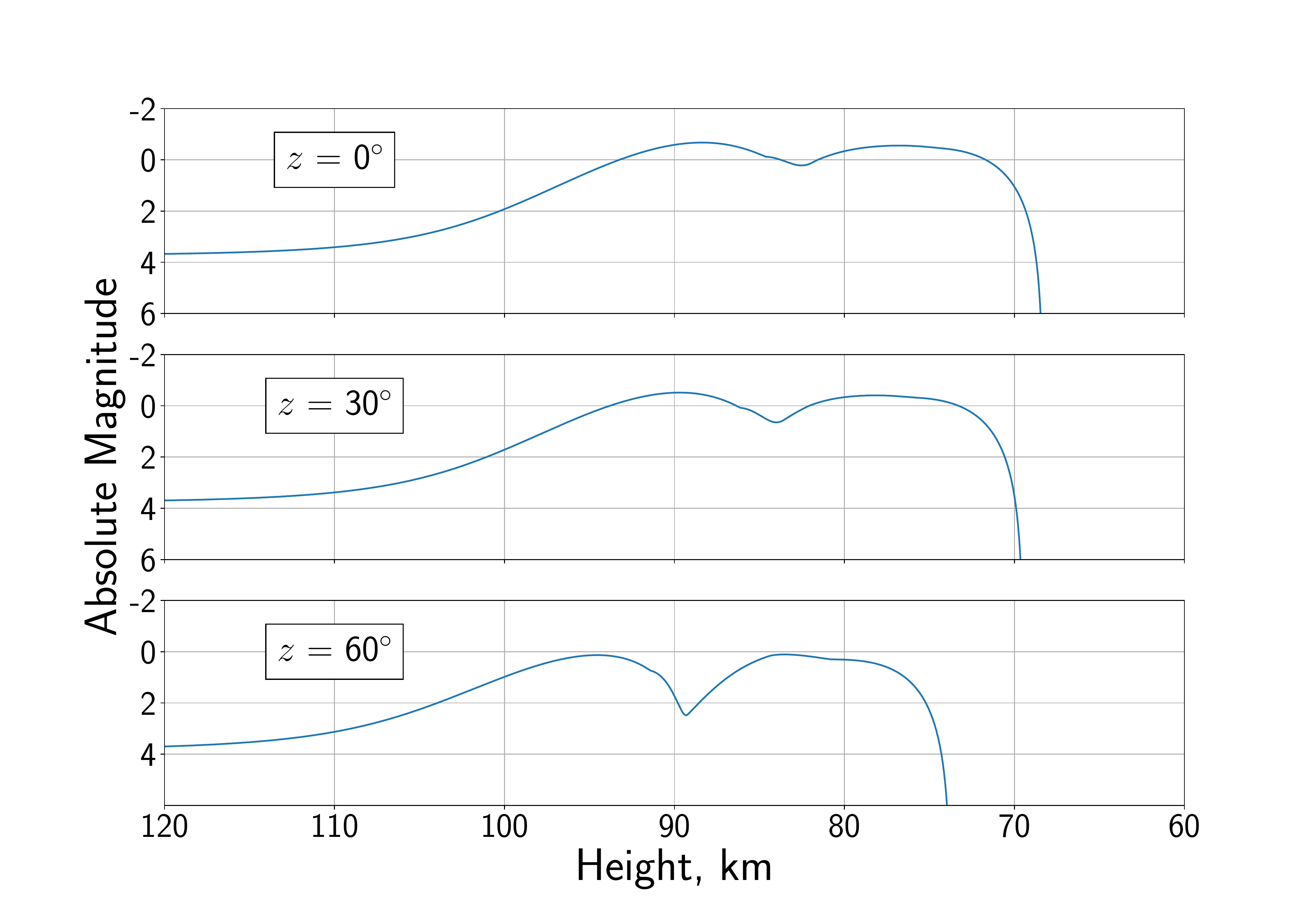}
    \includegraphics[scale=0.2]{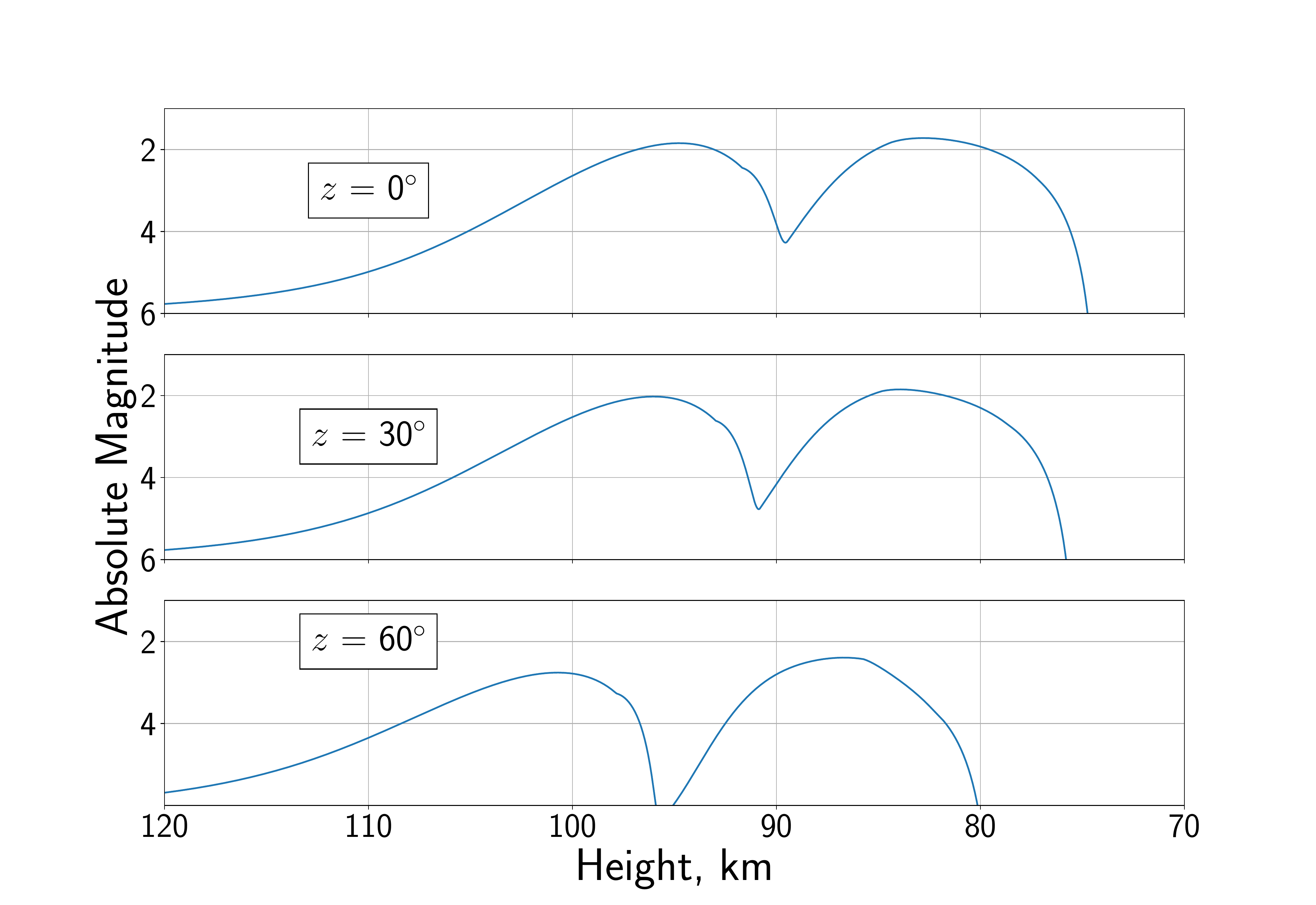}
    \caption{The effect of zenith angle on simulated meteor light curve shape. The meteors for both the left and right figures have the following properties: $\rho_1$ = 500 kg/m$^3$; $\rho_2$ = 8000 kg/m$^3$; Q$_1$ = 5 X 10$^5$ J/kg; Q$_2$ = 2 X 10$^6$ J/kg. The left figure has m$_1$ = m$_2$ = 10$^{-4}$ kg. The figure on the right has m$_1$ = m$_2$ = 10$^{-5}$ kg. Entry speed is 30 km/s.}
    \label{fig:zenith}
\end{figure*}

The effect of the zenith angle is shown in Fig.~\ref{fig:zenith}, as well as in Figures~\ref{fig:Q2_evolution} and~\ref{fig:rho2_evolution}. The zenith angle can change the shape of the first peak, shape of the second peak, depth of the trough, brightness of the event, and the beginning and ending heights. The mean zenith angle for our twenty-one events is 33.3$^\circ$: for our simulated meteors we used 45$^\circ$. \newline

Many of the half double peaked (in which either the first peak to trough or trough to second peak is at least 0.5 mag) simulated meteors are actually not double peaked -- they show a small hook (Fig.~\ref{fig:METSIM_3208}), which passes the requirement for one magnitude difference of at least 0.5 mag, but visually fails as a double peaked meteor. The second peak in Fig.~\ref{fig:METSIM_3208} would be lost in the noise of the system. This explains the large number of red plus symbols (+) in Fig.~\ref{fig:ratios}.

\subsection{Other variations of double peaked curves}
\begin{figure}
    \centering
    \includegraphics[width=\columnwidth]{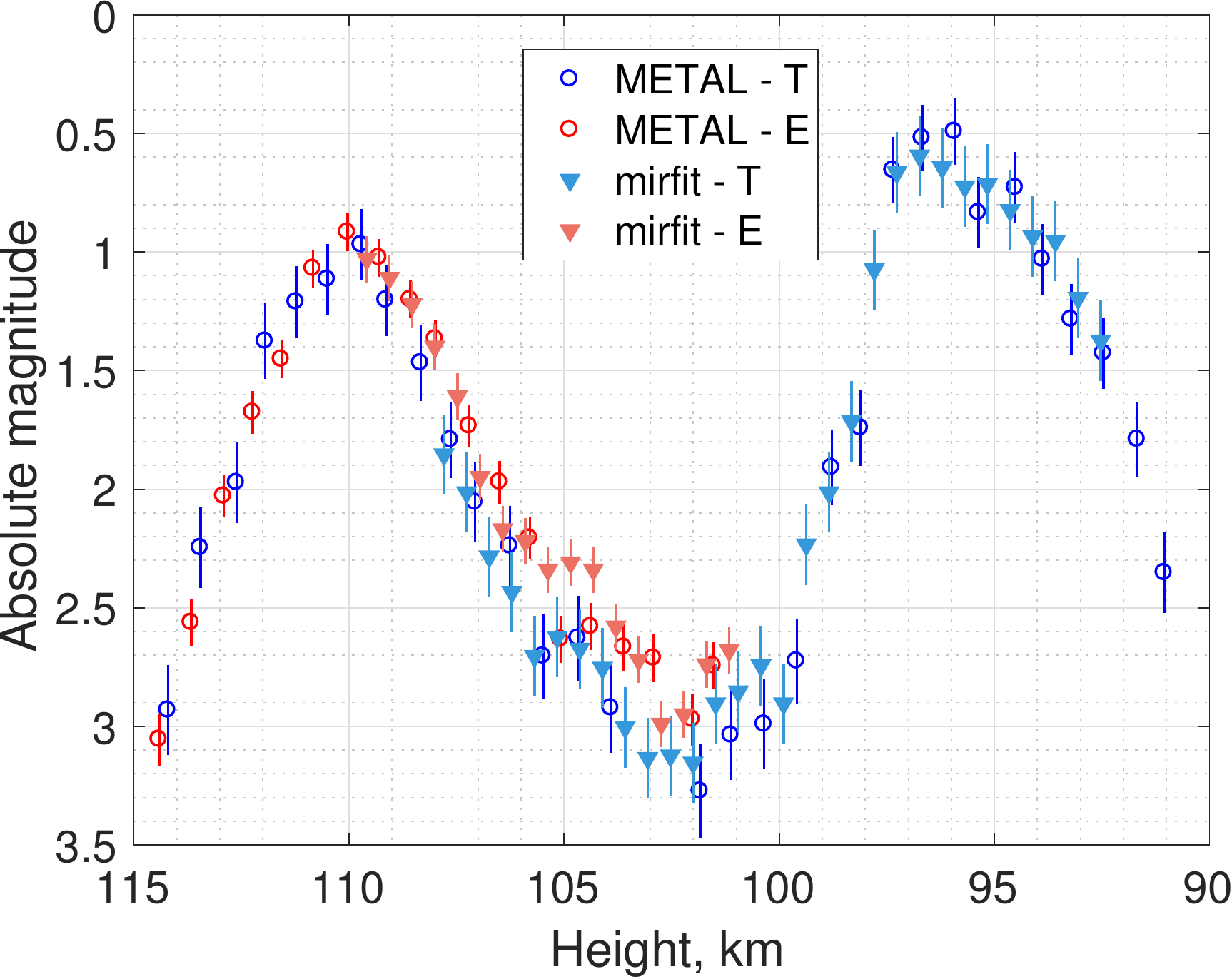}
    \caption{A smoothed peaked meteor light curve showing a greater decrease in brightness than most events, over a greater height range. The second peak was observed exclusively at the Tavistock station. This is meteor event 20111104\_0832320.}
    \label{fig:double_2}
\end{figure}

The events studied here can be classified into two broad categories, smooth peaked and sudden peaked. There are variations within those groups. Typically, the smooth peaked events show a decrease of half to one magnitude before the next peak begins to increase, as seen in Fig.~\ref{fig:double}. Some meteor events, however, show a much greater decrease in brightness, as seen in Fig.~\ref{fig:double_2}, or an unusual second peak shape, as shown in Fig.~\ref{fig:double_3}. Each meteor light curve studied can be found in the Appendix, with their corresponding mirfit brightness profile observations overplotted.

\begin{figure}
    \centering
    \includegraphics[width=\columnwidth]{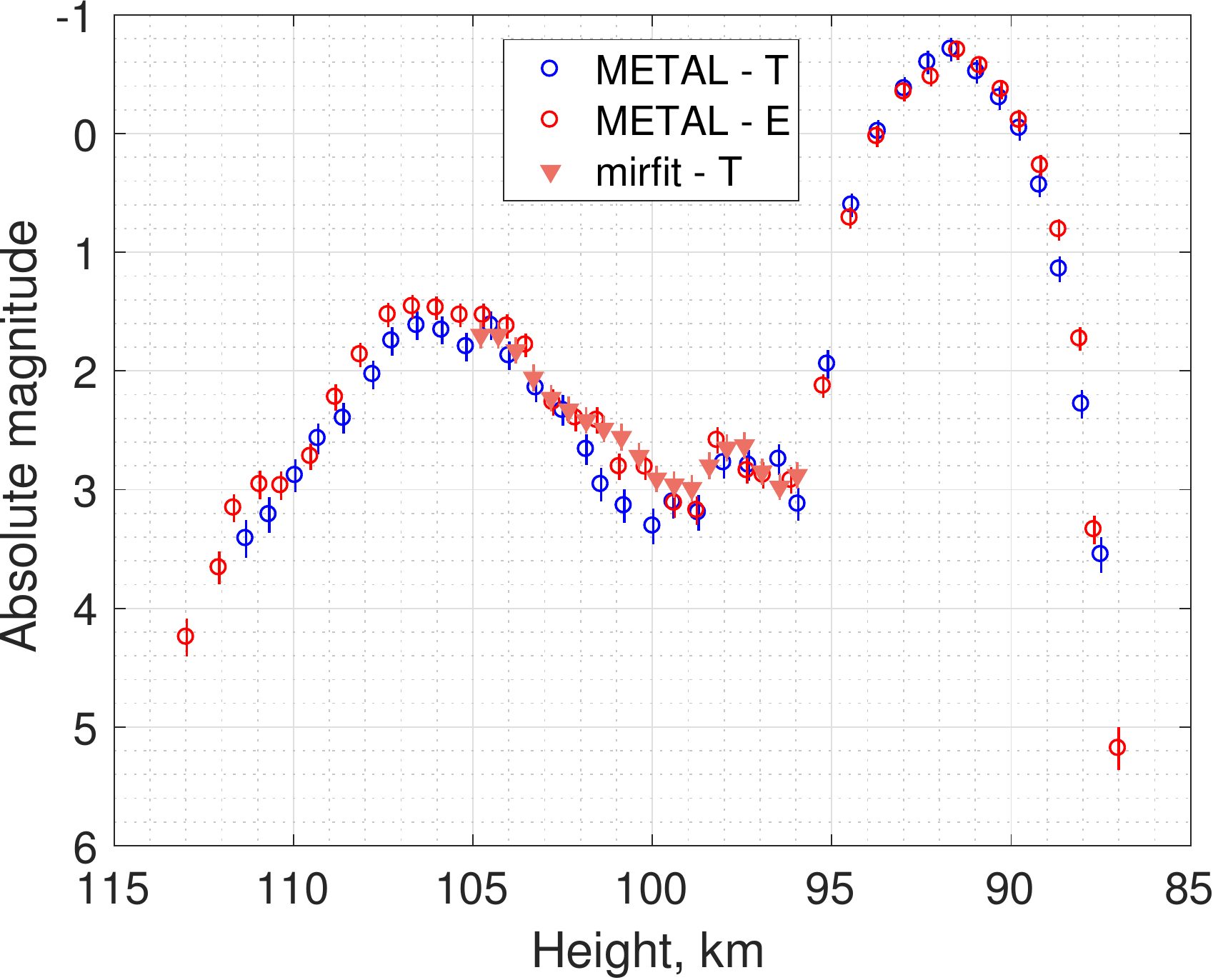}
    \caption{A smoothed peaked meteor light curve (event 20160808\_083545)} showing an unusual shape (and possibly a third peak nestled between the two distinct peaks). Only one narrow-field camera tracked this meteor, and it did not observe the second peak.
    \label{fig:double_3}
\end{figure}

\section{Conclusions}
We analysed 21 double peaked meteor light curves and found that five of those meteor events showed `sudden peaked' curves, and the remaining 16 showed `smooth peaked' light curves. The sudden peaked meteor events were on asteroidal orbits, and mostly showed fragmentation, whereas the smooth peaked curves were on cometary orbits and showed mostly faint wake in their narrow-field observations. The difference in origin between the two types of light curve is striking but difficult to explain; the differences in the appearance of the sudden peaked and smooth peaked curves may be related to the composition and structure of the meteoroids. This is particularly interesting in light of the results of \cite{Subasinghe2016}, who found no difference in the light curve shapes or fragmentation behaviour between dynamically asteroidal and cometary meteoroids. The sample size here is small, but the near complete separation of the two populations is still suggestive. 

The sudden peaked light curves might be explained by a sudden release of grains, as used in the analysis done by \cite{Roberts2014}, since many of those meteors showed fragmentation. On the other hand, this burst of fragmentation seems to be an unlikely mechanism for many of the smooth peaked events analysed here, as indicated by the time evolution of the narrow-field observations, which are consistent only with the shedding of small fragments as modelled in \cite{CampbellBrown2017}. Spectral observations were not collected for this study, but will be in the future; those observations will confirm whether differential ablation is occurring, and whether a particular elemental composition is responsible for each peak. Ablation modelling using only inhomogeneous compositions without fragmentation was able to produce double peaked light curves if the two components had densities or heats of ablation which differed by more than a factor of 2, but the width of the light curves tended to be larger than the observed widths, implying that some form of fragmentation must play a role, even when composition is the dominant effect. In the future, ablation modelling of meteor events with double peaked light curves using a two component model that includes fragmentation mechanisms may be attempted to investigate possible mechanisms, and physical properties such as grain sizes and distributions, structure and composition, and mass. 

\section*{Acknowledgements}

This work was supported by the NASA Meteoroid Environment Office [grant NNX11AB76A]. Thanks to  J. Gill, Z. Krzeminski, and D. Vida for their assistance with data collection and reduction. The authors thank J. Borovi\v{c}ka for helpful comments.


\bibliographystyle{mnras}
\bibliography{main} 



\appendix \label{appendix}

\section{Sudden peaked meteor events}

\begin{figure}
    \centering
    \includegraphics[width=0.95\columnwidth]{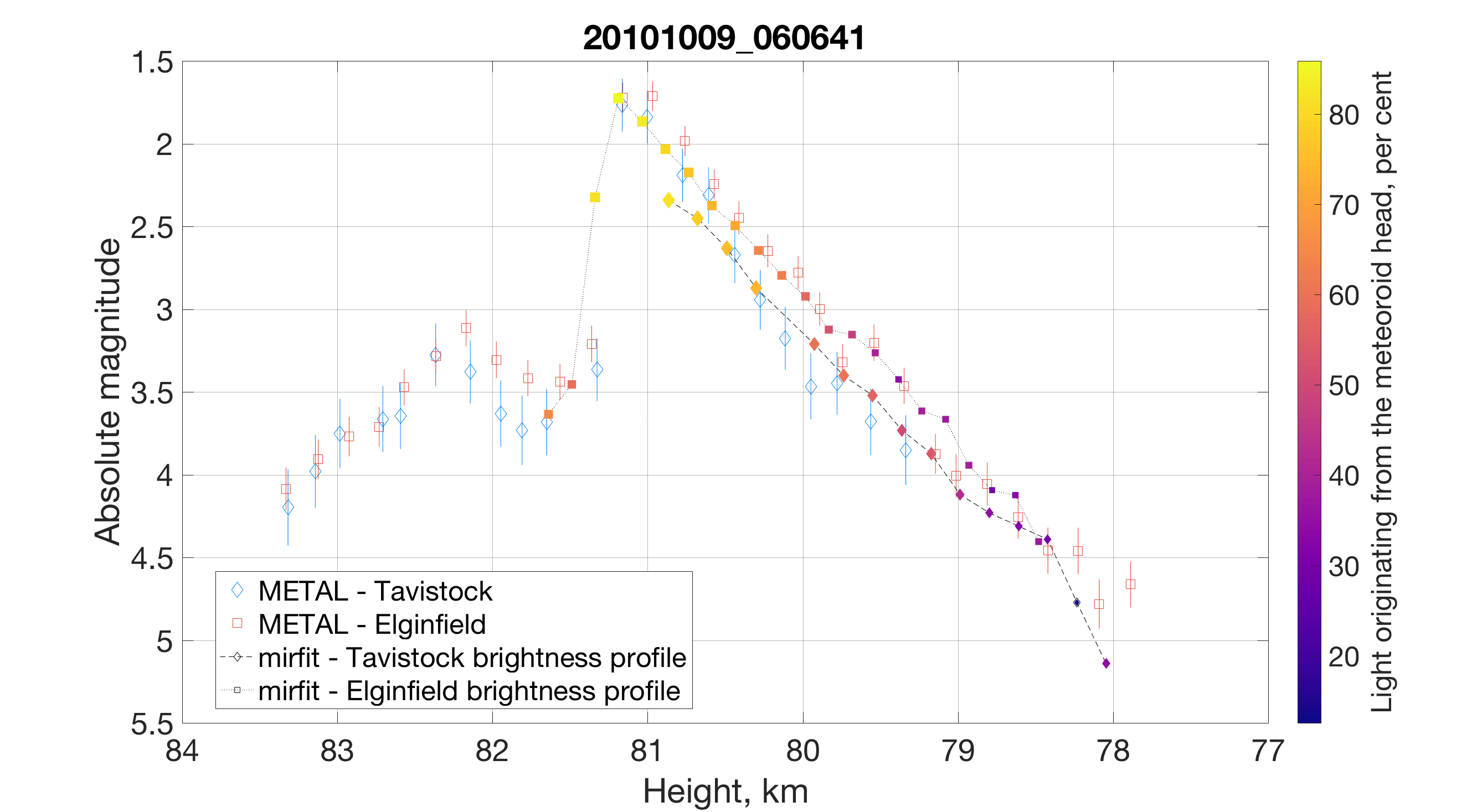}
    \caption{20101009\_060641}
\end{figure}

\begin{figure}
    \centering
    \includegraphics[width=0.95\columnwidth]{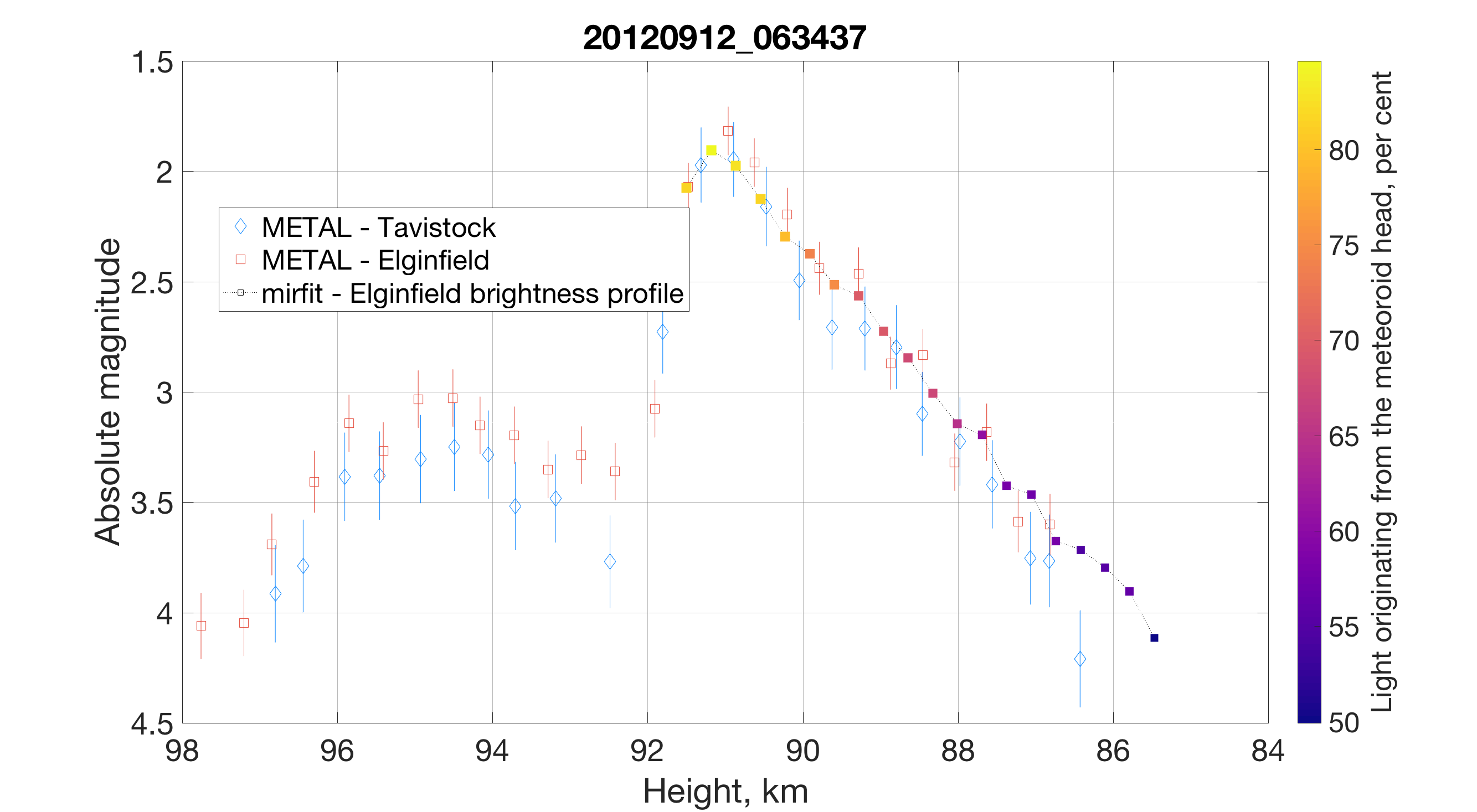}
    \caption{20120912\_063437}
\end{figure}

\begin{figure}
    \centering
    \includegraphics[width=0.95\columnwidth]{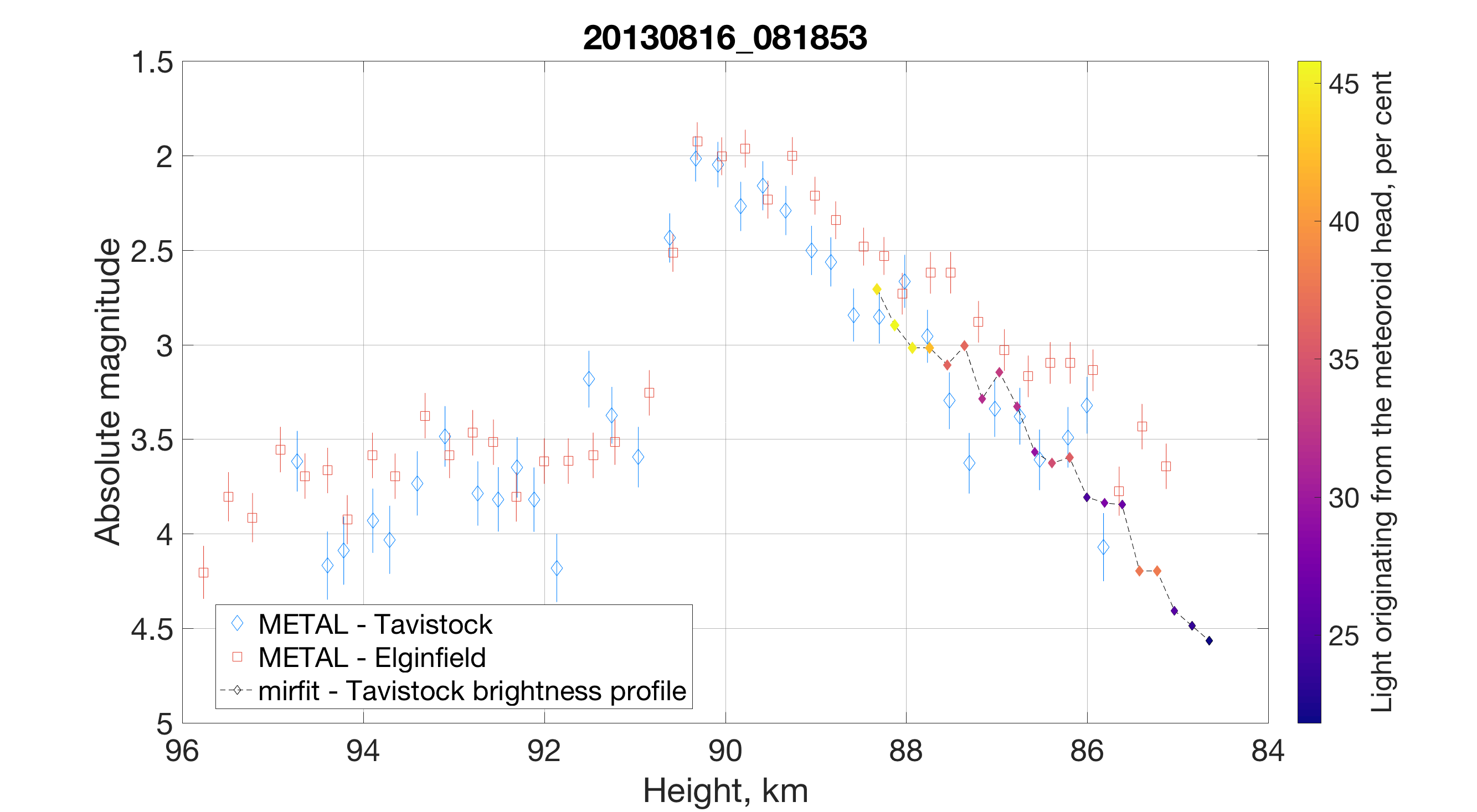}
    \caption{20130816\_081853}
    \label{fig:A3}
\end{figure}

\begin{figure}
    \centering
    \includegraphics[width=0.95\columnwidth]{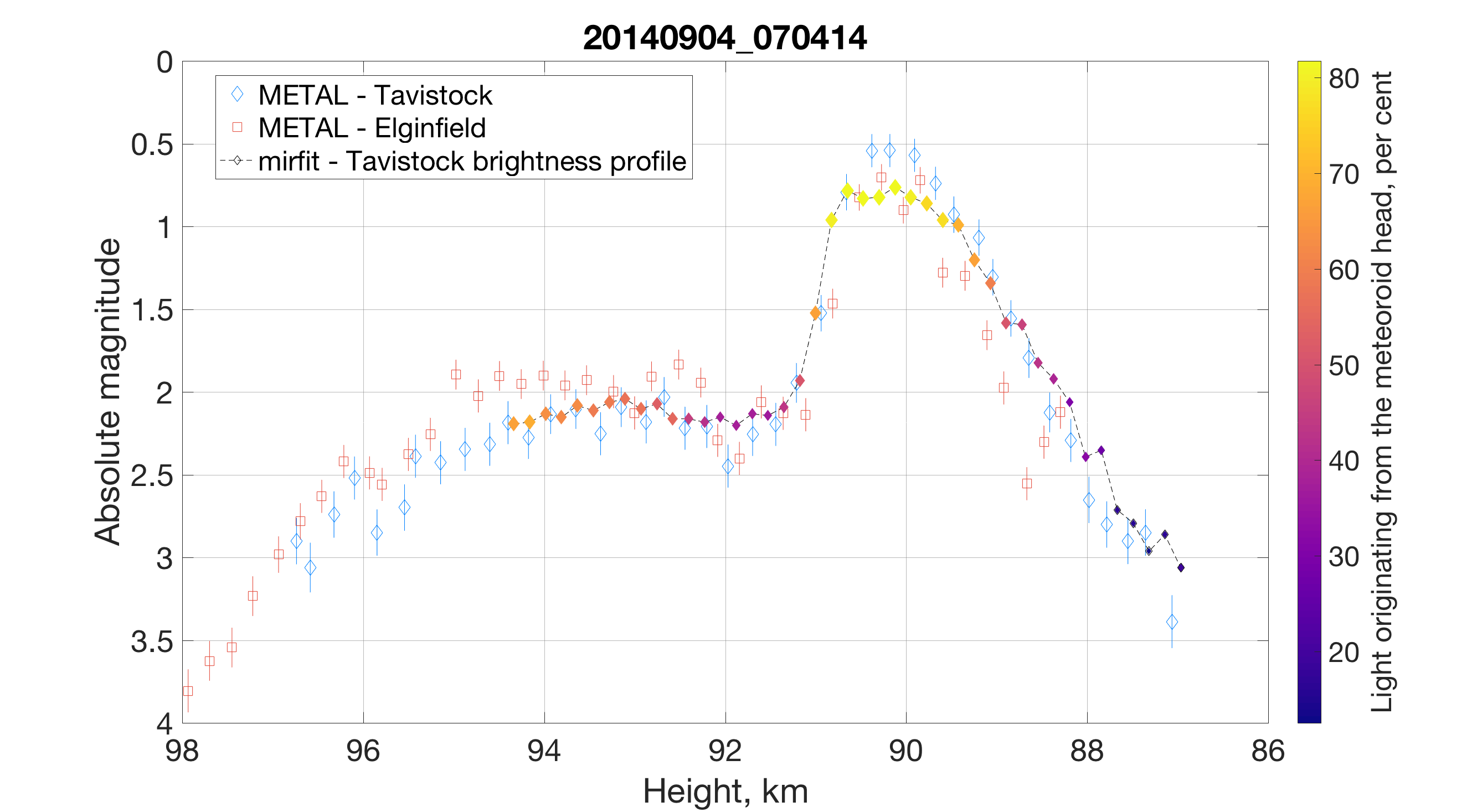}
    \caption{20140904\_070414}
    \label{fig:A4}
\end{figure}

\begin{figure}
    \centering
    \includegraphics[width=0.95\columnwidth]{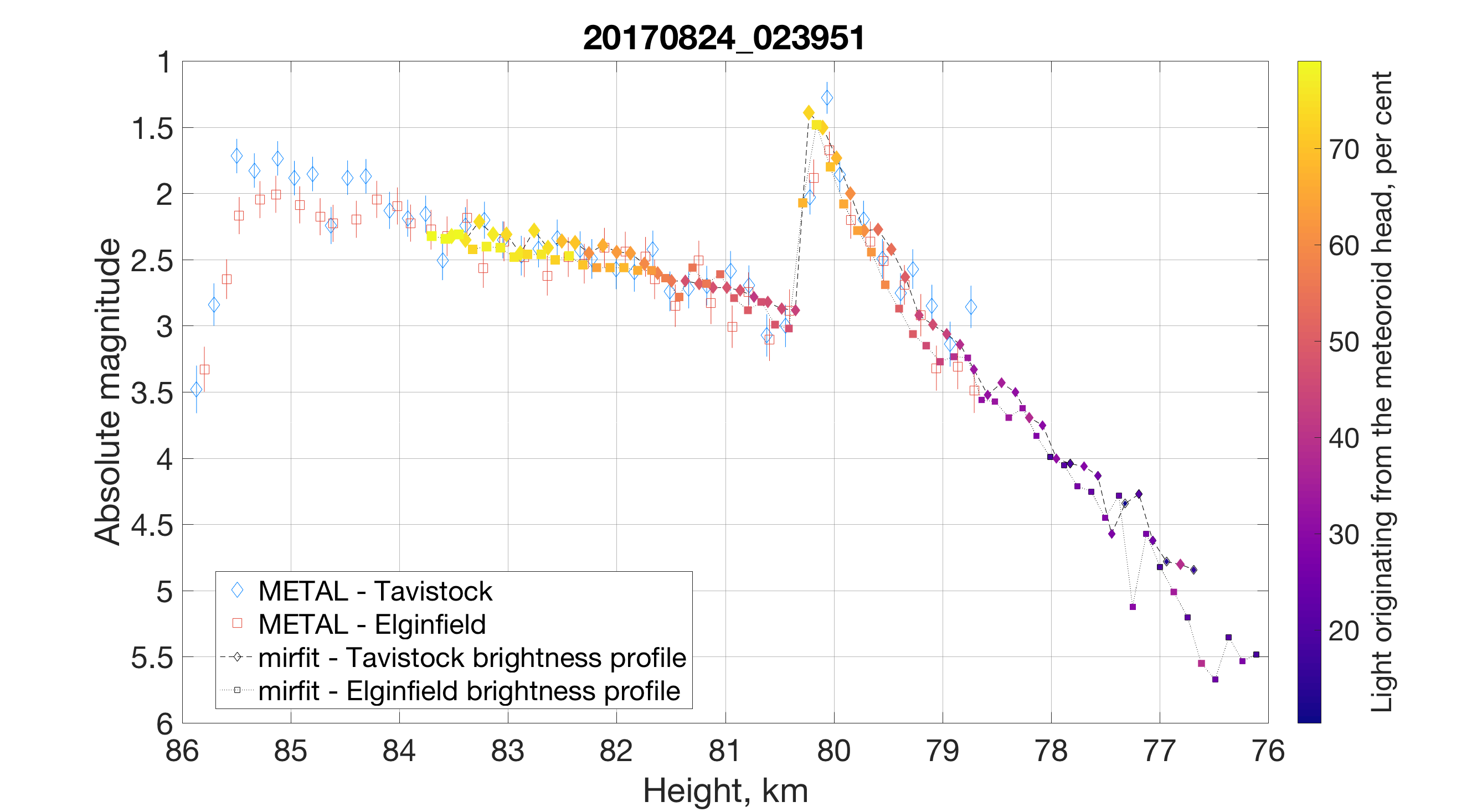}
    \caption{20170824\_023951}
\end{figure}

\section{Smooth peaked meteor events}

\begin{figure}
    \centering
    \includegraphics[width=0.95\columnwidth]{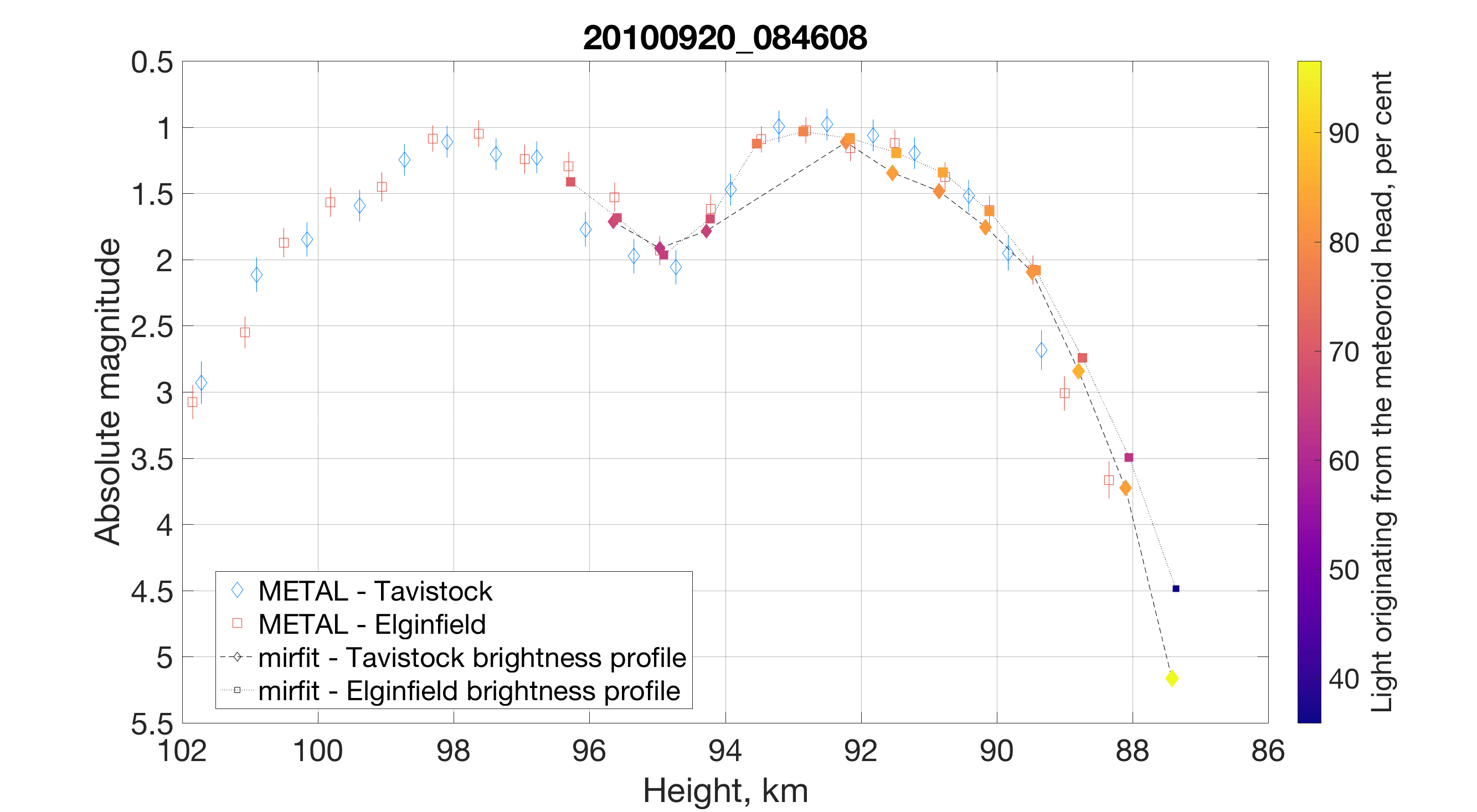}
    \caption{20100920\_084608}
\end{figure}

\begin{figure}
    \centering
    \includegraphics[width=0.95\columnwidth]{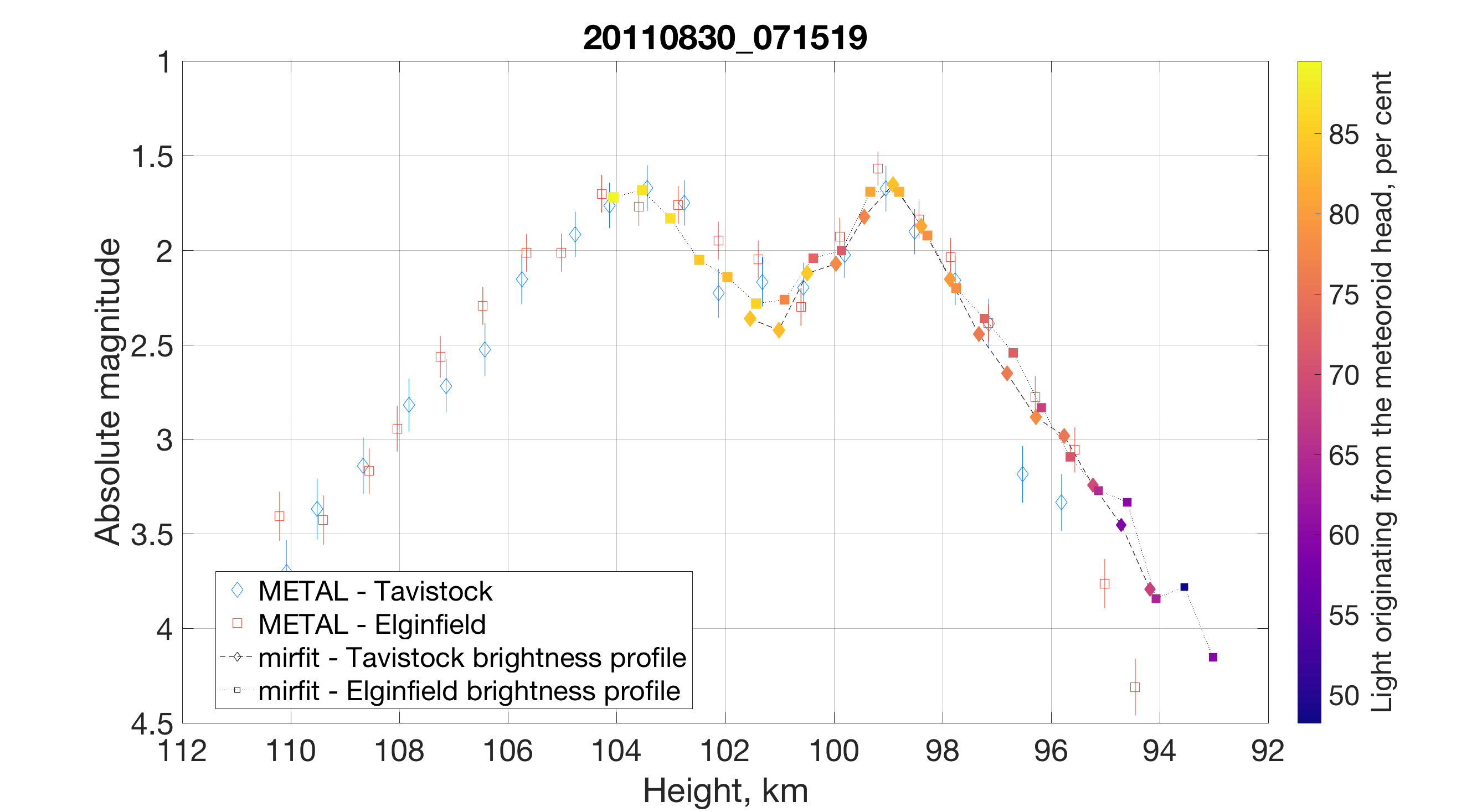}
    \caption{20110830\_071519}
\end{figure}

\begin{figure}
    \centering
    \includegraphics[width=0.95\columnwidth]{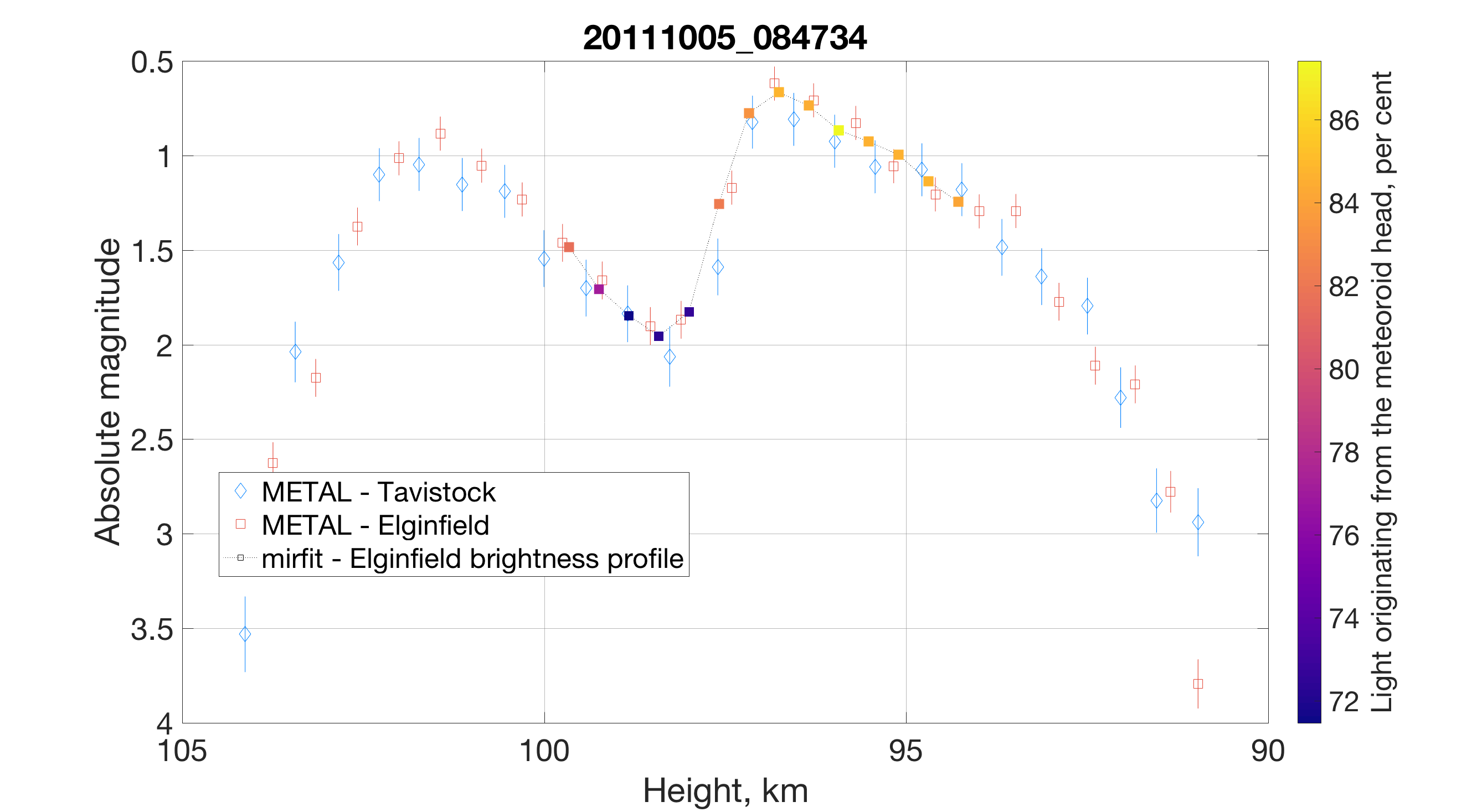}
    \caption{20111005\_084734}
\end{figure}

\begin{figure}
    \centering
    \includegraphics[width=0.95\columnwidth]{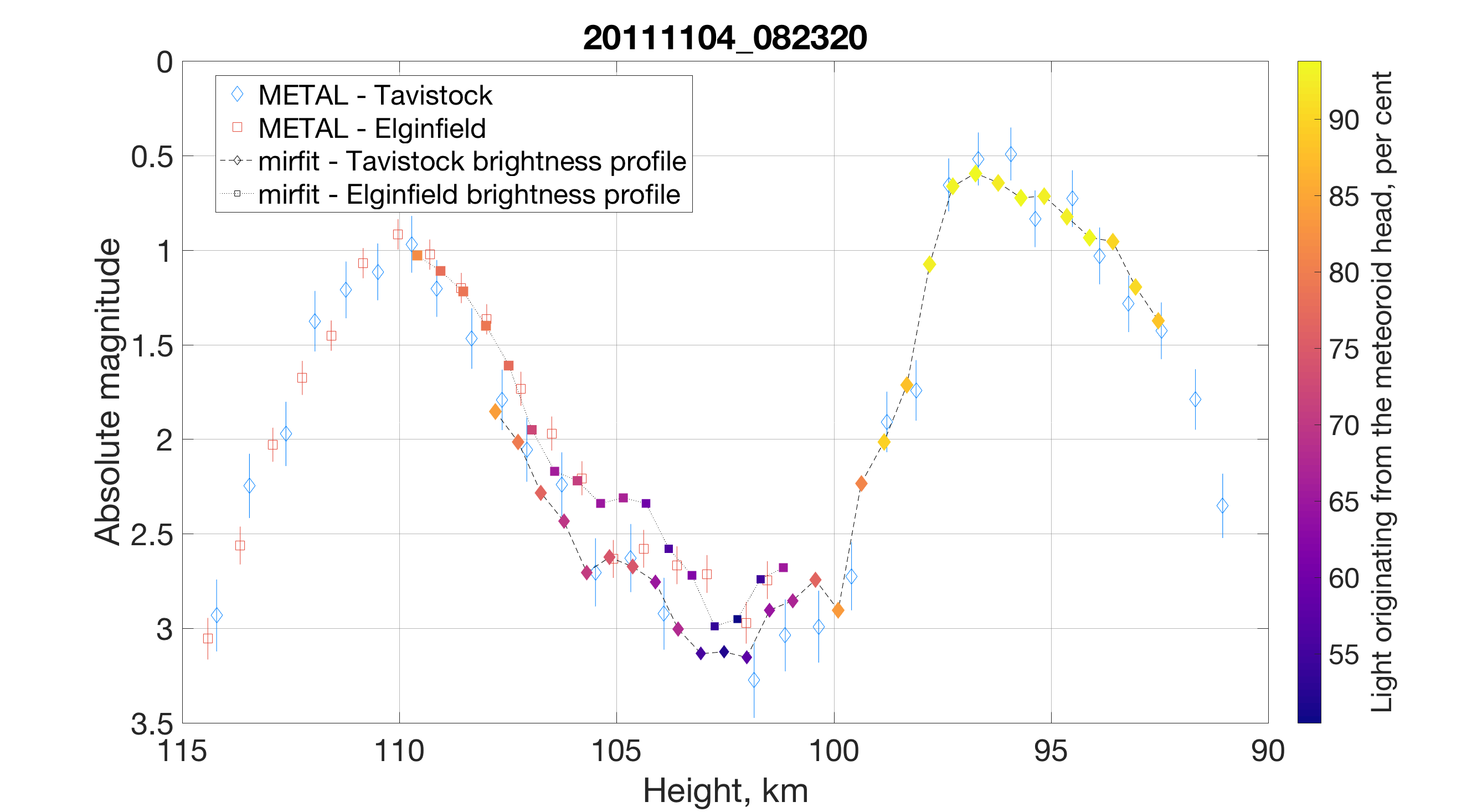}
    \caption{20111104\_082320}
\end{figure}

\begin{figure}
    \centering
    \includegraphics[width=0.95\columnwidth]{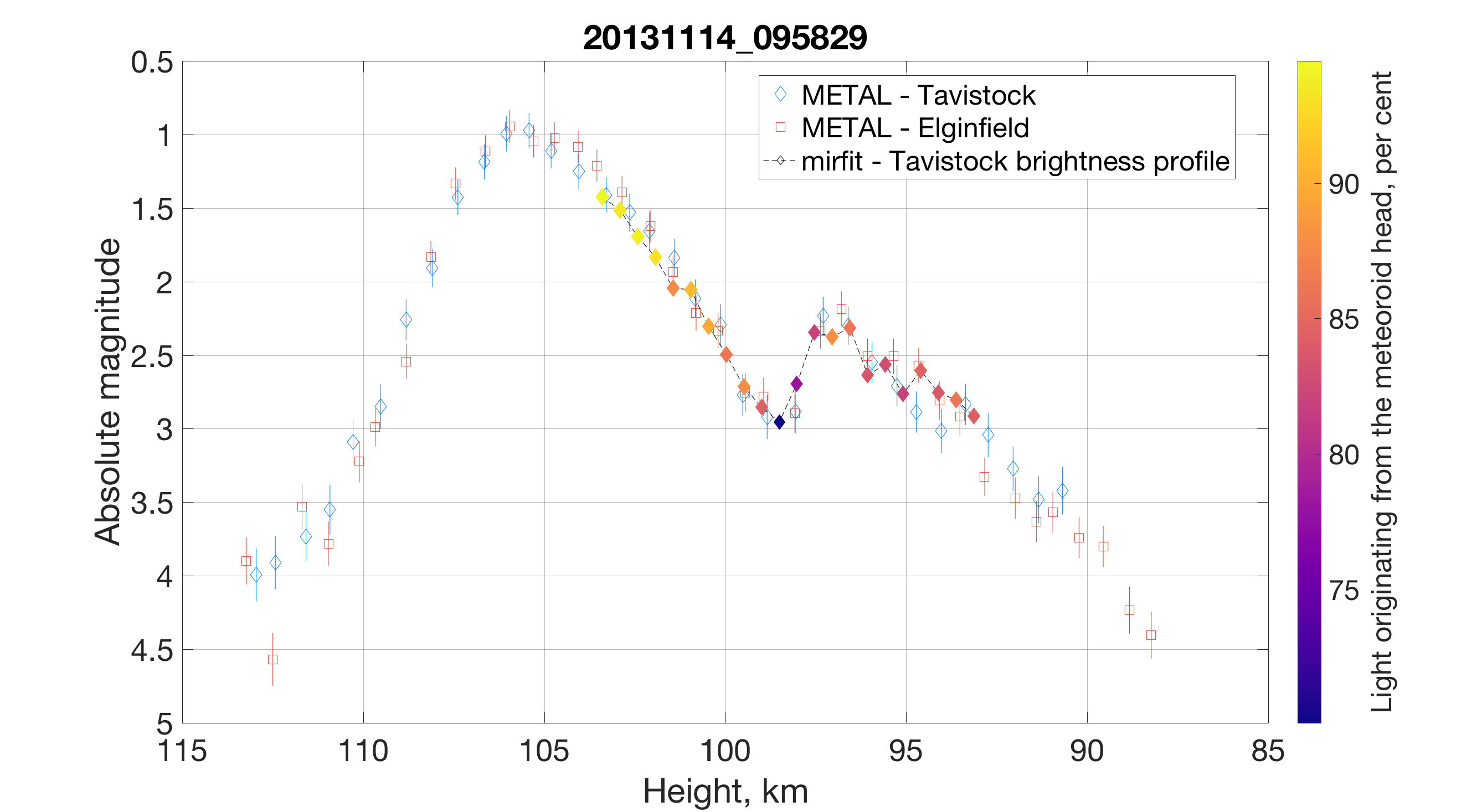}
    \caption{20131114\_095829}
\end{figure}

\begin{figure}
    \centering
    \includegraphics[width=0.95\columnwidth]{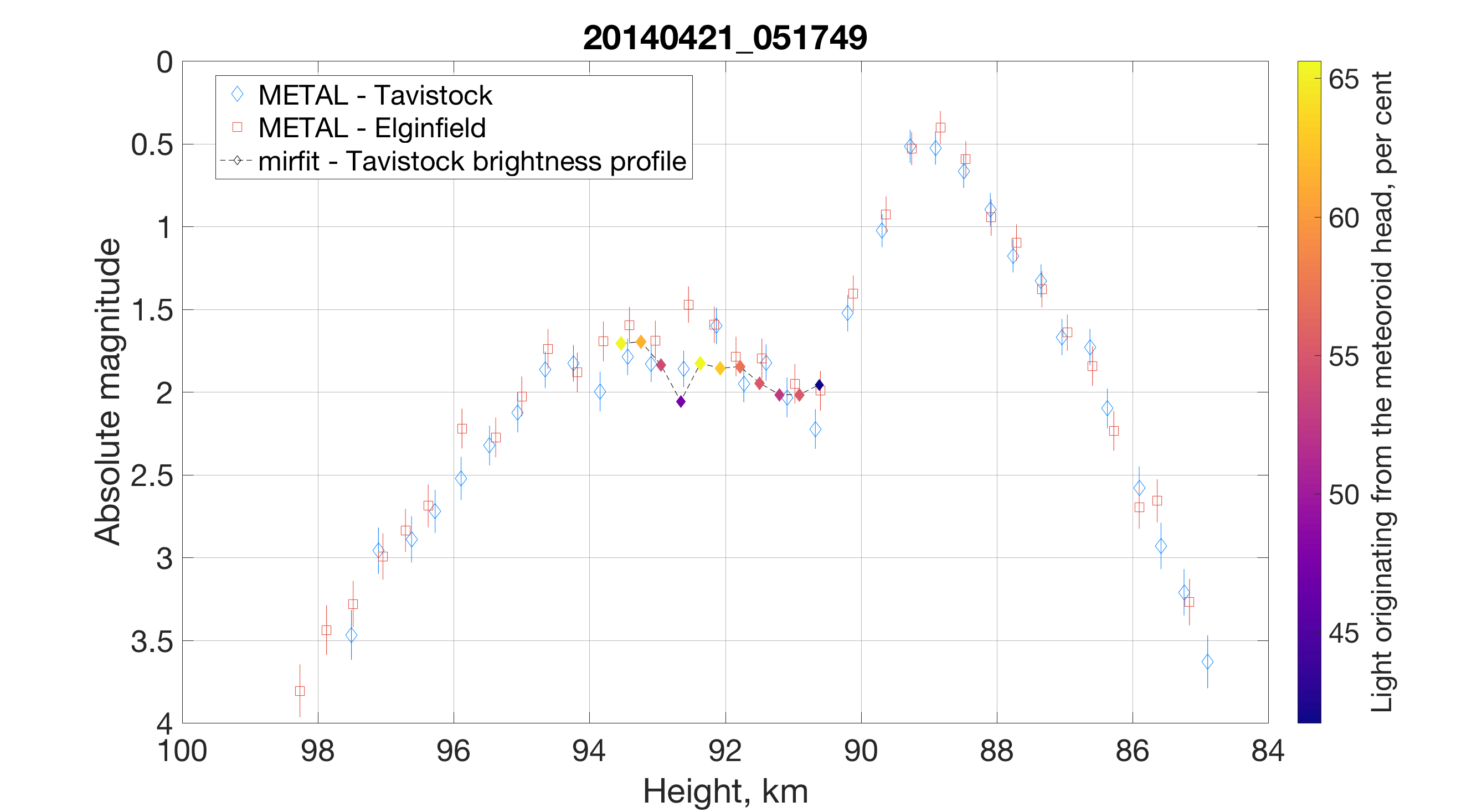}
    \caption{20140421\_051749}
\end{figure}

\begin{figure}
    \centering
    \includegraphics[width=0.95\columnwidth]{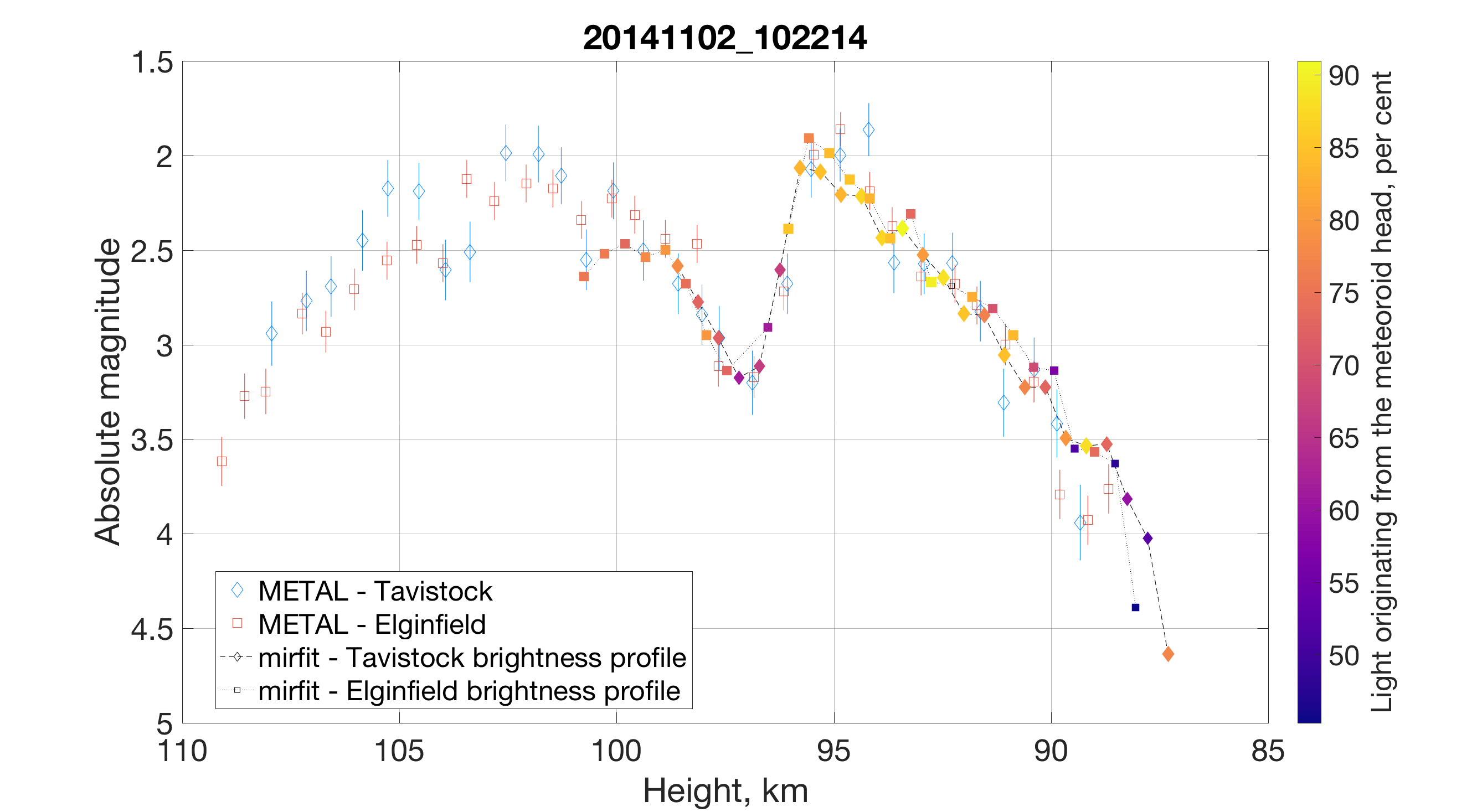}
    \caption{20141102\_102214}
\end{figure}

\begin{figure}
    \centering
    \includegraphics[width=0.95\columnwidth]{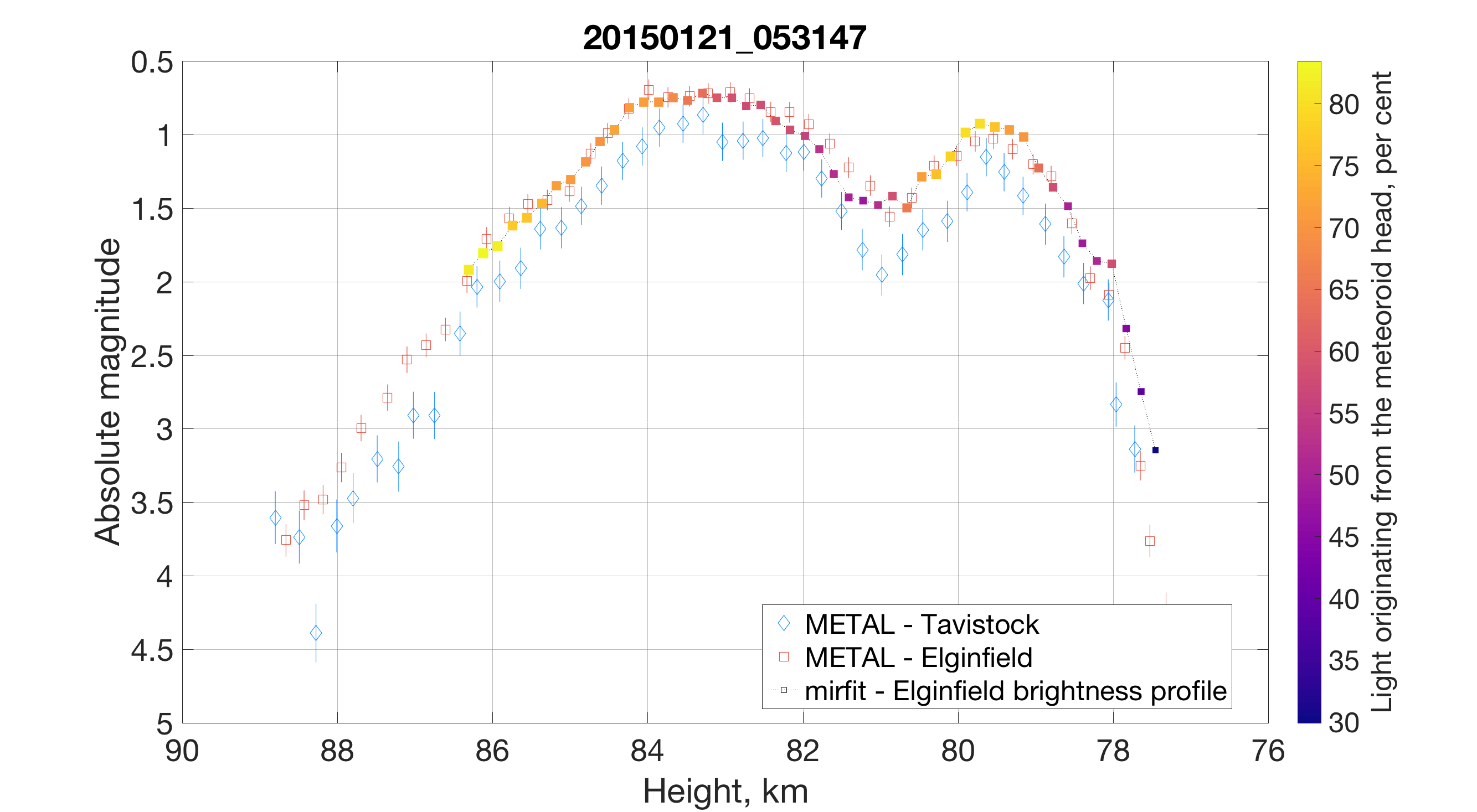}
    \caption{20150121\_053147}
\end{figure}

\begin{figure}
    \centering
    \includegraphics[width=0.95\columnwidth]{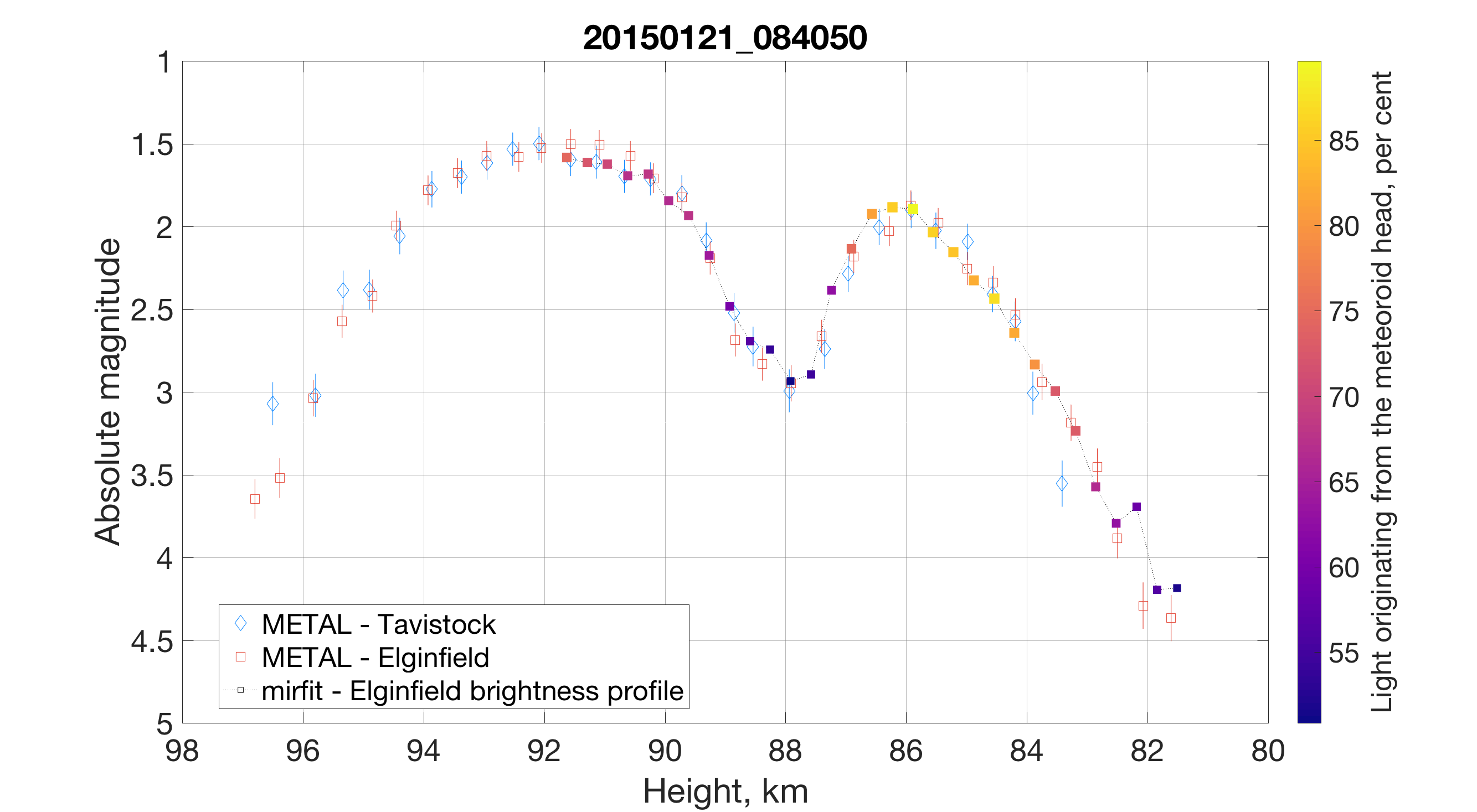}
    \caption{20150121\_084050}
\end{figure}

\begin{figure}
    \centering
    \includegraphics[width=0.95\columnwidth]{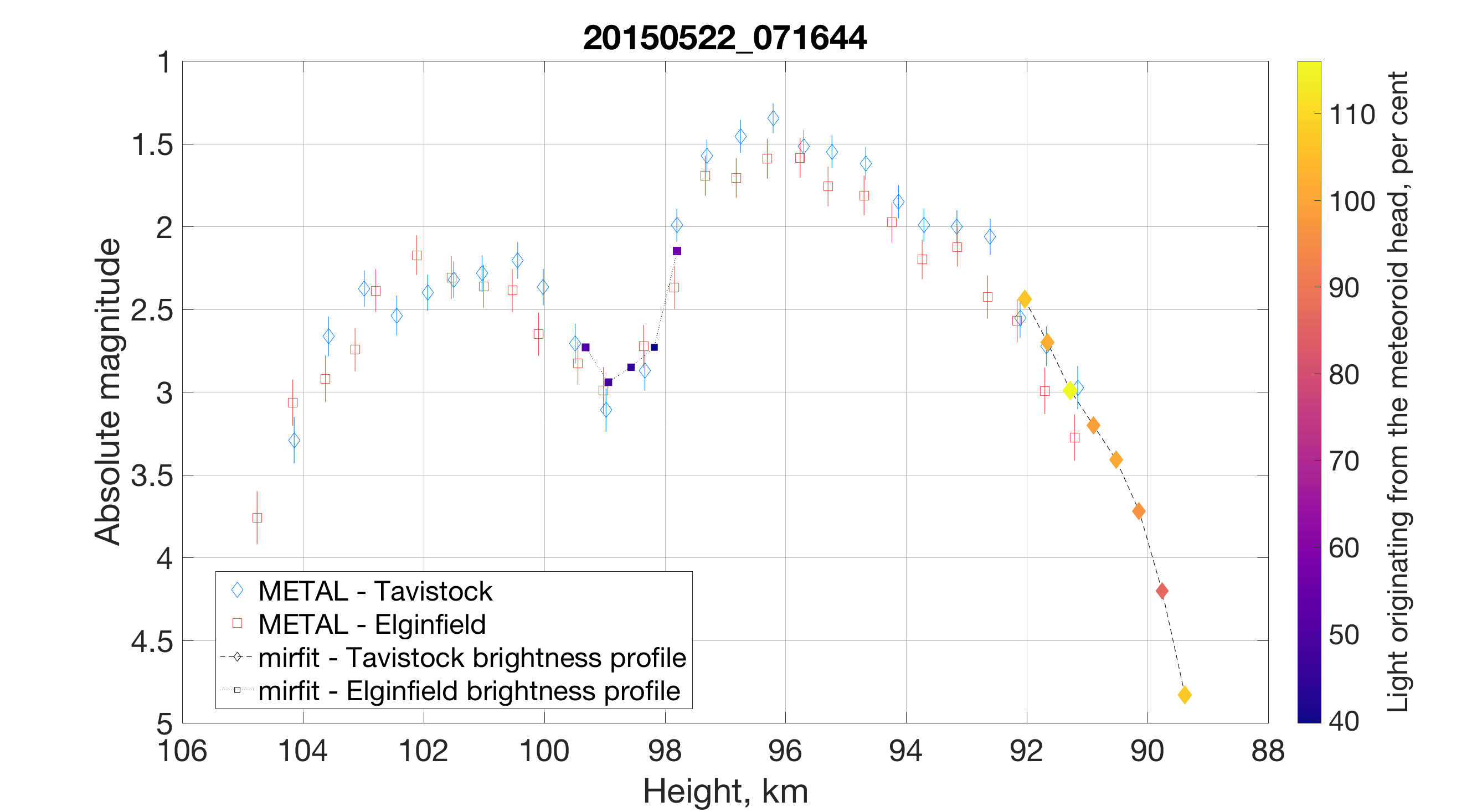}
    \caption{20150522\_071644}
\end{figure}

\begin{figure}
    \centering
    \includegraphics[width=0.95\columnwidth]{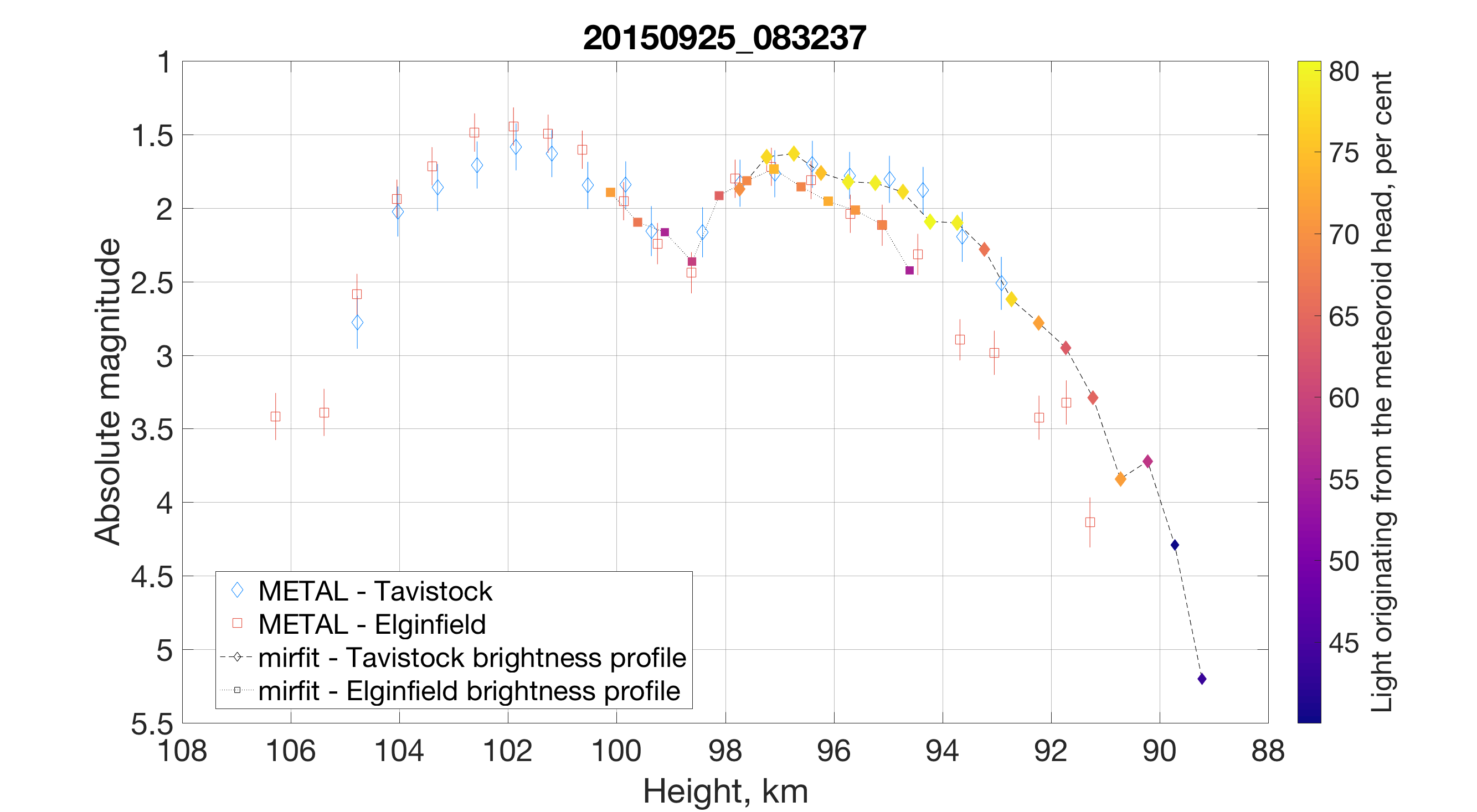}
    \caption{20150925\_083237}
\end{figure}

\begin{figure}
    \centering
    \includegraphics[width=0.95\columnwidth]{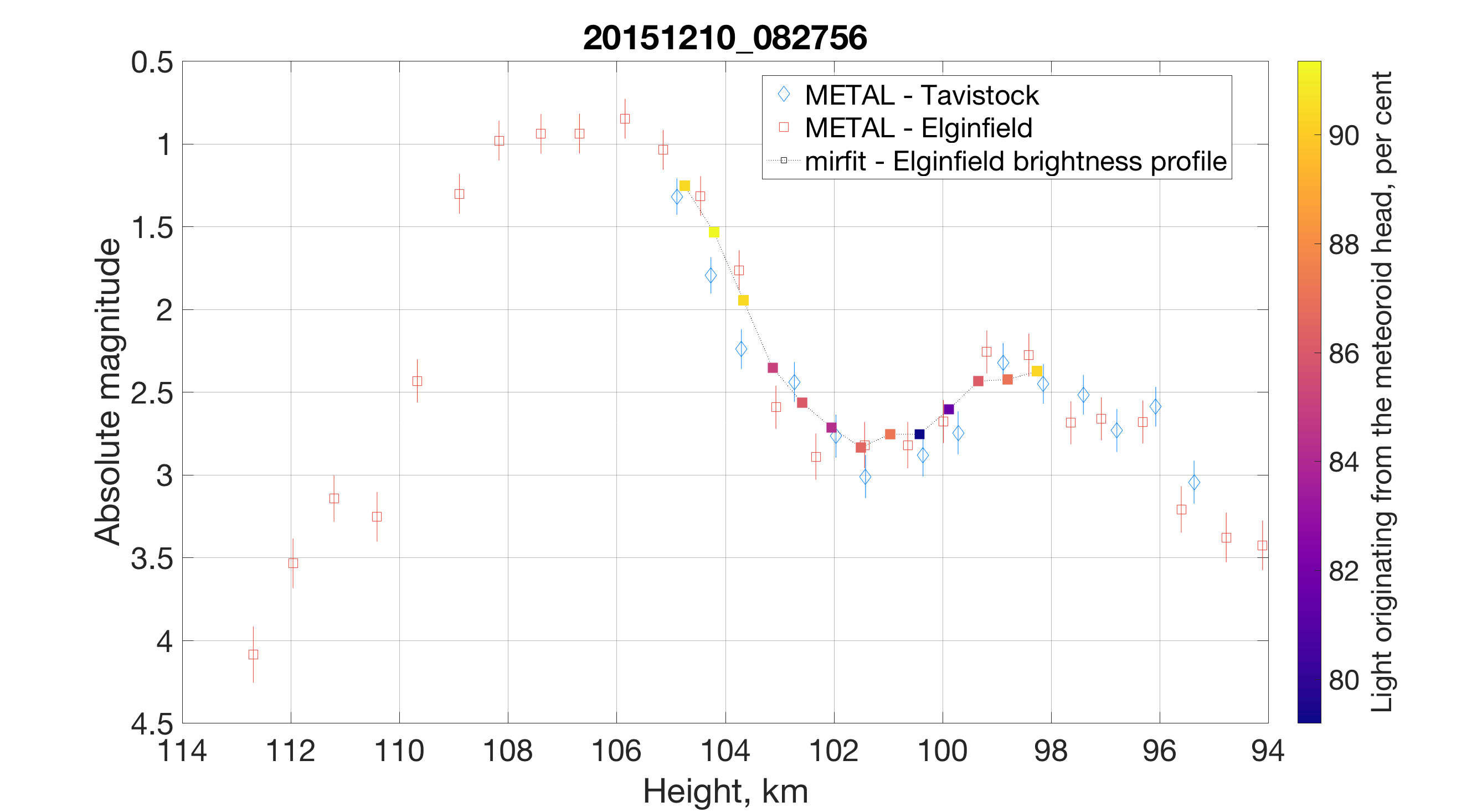}
    \caption{20151210\_082756}
\end{figure}

\begin{figure}
    \centering
    \includegraphics[width=0.95\columnwidth]{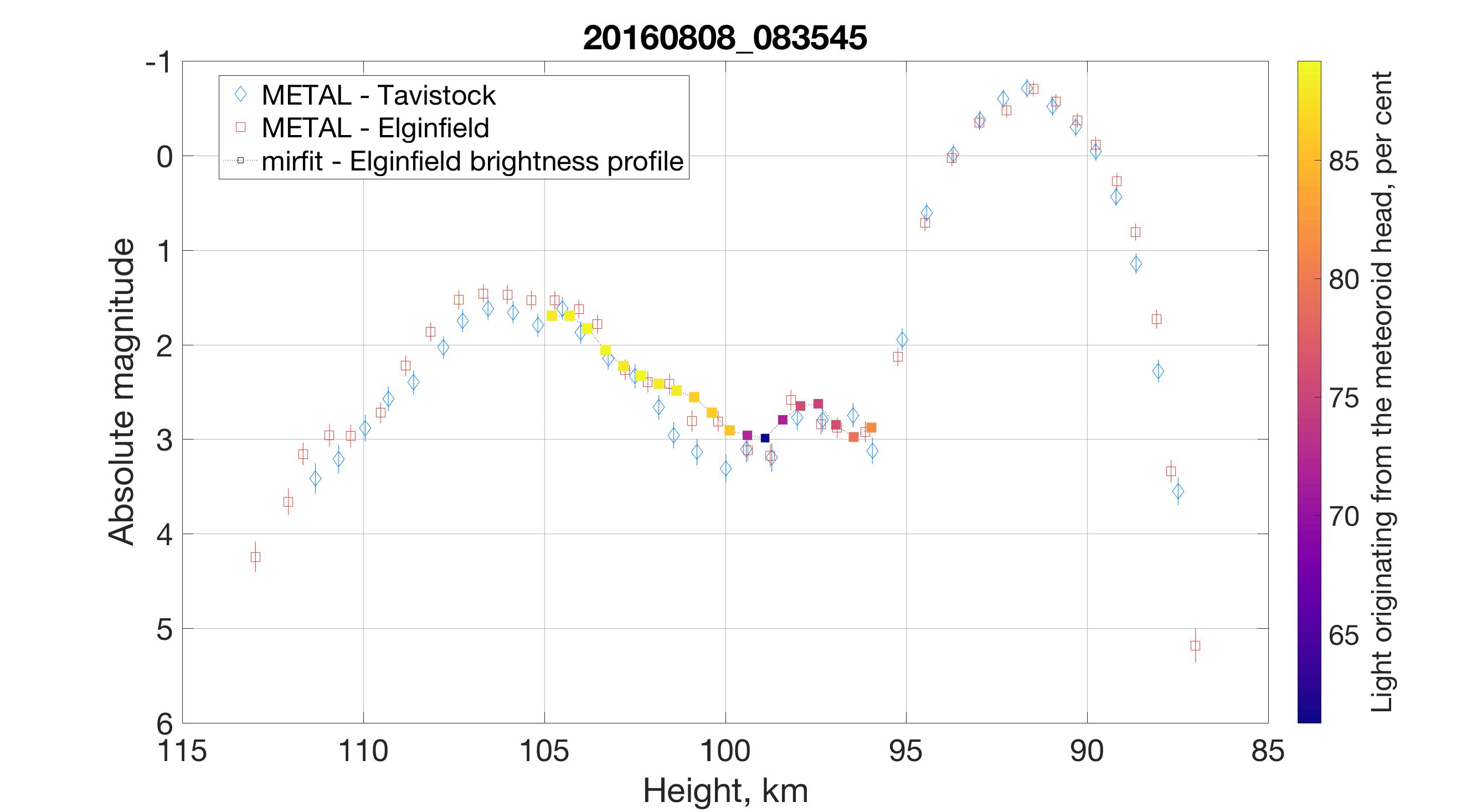}
    \caption{20160808\_083545}
\end{figure}

\begin{figure}
    \centering
    \includegraphics[width=0.95\columnwidth]{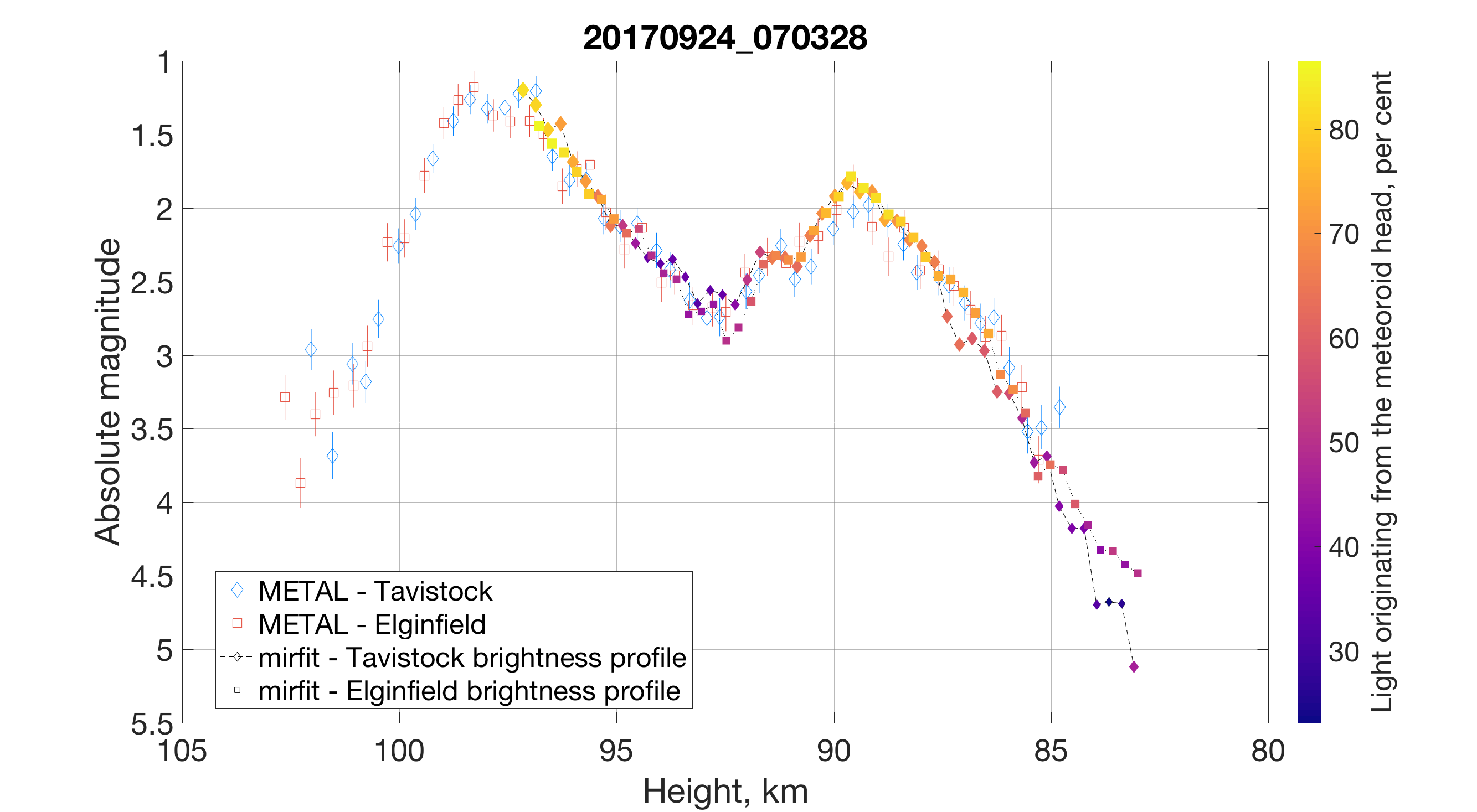}
    \caption{20170924\_070328}
\end{figure}

\begin{figure}
    \centering
    \includegraphics[width=0.95\columnwidth]{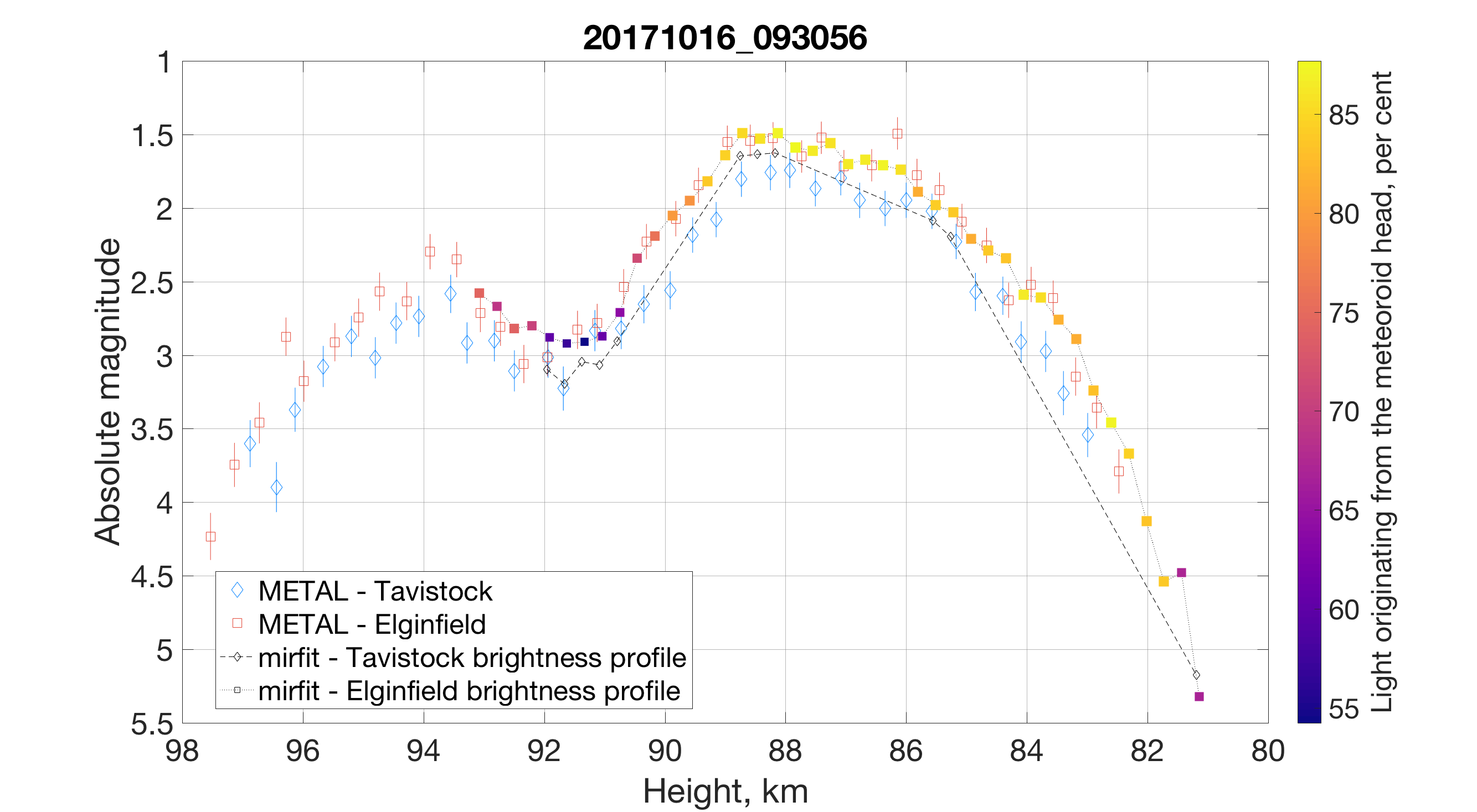}
    \caption{20171016\_093056}
\end{figure}

\begin{figure}
    \centering
    \includegraphics[width=0.95\columnwidth]{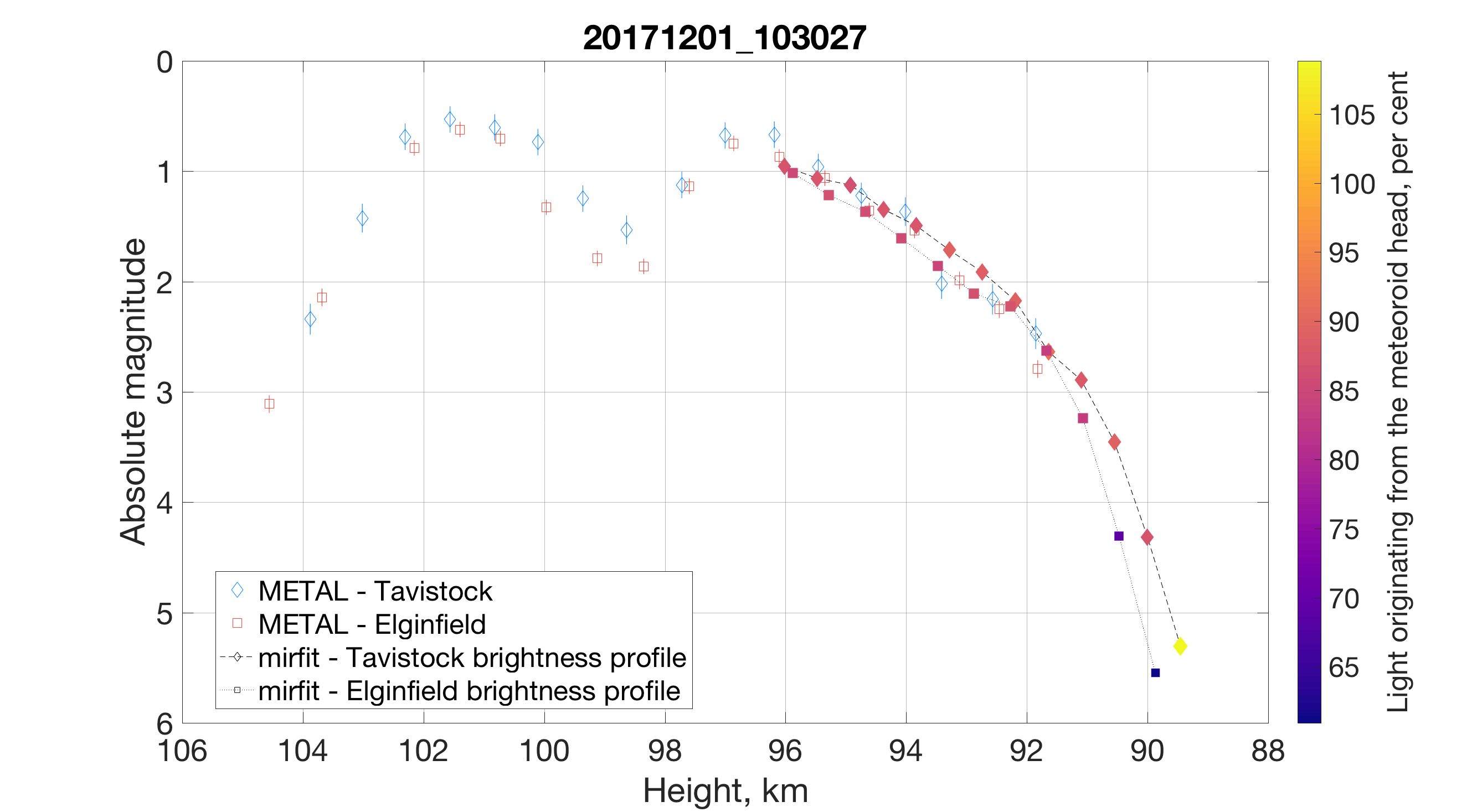}
    \caption{20171201\_103027}
\end{figure}


\bsp	
\label{lastpage}
\end{document}